\documentclass{article}
\usepackage[top = 1.0in, bottom = 1.0in, left = 1.0in, right = 1.0in]{geometry}
\usepackage[parfill]{parskip}    
\usepackage{graphicx}
\usepackage{amssymb}
\usepackage{amsmath}
\usepackage{color} 

\usepackage{authblk}
\usepackage[
	backend=bibtex,
	style=ieee, 
	sorting=none,				
	url=false, 					
	doi=false, 					
	isbn=false, 				
	eprint=false, 			    
	maxbibnames=9, 			    
	firstinits=true,
]{biblatex}
\usepackage[
	aboveskip=1pt,
	labelfont=bf,
	labelsep=period,
	justification=raggedright,
	singlelinecheck=off,
	font=small
]{caption}

\usepackage{afterpage}

\newcommand{\vect}[1]{\vec{#1}}

\newcommand{\rate}[3]{{#1}_{#2}^{#3}}

\newcommand{\deriv}[2][{}]{\frac{d #1}{d #2}}
\newcommand{\pderiv}[2][{}]{\frac{\partial #1}{\partial #2}}
\newcommand{\psecderiv}[2][{}]{\frac{\partial{^2} #1}{\partial #2{^2}}}

\newcommand{\fig}[1]{Figure~\ref{#1}}

\newcommand{\eq}[1]{Eq.~(\ref{#1})}

\newcommand{\beginsupplement}{
				\setcounter{section}{0} 
        \renewcommand{\thesection}{S\arabic{section}}%
        \setcounter{table}{0} 
        \renewcommand{\thetable}{S\arabic{table}}%
        \setcounter{figure}{0} 
        \renewcommand{\thefigure}{S\arabic{figure}}%
        \setcounter{equation}{0} 
        \renewcommand{\theequation}{S\arabic{equation}}%
     }

\title{Reconciling Kinetic and Equilibrium Models of Bacterial Transcription}

\author[1]{Muir Morrison}
\author[2]{Manuel Razo-Mejia}
\author[1, 2, *]{Rob Phillips}

\affil[1]{Department of Physics, California Institute of Technology, Pasadena,
CA 91125, USA}
\affil[2]{Division of Biology and Biological Engineering, California Institute
of Technology, Pasadena, CA 91125, USA}
\affil[*]{Correspondence: phillips@pboc.caltech.edu}

\date{}  
\begin{document}

\maketitle 

\addtocontents{toc}{\protect\setcounter{tocdepth}{-1}}

	\begin{refsegment}
    \defbibfilter{notother}{not segment=\therefsegment}
\begin{abstract}
The study of transcription remains one of the centerpieces of modern biology
with implications in settings from development to metabolism to evolution to
disease. Precision measurements using a host of different techniques including
fluorescence and sequencing readouts have raised the bar for what it means to
quantitatively understand transcriptional regulation. In particular our
understanding of the simplest genetic circuit is sufficiently refined both
experimentally and theoretically that it has become possible to carefully
discriminate between different conceptual pictures of how this regulatory system
works. This regulatory motif, originally posited by Jacob and Monod in the
1960s, consists of a single transcriptional repressor binding to a promoter site
and inhibiting transcription. In this paper, we show how seven distinct models
of this so-called simple-repression motif, based both on equilibrium and kinetic
thinking, can be used to derive the predicted levels of gene expression and
shed light on the often surprising past success of the equilbrium models. These
different models are then invoked to confront a variety of different data on mean,
variance and full gene expression distributions, illustrating the extent to which
such models can and cannot be distinguished, and suggesting a two-state model with a
distribution of burst sizes as the most potent of the seven for describing the
simple-repression motif.
\end{abstract}
\section{Introduction}

Gene expression presides over much of the most important dynamism of living
organisms. The level of expression of batteries of different genes is altered as
a result of spatiotemporal cues that integrate chemical, mechanical and other
types of signals. The original repressor-operator model conceived by Jacob and
Monod in the context of bacterial metabolism has now been transformed into the
much broader subject of gene regulatory networks in living organisms of all
kinds~\cite{Jacob1961, Britten1969, Ben-TabouDe-Leon2007}. One of the remaining
outstanding challenges to have emerged in the genomic era is our continued
inability to predict the regulatory consequences of different regulatory
architectures, i.e. the arrangement and affinity of binding sites for
transcription factors and RNA polymerases on the DNA. This challenge stems first
and foremost from our ignorance about what those architectures even are, with
more than 60\% of the genes even in an ostensibly well understood organism such
as {\it E. coli} having no regulatory insights at all~\cite{Rydenfelt2014-2,
Belliveau2018, Ghatak2019, Santos_Zavaleta2019}. But even once we have
established the identity of key transcription factors and their binding sites of
a given promoter architecture, there remains the predictive challenge of
understanding its input-output properties, an objective that can be met by a
myriad of approaches using the tools of statistical physics~\cite{Ackers1982,
Shea1985, Buchler2003, Vilar2003a, Vilar2003b, Bintu2005a, Bintu2005c,
Gertz2009, Sherman2012, Saiz2013, Ko1991, Peccoud1995, Record1996, Kepler2001,
Sanchez2008, Shahrezaei2008, Sanchez2011, Michel2010}. One route to such
predictive understanding is to focus on the simplest regulatory architecture and
to push the theory-experiment dialogue as far and as hard as it can be
pushed~\cite{Garcia2011, Phillips2019}. If we demonstrate that we can pass that
test by successfully predicting both the means and variance in gene expression
at the mRNA level,
then that provides a more solid foundation upon which to launch into more
complex problems - for instance, some of the previously unknown architectures
uncovered in~\cite{Belliveau2018} and~\cite{Ireland2020}.

To that end, in this paper we examine a wide variety of distinct models for the
simple repression regulatory architecture. This genetic architecture consists of
a DNA promoter regulated by a transcriptional repressor that binds to a single
binding site as developed in pioneering early work on the quantitative
dissection of transcription \cite{Oehler1994, Oehler1990}. All of the proposed
models coarse-grain away some of the important microscopic features of
this architecture that have been elucidated by generations of geneticists,
molecular biologists and biochemists.
One goal in exploring such coarse-grainings is to build towards the
future models of regulatory response that will be
able to serve the powerful predictive role needed to take synthetic biology from
a brilliant exercise in enlightened empiricism to a rational design framework as
in any other branch of engineering. More precisely, we want phenomenology in the
sense of coarse-graining away atomistic detail, but still retaining biophysical
meaning. For example, we are not satisfied with the strictly phenomenological
approach offered by the commonly used Hill functions. As argued
in~\cite{Frank2013}, Hill functions are ubiquitous precisely because they
coarse-grain away all biophysical details into inscrutable parameters. Studies
like~\cite{Razo-Mejia2018} have demonstrated that Hill functions are clearly
insufficient since each new situation requires a completely new set of
parameters. Such work requires a quantitative theory of how biophysical changes
at the molecular level propagate to input-output functions at the genetic
circuit level. In particular a key question
is: at this level of coarse-graining, what microscopic details do we need to
explicitly model, and how do we figure that out? For example, do we need to
worry about all or even any of the steps that individual RNA polymerases go
through each time they make a transcript? Turning the question around, can we
see any imprint of those processes in the available data? If the answer is no,
then those processes are irrelevant for our purposes. Forward modeling and
inverse (statistical inferential) modeling are necessary to tackle such
questions.

\begin{figure}[p]
\centering
\includegraphics[width=\textwidth]{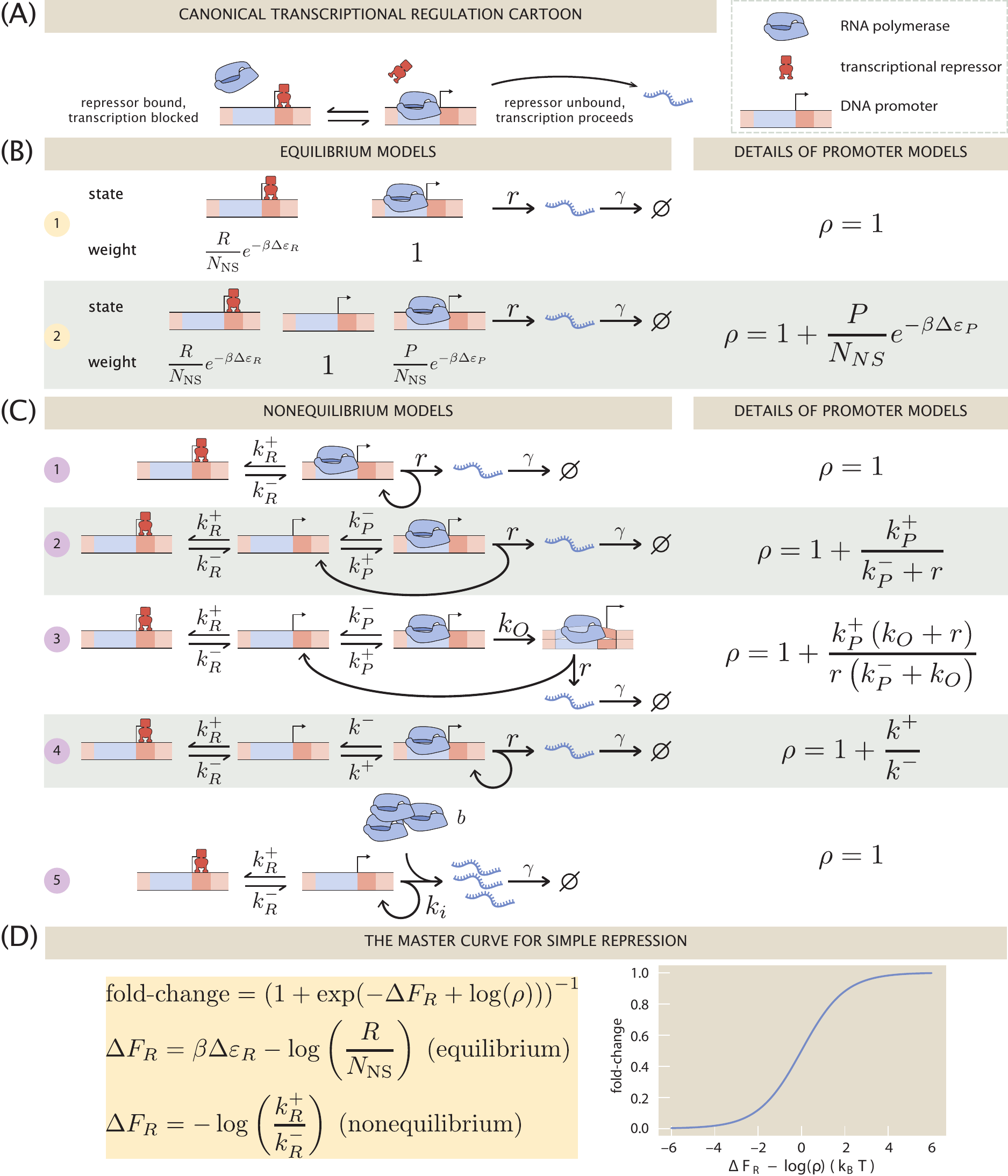}
\caption{\textbf{An overview of the simple repression motif at the level of
means.} (A) Schematic of the qualitative biological picture of the simple
repression genetic architecture. (B) and (C) A variety of possible
mathematicized cartoons of simple repression, along with the effective parameter
$\rho$ which subsumes all regulatory details of the architecture that do not
directly involve the repressor. (B) Simple repression models from an equilibrium
perspective. (C) Equivalent models cast in chemical kinetics language. (D) The
``master curve'' to which all cartoons in (B) and (C) collapse.}
\label{fig1:means_cartoons}
\end{figure}

Figure~\ref{fig1:means_cartoons}(A) shows the qualitative picture of simple
repression that is implicit in the repressor-operator model. An operator, the
binding site on the DNA for a repressor protein, may be found occupied by a
repressor, in which case transcription is blocked from occurring. Alternatively,
that binding site may be found unoccupied, in which case RNA polymerase (RNAP)
may bind and transcription can proceed. The key assumption we make in this
simplest incarnation of the repressor-operator model is that binding of
repressor and RNAP in the promoter region of interest is exclusive, meaning that
one or the other may bind, but never may both be simultaneously bound. It is
often imagined that when the repressor is bound to its operator, RNAP is
sterically blocked from binding to its promoter sequence. Current evidence
suggests this is sometimes, but not always the case, and it remains an
interesting open question precisely how a repressor bound far upstream is able
to repress transcription~\cite{Rydenfelt2014-2}. Suggestions include
``action-at-a-distance'' mediated by kinks in the DNA, formed when the repressor
is bound, that prevent RNAP binding. Nevertheless, our modeling in this work is
sufficiently coarse-grained that we simply assume exclusive binding and leave
explicit accounting of these details out of the problem.

\afterpage{\clearpage}

The logic of the remainder of the paper is as follows. In
section~\ref{section_02_means}, we show how both thermodynamic models and
kinetic models based upon the chemical master equation all culminate in the same
underlying functional form for the fold-change in the average level of gene
expression as shown in Figure~\ref{fig1:means_cartoons}(D).
Section~\ref{sec:beyond_means} goes beyond an analysis of the mean gene
expression by asking how the same models presented in
Figure~\ref{fig1:means_cartoons}(C) can be used to explore noise in gene
expression. To make contact with experiment, all of these models must make a
commitment to some numerical values for the key parameters found in each such
model. Therefore in Section~\ref{section_04_bayesian_inference} we explore the
use of Bayesian inference to establish these parameters and to rigorously answer
the question of how to discriminate between the different models.


\section{Mean Gene Expression}\label{section_02_means}

As noted in the previous section, there are two broad classes of models in play
for computing the input-output functions of regulatory architectures as shown in
Figure~\ref{fig1:means_cartoons}. In both classes of model, the promoter is
imagined to exist in a discrete set of states of occupancy, with each such state
of occupancy accorded its own rate of transcription -- including no 
transcription for many of these states. The models are probabilistic with each
state assigned some probability and the overall rate of transcription given by 
\begin{equation}
\mbox{average rate of transcription} = \sum_i r_i p_i,
\label{eq:transcrip_prop_pbound}
\end{equation}
where $i$ labels the distinct states, $p_i$ is the probability of the
$i^{\text{th}}$ state, and $r_i$ is the rate of transcription of that state.
Ultimately, the different models differ along several key axes: what states to
consider and how to compute the probabilities of those states.

The first class of models that are the focus of the present section on
predicting the mean level of gene expression, sometimes known as thermodynamic
models, invoke the tools of equilibrium statistical mechanics to compute the
probabilities~\cite{Ackers1982, Shea1985, Buchler2003, Vilar2003a, Vilar2003b,
Bintu2005a, Bintu2005c, Gertz2009, Sherman2012, Saiz2013}. As seen in
Figure~\ref{fig1:means_cartoons}(B), even within the class of thermodynamic
models, we can make different commitments about the underlying microscopic
states of the promoter.  Indeed, the list of options considered here does not at
all exhaust the suite of different microscopic states we can assign to the
promoter.

The second class of models that allow us to access the mean gene expression use
chemical master equations to compute the probabilities of the different
microscopic states ~\cite{Ko1991, Peccoud1995, Record1996, Kepler2001,
Sanchez2008, Shahrezaei2008, Sanchez2011, Michel2010}. As seen in
Figure~\ref{fig1:means_cartoons}(C), we consider a host of different
nonequilibrium models, each of which will have its own result for both the mean
(this section) and noise (next section) in gene expression.


\subsection{Fold-changes are indistinguishable across models}
As a first stop on our search for the ``right'' model of simple repression, let
us consider what we can learn from theory and experimental measurements on the
average level of gene expression in a population of cells. One experimental
strategy that has been particularly useful (if incomplete since it misses out on
gene expression dynamics) is to measure the fold-change in mean expression. The
fold-change is defined as
\begin{equation}
\text{fold-change}
= \frac{\langle \text{gene expression with repressor present} \rangle}
        {\langle \text{gene expression with repressor absent} \rangle}
= \frac{\langle m (R > 0) \rangle}{\langle m (R = 0) \rangle}
\label{eq:fc_def}
\end{equation}
where angle brackets $\left\langle \cdot \right\rangle$ denote the average over
a population of cells and mean mRNA $\langle m\rangle$ is viewed as a function
of repressor copy number $R$. What this means is that the fold-change in gene
expression is a relative measurement of the effect of the transcriptional 
repressor ($R > 0$) on the gene expression level compared to an unregulated
promoter ($R = 0$). The second equality in Eq.~\ref{eq:fc_def} follows from assuming
that the translation efficiency, i.e., the number of proteins translated per
mRNA, is the same in both conditions. In other words, we assume that mean
protein level is proportional to mean mRNA level, and that the proportionality
constant is the same in both conditions and therefore cancels out in the ratio.
This is reasonable since the cells in the two conditions are identical except
for the presence of the transcription factor, and the model assumes that the
transcription factor has no direct effect on translation.

Fold-change has proven a very convenient observable in past
work~\cite{Garcia2011a, Brewster2014, Razo-Mejia2018, Chure2019}. Part of its
utility in dissecting transcriptional regulation is its ratiometric nature,
which removes many secondary effects that are present when making an absolute
gene expression measurement. Also, by measuring otherwise identical cells with
and without a transcription factor present, any biological noise common to both
conditions can be made to cancel away.

\fig{fig1:means_cartoons} depicts a smorgasbord of mathematicized cartoons for
simple repression using both thermodynamic and kinetic models that have appeared
in previous literature. For each cartoon, we calculate the fold-change in mean
gene expression as predicted by that model, deferring some algebraic details to
Appendix~\ref{sec:non_bursty}. What we will find is that all cartoons collapse
to a single master curve, shown in \fig{fig1:means_cartoons}(D), which contains
just two parameters. We label the parameters $\Delta F_R$, an effective free
energy parametrizing the repressor-DNA interaction, and $\rho$, which subsumes
all details of transcription in the absence of repressors. We will offer some
intuition for why this master curve exists and discuss why at the level of the
mean expression, we are unable to discriminate ``right'' from ``wrong'' cartoons
given only measurements of fold-changes in expression.

\subsubsection{The Two-State Equilibrium Model}
In this simplest model, depicted as (1) in Figure~\ref{fig1:means_cartoons}(B),
the promoter is idealized as existing in one of two states, either repressor
bound or repressor unbound. The rate of transcription
is assumed to be proportional to the fraction of time spent
in the repressor unbound state.
From the relative statistical weights listed in
Figure~\ref{fig1:means_cartoons}, the probability $p_U$ of being in the
unbound state is
\begin{equation}
p_U = \left(1 + \frac{R}{N_{NS}} e^{-\beta\Delta\varepsilon_R}\right)^{-1}.
\end{equation}
The mean rate of transcription is then given by $r p_U$, as assumed by
Eq.~\ref{eq:transcrip_prop_pbound}. The mean number of mRNA is
set by the balance of average mRNA transcription and degradation rates,
so it follows that the mean mRNA level is given by
\begin{equation}
\langle m \rangle = \frac{r}{\gamma}
        \left(1 + \frac{R}{N_{NS}} e^{-\beta\Delta\varepsilon_R}\right)^{-1},
\end{equation}
where $r$ is the transcription rate from the repressor unbound state, $\gamma$
is the mRNA degradation rate, $R$ is repressor copy number, $N_{NS}$ is the
number of nonspecific binding sites in the genome where repressors spend most
of their time when not bound to the operator, $\beta \equiv 1/k_BT$, and
$\Delta\varepsilon_R$ is the binding energy of a repressor to its operator site.
The derivation of this result is deferred to Appendix~\ref{sec:non_bursty}.

The fold-change is  found as the ratio of mean mRNA with and without repressor
as introduced in Eq.~\ref{eq:fc_def}. Invoking that definition results in
\begin{equation}
\text{fold-change}
= \left(1 + \frac{R}{N_{NS}} e^{-\beta\Delta\varepsilon_R}\right)^{-1},
\end{equation}
which matches the form of the master curve in
Figure~\ref{fig1:means_cartoons}(D) with $\rho=1$ and $\Delta F_R = 
\beta\Delta\varepsilon_r - \log (R / N_{NS})$.

In fact it was noted in~\cite{Chure2019} that this two-state model can be viewed
as the coarse-graining of any equilibrium promoter model in which no
transcriptionally active states have transcription factor bound, or put
differently, when there is no overlap between
transcription factor bound states and transcriptionally
active states. We will see this explicitly in the 3-state equilibrium model
below, but perhaps surprising is that an analogous result carries over even to
the nonequilibrium models we consider later.

\subsubsection{The Three-State Equilibrium Model}
Compared to the previous model, here we fine-grain the repressor unbound state
into two separate states: empty, and RNAP bound as shown in (2) in
Figure~\ref{fig1:means_cartoons}(B). This picture was used in~\cite{Garcia2011a}
as we use it here, and in~\cite{Razo-Mejia2018} and~\cite{Chure2019} it was
generalized to incorporate small-molecule inducers that bind the repressor. 
The effect of this generalization is, roughly speaking,
simply to rescale $R$ from the total number of repressors to a smaller
effective number of available repressors which are unbound by inducers.
We point out that the same generalization can be incorporated quite easily
into any of our models in Figure~\ref{fig1:means_cartoons} by simply
rescaling the repressor copy number $R$ in the equilibrium models, or
equivalently $k_R^+$ in the nonequilibrium models.

The mean mRNA copy number, as derived in Appendix~\ref{sec:non_bursty}
from a similar enumeration of states and weights as the previous model, is
\begin{equation}
\langle m \rangle = \frac{r}{\gamma}
\frac{\frac{P}{N_{NS}} e^{-\beta\Delta\varepsilon_P}}
        {
        1 + \frac{R}{N_{NS}} e^{-\beta\Delta\varepsilon_R}
        + \frac{P}{N_{NS}} e^{-\beta\Delta\varepsilon_P}
        },
\end{equation}
where the new variables are $\Delta\varepsilon_P$, the difference in RNAP
binding energy to its specific site (the promoter) relative to an average
nonspecific background site, and the RNAP copy number, $P$. The fold-change
again follows immediately as
\begin{align}
\text{fold-change}
&= \frac{\frac{P}{N_{NS}} e^{-\beta\Delta\varepsilon_P}}
        {
        1 + \frac{R}{N_{NS}} e^{-\beta\Delta\varepsilon_R}
        + \frac{P}{N_{NS}} e^{-\beta\Delta\varepsilon_P}
        }
\frac{1 + \frac{P}{N_{NS}} e^{-\beta\Delta\varepsilon_P}}
        {\frac{P}{N_{NS}} e^{-\beta\Delta\varepsilon_P}}
\\
&= \left(
1 + \frac{\frac{R}{N_{NS}} e^{-\beta\Delta\varepsilon_R}}
        {1 + \frac{P}{N_{NS}} e^{-\beta\Delta\varepsilon_P}}
\right)^{-1}
\\
&= (1 + \exp(-\Delta F_R - \log\rho))^{-1},
\end{align}
with $\Delta F_R = \beta\Delta\varepsilon_R - \log(R/N_{NS})$ and $\rho = 1 +
\frac{P}{N_{NS}}\mathrm{e}^{-\beta\Delta\varepsilon_P}$ as shown
in~\fig{fig1:means_cartoons}(B). Thus far, we see that the two thermodynamic
models, despite making different coarse-graining commitments, result in the same
functional form for the fold-change in mean gene expression.  We now explore how
kinetic models fare when faced with computing the same observable.

\subsubsection{The Poisson Promoter Nonequilibrium Model}
For our first kinetic model, we  imitate the states considered in the Two-State
Equilibrium Model and consider the simplest possible picture with only two
states, repressor bound and unbound. This is exactly the model used for the main
results of~\cite{Jones2014}. In this picture, repressor association and
dissociation rates from its operator site, $k_R^+$ and $k_R^-$, respectively,
govern transitions between the two states. When the system is in the unbound
state, transcription initiates at rate $r$, which represents a coarse-graining
of all the downstream processes into a single effective rate. mRNA is degraded
at rate $\gamma$ as already exploited in the previous models.

Let $p_R(m,t)$ denote the joint probability of finding the system in the
repressor bound state $R$ with $m$ mRNA molecules present at time $t$. Similarly
define $p_U(m,t)$ for the repressor unbound state $U$. This model is governed
by coupled master equations giving the time evolution of $p_R(m,t)$ and
$p_U(m,t)$~\cite{Sanchez2008, Sanchez2011, Phillips2019} which we can write as
\begin{align}
\begin{split}
\deriv{t}p_R(m,t) =& 
- \overbrace{k_R^- p_R(m,t)}^{R \rightarrow U}
+ \overbrace{k_R^+ p_U(m,t)}^{U \rightarrow R}
+ \overbrace{(m+1)\gamma p_R(m+1,t)}^{m + 1 \rightarrow m}
- \overbrace{\gamma p_R(m,t)}^{m \rightarrow m - 1}
\\
\deriv{t}p_U(m,t) =&\; 
\overbrace{k_R^- p_R(m,t)}^{R \rightarrow U}
- \overbrace{k_R^+ p_U(m,t)}^{U \rightarrow R}
+ \overbrace{rp_U(m-1,t)}^{m-1 \rightarrow m}
- \overbrace{rp_U(m,t)}^{m \rightarrow m + 1}
\\
&+ \overbrace{(m+1)\gamma p_U(m+1,t)}^{m + 1 \rightarrow m}
- \overbrace{\gamma p_U(m,t)}^{m \rightarrow m - 1},
\label{eq:poisson_promoter_cme}
\end{split}
\end{align}
where each term on the right corresponds to a transition between two states of
the promoter as indicated by the overbrace label. In each equation, the first
two terms describe transitions between promoter states due to repressors
unbinding and binding, respectively. The final two terms describe degradation of
mRNA, decreasing the copy number by one, and the terms with coefficient $r$
describe transcription initiation increasing the mRNA copy number by one.
We direct the reader to Appendix~\ref{sec:cme_from_cartoon} for a careful
treatment showing how the form of this master equation follows from the
corresponding cartoon in Figure~\ref{fig1:means_cartoons}.

We can greatly simplify the notation, which will be especially useful for the
more complicated models yet to come, by re-expressing the master equation in
vector form~\cite{Phillips2012}. The promoter states are collected into a vector
and the rate constants are collected into matrices as
\begin{equation}
\vec{p}(m) = \begin{pmatrix} p_R(m) \\ p_U(m) \end{pmatrix},\
\mathbf{K} = \begin{pmatrix} -k_R^- & k_R^+ \\ k_R^- & -k_R^+ \end{pmatrix},\
\mathbf{R} = \begin{pmatrix} 0 & 0 \\ 0 & r \end{pmatrix},\
\label{eq:2state_cme_matrices}
\end{equation}
so that the master equation may be condensed as
\begin{equation}
\deriv{t}\vec{p}(m,t) =
\left( \mathbf{K} - \mathbf{R} - \gamma m \mathbf{I} \right) \vec{p}(m,t)
                + \mathbf{R} \vec{p}(m-1,t)
                + \gamma (m+1) \mathbf{I} \vec{p}(m+1,t),
\label{eq:2state_rep_cme}
\end{equation}
where $\mathbf{I}$ is the identity matrix. Taking steady state by setting time
derivatives to zero, the mean mRNA can be found to be
\begin{equation}
\langle m \rangle = \frac{r}{\gamma}
        \left(1 + \frac{k_R^+}{k_R^-}\right)^{-1},
\label{eq:mean_m_model1}
\end{equation}
with the algebra details again deferred to Appendix~\ref{sec:non_bursty}. Recall
$k_R^+$ is proportional to the repressor copy number, so in computing
fold-change, absence of repressor corresponds to $k_R^+\rightarrow0$. Therefore
fold-change in this model is simply
\begin{equation}
\text{fold-change} = \left(1 + \frac{k_R^+}{k_R^-}\right)^{-1},
\end{equation}
again matching the master curve of~\fig{fig1:means_cartoons}(D) with $\rho=1$.

\subsubsection{Nonequilibrium Model Two - RNAP Bound and Unbound States}
Our second kinetic model depicted in Figure~\ref{fig1:means_cartoons}(C) mirrors
the second equilibrium model of Figure~\ref{fig1:means_cartoons}(B) by
fine-graining  the repressor unbound state of nonequilibrium model 1, resolving
it into an empty promoter state and an RNAP-bound state. Note in this picture,
in contrast with model 4 below, transcription initiation is accompanied by a
promoter state change, in keeping with the interpretation as RNAP-bound and
empty states: if an RNAP successfully escapes the promoter and proceeds to
elongation of a transcript, clearly it is no longer bound at the promoter.
Therefore another RNAP must bind before another transcript can be initiated.

The master equation governing this model is analogous
to Eqs.~\ref{eq:2state_cme_matrices}-\ref{eq:2state_rep_cme} for model 1 above. The
main subtlety arises since transcription initiation accompanies a promoter state
change. This can be understood by analogy to $\mathbf{K}$. The off-diagonal and
diagonal elements of $\mathbf{K}$ correspond to transitions arriving at or
departing from, respectively, the promoter state of interest. If transcription
initiation is accompanied by promoter state changes, we must have separate
matrices for arriving and departing transcription events since the arriving and
departing transitions have different initial copy numbers of mRNA, unlike for
$\mathbf{K}$ where they are the same (see Appendix~\ref{sec:non_bursty}). The
master equation for this model is
\begin{equation}
\deriv{t}\vec{p}(m,t) =
\left( \mathbf{K} - \mathbf{R_D} - \gamma m \mathbf{I} \right) \vec{p}(m,t)
                + \mathbf{R_A} \vec{p}(m-1,t) +
                \gamma (m+1) \mathbf{I} \vec{p}(m+1,t),
\label{eq:3state_rep_cme}
\end{equation}
with the state vector and promoter transition matrix defined as
\begin{equation}
\vec{p}(m) = \begin{pmatrix} p_R(m) \\ p_E(m) \\ p_P(m) \end{pmatrix},\
\mathbf{K} = \begin{pmatrix} -k_R^- & k_R^+ & 0 \\
                        k_R^- & -k_R^+ -k_P^+ & k_P^- \\
                        0 & k_P^+ & -k_P^- 
                \end{pmatrix},
\label{eq:3state_cme_matrices_pt1}
\end{equation}
and the initiation matrices given by
\begin{equation}
\mathbf{R_A} = \begin{pmatrix}
                0 & 0 & 0 \\ 
                0 & 0 & r \\ 
                0 & 0 & 0
                \end{pmatrix},\
\mathbf{R_D} = \begin{pmatrix}
                0 & 0 & 0 \\ 
                0 & 0 & 0 \\ 
                0 & 0 & r
                \end{pmatrix}.
\label{eq:3state_cme_matrices_pt2}
\end{equation}
The elements of $\vec{p}(m)$ encode the probabilities of having $m$ mRNA present
along with the promoter having repressor bound ($R$), being empty ($E$), or
having RNAP bound ($P$), respectively. $\mathbf{R_A}$ describes probability flux
\textit{arriving} at the state $\vec{p}(m)$ from a state with one fewer mRNA,
namely $\vec{p}(m-1)$, and $\mathbf{R_D}$ describes probability flux
\textit{departing} from the state $\vec{p}(m)$ for a state with one more mRNA,
namely $\vec{p}(m+1)$. $\mathbf{K}$ is closely analogous to model 1.

Mean mRNA at steady state is found analogously to model 1, with the result
\begin{equation}
\langle m\rangle = \frac{r}{\gamma}
        \frac{k_R^- k_P^+}
        {k_R^- k_P^+ + k_R^- (k_P^- + r) + k_R^+ (k_P^- + r)},
\label{eq:model2_meanm}
\end{equation}
and with details again deferred to Appendix~\ref{sec:non_bursty}.
Fold-change is again found from the ratio prescribed by Eq.~\ref{eq:fc_def}, from
which we have
\begin{align}
\text{fold-change}
&=      \frac{k_R^- k_P^+}
        {k_R^- k_P^+ + k_R^- (k_P^- + r) + k_R^+ (k_P^- + r)}
        \frac{k_P^+ + k_P^- + r}{k_P^+}
\\
&=      \left(1 + \frac{k_R^+}{k_R^-}
                \frac{k_P^- + r}{k_P^+ + k_P^- + r}
        \right)^{-1}
\\
&=      \left(1 + \frac{k_R^+}{k_R^-}
        \left(1 + \frac{k_P^+}{k_P^- + r}\right)^{-1}
        \right)^{-1},
\end{align}
which follows the master curve of~\fig{fig1:means_cartoons}(D) with $\rho = 1 +
k_P^+/(k_P^- + r)$ as claimed.

\subsubsection{Nonequilibrium Model Three - Multistep Transcription Initiation
and Escape}
One might reasonably complain that the first two ``nonequilibrium'' models we
have considered are straw men. Their steady states necessarily satisfy detailed
balance which is equivalent to thermodynamic equilibrium. Why is this the case?
At steady state there is by definition no net probability flux in or out of each
promoter state, but since the promoter states form a linear chain, there is only
one way in or out of the repressor bound and RNAP bound states, implying each
edge must actually have a net zero probability flux, which is the definition of
detailed balance (usually phrased as equality of forward and reverse transition
fluxes).

Now we consider model 3 in Figure~\ref{fig1:means_cartoons}(C) which allows the
possibility of true nonequilibrium steady-state fluxes through the promoter
states. We point out that this model was considered previously
in~\cite{Mitarai2015} where a comparison was made with model 1 as used
in~\cite{Jones2014}. The authors of~\cite{Mitarai2015} argued that the
additional complexity is essential to properly account for the noise in the mRNA
distribution. We will weigh in on both models later when we consider observables
beyond fold-change.

The master equation governing this model is identical in form to model 2 above,
namely
\begin{equation}
\deriv{t}\vec{p}(m,t) =
\left( \mathbf{K} - \mathbf{R_D} - \gamma m \mathbf{I} \right) \vec{p}(m,t)
                + \mathbf{R_A} \vec{p}(m-1,t) +
                \gamma (m+1) \mathbf{I} \vec{p}(m+1,t),
\end{equation}
but with a higher-dimensional state space and different matrices. The
state vector and promoter transition matrix are now
\begin{equation}
\vec{p}(m) = \begin{pmatrix} p_R(m) \\ p_E(m) \\
                             p_C(m) \\ p_O(m)\end{pmatrix},\
\mathbf{K} = \begin{pmatrix} -k_R^- & k_R^+ & 0 & 0\\
                        k_R^- & -k_R^+ -k_P^+ & k_P^- & 0 \\
                        0 & k_P^+ & -k_P^- - k_O & 0 \\
                        0 & 0 & k_O & 0
                \end{pmatrix},
\end{equation}
with the four promoter states, in order, being repressor bound ($R$), empty
($E$), RNAP closed complex ($C$), and RNAP open complex ($O$). Besides
increasing dimension by one, the only new feature in $\mathbf{K}$ is the
rate $k_O$, representing the rate of open complex formation from
the closed complex, which we assume for simplicity to be irreversible
in keeping with some~\cite{Mitarai2015} but not all~\cite{DeHaseth1998}
past literature. The initiation matrices are given by
\begin{equation}
\mathbf{R_A} = \begin{pmatrix}
        0 & 0 & 0 & 0 \\ 
        0 & 0 & 0 & r \\ 
        0 & 0 & 0 & 0 \\ 
        0 & 0 & 0 & 0
                \end{pmatrix},\
\mathbf{R_D} = \begin{pmatrix}
        0 & 0 & 0 & 0 \\ 
        0 & 0 & 0 & 0 \\ 
        0 & 0 & 0 & 0 \\ 
        0 & 0 & 0 & r
                \end{pmatrix},
\end{equation}
again closely analogous to nonequilibrium model 2.

The expression for mean mRNA is substantially more complicated now, as worked
out in Appendix~\ref{sec:non_bursty} where we find
\begin{equation}
\langle m\rangle = \frac{r}{\gamma}
        \frac{k_R^- k_P^+ k_O}
        {k_R^- [(k_P^+ (k_O + r) + r(k_P^- + k_O)] + k_R^+ r(k_P^- + k_O)},
\label{eq:model3_mean_m}
\end{equation}
which can be simplified to
\begin{equation}
\langle m\rangle
= \frac{r}{\gamma}
\frac{\frac{k_P^+ k_O}{r(k_O + k_P^-)}}
        {1 + \frac{k_P^+ (k_O + r)}{r(k_O + k_P^-)} + \frac{k_R^+}{k_R^-}}.
\end{equation}
The strategy is to isolate the terms involving the repressor, so that now the
fold-change is seen to be simply
\begin{align}
\text{fold-change}
&= \frac{\frac{k_P^+ k_O}{r(k_O + k_P^-)}}
        {1 + \frac{k_P^+ (k_O + r)}{r(k_O + k_P^-)} + \frac{k_R^+}{k_R^-}}
        \frac{1 + \frac{k_P^+ (k_O + r)}{r(k_O + k_P^-)}}
                {\frac{k_P^+ k_O}{r(k_O + k_P^-)}}
\\
&= \left(
        1 + \frac{k_R^+}{k_R^-}
        \left(1 + \frac{k_P^+ (k_O + r)}{r(k_O + k_P^-)}\right)^{-1}
\right)^{-1},
\end{align}
surprisingly reducing to the master curve of~\fig{fig1:means_cartoons}(D) once
again, with $\rho = 1 + \frac{k_P^+ (k_O + r)}{r(k_O + k_P^-)}$.

This example hints that an arbitrarily fine-grained model of downstream
transcription steps may still be collapsed to the form of the master curve for
the means given in Figure~\ref{fig1:means_cartoons}(D), so long as the repressor
binding is exclusive with transcriptionally active states. We offer this as a
conjecture, and we suspect that a careful argument using the King-Altman diagram
method~\cite{King1956, Hill1966} might furnish a ``proof.'' Our focus here is
not on full generality but rather to survey an assortment of plausible models
for simple repression  that have been proposed in the literature.

\subsubsection{Nonequilibrium Model Four - ``Active'' and ``Inactive'' States}
Model 4 in Figure~\ref{fig1:means_cartoons}(C) is at the core of the theory
in~\cite{Razo-Mejia2020}. At a glance the cartoon for this model may appear very
similar to model 2, and mathematically it is, but the interpretation is rather
different. In model 2, we interpreted the third state literally as an RNAP-bound
promoter and modeled initiation of a transcript as triggering a promoter state
change, making the assumption that an RNAP can only
make one transcript at a time. In contrast, in the present model the promoter
state does \textit{not} change when a transcript is initiated. So we no longer
interpret these states as literally RNAP bound and unbound but instead as
coarse-grained ``active'' and ``inactive'' states, the details of which we leave
unspecified for now. We will comment more on this model below when we discuss
Fano factors of models.

Mathematically this model is very similar to models 1 and 2. Like model 1, the
matrix $R$ describing transcription initiation is diagonal, namely
\begin{equation}
\mathbf{R} = \begin{pmatrix}
                0 & 0 & 0 \\ 
                0 & 0 & 0 \\ 
                0 & 0 & r
        \end{pmatrix}.
\end{equation}
The master equation takes verbatim the same form as it did for model 1,
Eq.~\ref{eq:2state_rep_cme}. Meanwhile the promoter transition
matrix $\mathbf{K}$ is the same as Eq.~\ref{eq:3state_cme_matrices_pt1}
from model 2, although we relabel the
rate constants from $k_P^\pm$ to $k^\pm$ to reiterate that these are not simply
RNAP binding and unbinding rates.

Carrying out the algebra, the mean mRNA can be found to be
\begin{equation}
\langle m\rangle = \frac{r}{\gamma}
\frac{k_R^- k^+}
{k_R^- k^+ + k_R^- k^- + k_R^+ k^-},
\end{equation}
and the fold-change readily follows,
\begin{align}
\text{fold-change}
&=      \frac{k_R^- k^+}{k_R^- k^+ + k_R^- k^- + k_R^+ k^-}
        \frac{k_R^- k^+ + k_R^- k^-}{k_R^- k^+}
\\
&=      \left(1 + \frac{k_R^+}{k_R^-}
                \left(1 + \frac{k^+}{k^-}\right)^{-1}
        \right)^{-1},
\end{align}
from which we see $\rho = 1 + k^+/k^-$ as shown in~\fig{fig1:means_cartoons}(C).

\subsubsection{Nonequilibrium Model Five - Bursty Promoter}
The final model we consider shown in Figure~\ref{fig1:means_cartoons}(C) is an
intuitive analog to model 1, with just two states, repressor bound or unbound,
and transition rates between them of $k_R^+$ and $k_R^-$. In model 1, when in
the unbound state, single mRNA transcripts are produced as a Poisson process
with some characteristic rate $r$. The current model by contrast produces, at
some Poisson rate $k_i$, \textit{bursts} of mRNA transcripts.
The burst sizes are assumed to be geometrically distributed with
a mean burst size $b$, which we will motivate in Section~\ref{sec:beyond_means}
when we derive this model as a certain limiting case of model 4.

From this intuitive picture and by analogy to model 1, then, it should be
plausible that the mean mRNA level is
\begin{equation}
\langle m\rangle = \frac{k_i b}{\gamma}
        \left(1 + \frac{k_R^+}{k_R^-}\right)^{-1},
\end{equation}
which will turn out to be correct from a careful calculation. For now, we simply
note that just like model 1, the fold-change becomes
\begin{equation}
\text{fold-change} = \left(1 + \frac{k_R^+}{k_R^-}\right)^{-1}
\end{equation}
with $\rho=1$ also like model 1.
We will also see later how this model emerges as a natural limit of model 4.

\subsection{Discussion of Results Across Models for Fold-Changes in Mean
Expression}
The key outcome of our analysis of the models in
Figure~\ref{fig1:means_cartoons} is the existence of a master curve shown
in Figure~\ref{fig1:means_cartoons}(D) to which the fold-change predictions
of all the models collapse. This master curve is parametrized by only two
effective parameters: $\Delta F_R$, which characterizes the number of repressors
and their binding strength to the DNA, and $\rho$, which characterizes all other
features of the promoter architecture. The key assumption underpinning this
result is that no transcription occurs when a repressor is bound to its
operator. Note, however, that we are agnostic about the molecular mechanism
which achieves this; steric effects are one plausible mechanism, but, for
instance, ``action-at-a-distance'' mediated by kinked DNA due to repressors
bound tens or hundreds of nucleotides upstream of a promoter is plausible as
well.

Why does the master curve of Figure~\ref{fig1:means_cartoons}(D) exist at all?
This brings to mind the deep questions posed in,
e.g.,~\cite{Frank2013} and~\cite{Frank2014a}, suggesting we consider multiple
plausible models of a system and search for their common patterns to tease
out which broad features are and are not important.
In our case, the key feature seems to be the
exclusive nature of repressor and RNAP binding, which allows the parameter
describing the repressor, $\Delta F_R$, to cleanly separate from all other
details of the promoter architecture, which are encapsulated in $\rho$.
Arbitrary nonequilibrium behavior can occur on the rest of the
promoter state space, but it may all be swept up in the effective
parameter $\rho$, to which the repressor makes no contribution.
We point the interested reader to~\cite{Gunawardena2012}
and~\cite{Ahsendorf2014} for an interesting analysis of similar problems
using a graph-theoretic language.

As suggested in~\cite{Chure2019}, we believe this master curve should generalize
to architectures with multiple repressor binding sites, as long as the
exclusivity of transcription factor binding and transcription initiation is
maintained. The interpretation of $\Delta F_R$ is then of an effective free
energy of all repressor bound states. In an equilibrium picture this is simply
given by the log of the sum of Boltzmann weights of all repressor bound states,
which looks like the log of a partition function of a subsystem. In a
nonequilibrium picture, while we can still mathematically gather terms and give
the resulting collection the label $\Delta F_R$, it is unclear if the physical
interpretation as an effective free energy makes sense. The problem is that free
energies cannot be assigned unambiguously to states out of equilibrium because
the free energy change along a generic path traversing the state space is path
dependent, unlike at equilibrium. A consequence of this is that, out of
equilibrium, $\Delta F_R$ is no longer a simple sum of Boltzmann weights.
Instead it resembles a restricted sum of King-Altman diagrams~\cite{King1956,
Hill1966}. Following the work of Hill~\cite{Hill1989}, it may yet be possible to
interpret this expression as an effective free energy, but this remains unclear
to us. We leave this an open problem for future work.

If we relax the requirement of exclusive repressor-RNAP binding, one could
imagine models in which repressor and RNAP doubly-bound states are allowed,
where the repressor's effect is to \text{reduce} the transcription rate rather
than setting it to zero. Our results do not strictly apply to such a model,
although we note that if the repressor's reduction of the transcription rate is
substantial, such a model might still be well-approximated by one of the models
in~\fig{fig1:means_cartoons}.

One may worry that our ``one curve to rule them all'' is a mathematical
tautology. In fact we \textit{agree} with this criticism if $\Delta F_R$ is
``just a fitting parameter'' and cannot be meaningfully interpreted as a real,
physical free energy. An analogy to Hill functions is appropriate here. One of
their great strengths and weaknesses, depending on the use they are put to, is
that their parameters coarse-grain many details and are generally not
interpretable in terms of microscopic models, for deep reasons discussed at
length in~\cite{Frank2013}. By contrast, our master curve claims to have the
best of both worlds: a coarse-graining of all details besides the repressor into
a single effective parameter $\rho$, while simultaneously retaining an
interpretation of $\Delta F_R$ as a physically meaningful and interpretable free
energy. Our task, then, is to prove or disprove this claim.

How do we test this and probe the theory with fold-change measurements? There is
a fundamental limitation in that the master curve is essentially a one-parameter
function of $\Delta F_R + \log\rho$. Worse, there are many \textit{a priori}
plausible microscopic mechanisms that could contribute to the value of $\rho$,
such as RNAP binding and escape kinetics~\cite{DeHaseth1998, Mitarai2015},
and/or supercoiling accumulation and release~\cite{Chong2014, Sevier2016},
and/or, RNAP clusters analogous to those observed in
eukaryotes~\cite{Cisse2013, Cho2016} and recently also observed in
bacteria~\cite{Ladouceur2020}. Even if $\Delta F_R$ is measured to high
precision, inferring the potential microscopic contributions to $\rho$, buried
inside a log no less, from fold-change measurements seems beyond reach. As a
statistical inference problem it is entirely nonidentifiable, in the language
of~\cite{Gelman2013}, Section 4.3.

If we cannot simply infer values of $\rho$ from measurements of fold-change, can
we perturb some of the parameters that make up $\rho$ and measure the change?
Unfortunately we suspect this is off-limits experimentally: most of the
potential contributors to $\rho$ are global processes that affect many or all
genes. For instance, changing RNAP association rates by changing RNAP copy
numbers, or changing supercoiling kinetics by changing topoisomerase copy
numbers, would massively perturb the entire cell's physiology and confound any
determination of $\rho$.

One might instead imagine a bottom-up modeling approach, where we mathematicize
a model of what we hypothesize the important steps are and are not, use \textit{in vitro}
data for the steps deemed important, and \textit{predict} what $\rho$ should be.
But again, because of the one-parameter nature of the master curve, many
different models will likely make indistinguishable predictions, and without any
way to experimentally perturb \textit{in vivo}, there is no clear way
to test whether the modeling assumptions are correct.

In light of this, we prefer the view that parameters and rates are not directly
comparable between cartoons in~\fig{fig1:means_cartoons}. Rather, parameters in
the simpler cartoons represent coarse-grained combinations of parameters in the
finer-grained models. For instance, by equating $\rho$ between any two models,
one can derive various possible correspondences between the two models'
parameters. Note that these correspondences are clearly not unique, since many
possible associations could be made. It then is a choice as to what microscopic
interpretations the model-builder prefers for the parameters in a particular
cartoon, and as to which coarse-grainings lend intuition and which seem
nonsensical. Indeed, since it remains an open question what microscopic features
dominate $\rho$ (as suggested above, perhaps RNAP binding and escape
kinetics~\cite{DeHaseth1998, Mitarai2015}, or supercoiling accumulation and
release~\cite{Chong2014, Sevier2016}, or, something more exotic like RNAP
clusters~\cite{Cisse2013, Cho2016, Ladouceur2020}), we are hesitant to put too
much weight on any one microscopic interpretation of model parameters that make
up $\rho$.

One possible tuning knob to probe $\rho$ that would not globally perturb the
cell's physiology is to manipulate RNAP binding sites. Work such
as~\cite{Kinney2010} has shown that models of sequence-dependent RNAP affinity can be
inferred from data, and the authors of~\cite{Brewster2012} showed that the model
of~\cite{Kinney2010} has predictive power by using the model to \textit{design}
binding sites of a desired affinity. But for our purposes, this begs the
question: the authors of~\cite{Kinney2010} \textit{assumed} a particular model
(essentially our 3-state equilibrium model but without the repressor), so it is
unclear how or if such methods can be turned around to \textit{compare}
different models of promoter function.

Another possible route to dissect transcription details without a global
perturbation would be to use phage polymerase with phage-specific promoters.
While such results would carry some caveats, e.g., whether the repression of
the phage polymerase is a good analog to the repression of the native RNAP,
it could nevertheless be worthy of consideration.

We have already pointed out that the master curve of~\fig{fig1:means_cartoons}
is essentially a one-parameter model, the one parameter being $\Delta F_R +
\log\rho$. By now the reader may be alarmed as to how can we even determine
$\Delta F_R$ and $\rho$ independently of each other, never mind shedding a lens
on the internal structure of $\rho$ itself. A hint is provided by the weak
promoter approximation, invoked repeatedly in prior studies~\cite{Bintu2005c,
Garcia2011a, Razo-Mejia2018} of simple repression using the 3-state equilibrium
model in~\fig{fig1:means_cartoons}(B). In that picture, the weak promoter
approximation means $\frac{P}{N_{NS}}\exp(-\beta\Delta\varepsilon_P) \ll 1$,
meaning therefore $\rho\approx1$.  This approximation can be well justified on
the basis of the number of RNAP and $\sigma$ factors per cell
and the strength of binding of RNAP to
DNA at weak promoters. This is suggestive, but how can we be sure that $\rho$ is
not, for instance, actually $10^2$ and that $\Delta F_R$ hides a compensatory
factor? A resolution is offered by an independent inference of $\rho$ in the
absence of repressors. This was done in~\cite{Razo-Mejia2020} by fitting
nonequilibrium model 4 in~\fig{fig1:means_cartoons}(C), with zero repressor
(looking ahead, this is equivalent to model 4
in Figure~\ref{fig2:constit_cartoons}(A)),
to single-cell mRNA counts data from~\cite{Brewster2014}. This provided a
determination of $k^+$ and $k^-$, from which their ratio is estimated to be no
more than a few $10^{-1}$ and possibly as small as $10^{-2}$.

The realization that $\rho\approx1$ to an excellent approximation,
\textit{independent} of which model in~\fig{fig1:means_cartoons} one prefers,
goes a long way towards explaining the surprising success of equilibrium models
of simple repression. Even though our 2- and 3-state models get so many details
of transcription wrong, it does not matter because fold-change is a cleverly
designed ratio. Since $\rho$ subsumes all details except the repressor, and
$\log\rho\approx0$, fitting these simple models to fold-change measurements can
still give a surprisingly good estimate of repressor binding energies. So the
ratiometric construction of fold-change fulfills its intended purpose of
canceling out all features of the promoter architecture except the repressor
itself. Nevertheless it is perhaps surprising how effectively it does so:
\textit{a priori}, one might not have expected $\rho$ to be quite so close to 1.

We would also like to highlight the relevance of~\cite{Landman2019} here.
Landman et.\ al.\ reanalyzed and compared \textit{in vivo} and
\textit{in vitro} data on the lacI repressor's binding affinity to its
various operator sequences. (The \textit{in vivo}
data was from, essentially, fitting our master curve to expression
measurements.) They find broad agreement between the
\textit{in vitro} and \textit{in vivo}
values. This reinforces the suspicion that the equilibrium $\Delta\varepsilon_R$
repressor binding energies do in fact represent real physical free energies.
Again, \textit{a priori} this did not have to be the case, even knowing that
$\rho\approx1$.

In principle, if $\Delta F_R$ can be measured to sufficient precision, then
deviations from $\rho=1$ become a testable matter of experiment. In practice, it
is probably unrealistic to measure repressor rates $k_R^+$ or $k_R^-$ or
fold-changes in expression levels (and hence $\Delta\varepsilon_R$) precisely
enough to detect the expected tiny deviations from $\rho=1$. We can estimate the
requisite precision in $\Delta F_R$ to resolve a given $\Delta\rho$ by noting,
since $\rho\approx1$, that $\log(1+\Delta\rho)\approx \Delta\rho$, so
$\Delta(\Delta F_R) \approx \Delta\rho$. Suppose we are lucky and $\Delta\rho$
is $\sim0.1$, on the high end of our range estimated above. A
determination of $\Delta\varepsilon_R/k_BT$ with an uncertainty of barely 0.1
was achieved in the excellent measurements of~\cite{Razo-Mejia2018}, so this
requires a very difficult determination of $\Delta F_R$ for a very crude
determination of $\rho$, which suggests, to put it lightly, this is not a
promising path to pursue experimentally. It is doubtful that inference of
repressor kinetic rates would be any easier.

Moving forward, we have weak evidence supporting the interpretation of $\Delta
F_R$ as a physically real free energy~\cite{Landman2019} and other work casting
doubt~\cite{Hammar2014}. How might we resolve the confusion? If there is no
discriminatory power to test the theory and distinguish the various models with
measurements of fold-changes in means, how do we probe the theory? Clearly to
discriminate between the nonequilibrium models in~\fig{fig1:means_cartoons}, we
need to go beyond means to ask questions about kinetics, noise and even full
distributions of mRNA copy numbers over a population of cells. If the
``one-curve-to-rule-them-all'' is more than a mathematical tautology, then the
free energy of repressor binding inferred from fold-change measurements should
agree with repressor binding and unbinding rates. In other words, the
equilibrium and nonequilibrium definitions of $\Delta F_R$ must agree, meaning
\begin{equation}
\Delta F_R = \beta\Delta\varepsilon_R - \log(R/N_{NS})
        = - \log(k_R^+/k_R^-),
\label{eq:deltaFR_eq_noneq_equiv}
\end{equation}
must hold, where $\beta\Delta\varepsilon_R$ is inferred from the master curve
fit to fold-change measurements, and $k_R^+$ and $k_R^-$ are inferred in some
orthogonal manner. Single molecule measurements such as from~\cite{Hammar2014}
have directly observed these rates, and in the remainder of this work we explore
a complementary approach: inferring repressor binding and unbinding rates
$k_R^+$ and $k_R^-$ from single-molecule measurements of mRNA population
distributions.
\section{Beyond Means in Gene Expression}
\label{sec:beyond_means}
 
In this section, our objective is to explore the same models considered in the
previous section, but now with reference to the the question of how well they
describe the distribution of gene expression levels, with special reference to
the variance in these distributions. To that end, we repeat the same pattern as
in the previous section by examining the models one by one. In particular
we will focus on the Fano factor, defined as the variance/mean. This metric
serves as a powerful discriminatory tool from the null model that the
steady-state mRNA distribution must be Poisson, giving a Fano factor of one.

\subsection{Kinetic models for unregulated promoter noise}

\begin{figure}
\centering
\includegraphics[width=\textwidth]{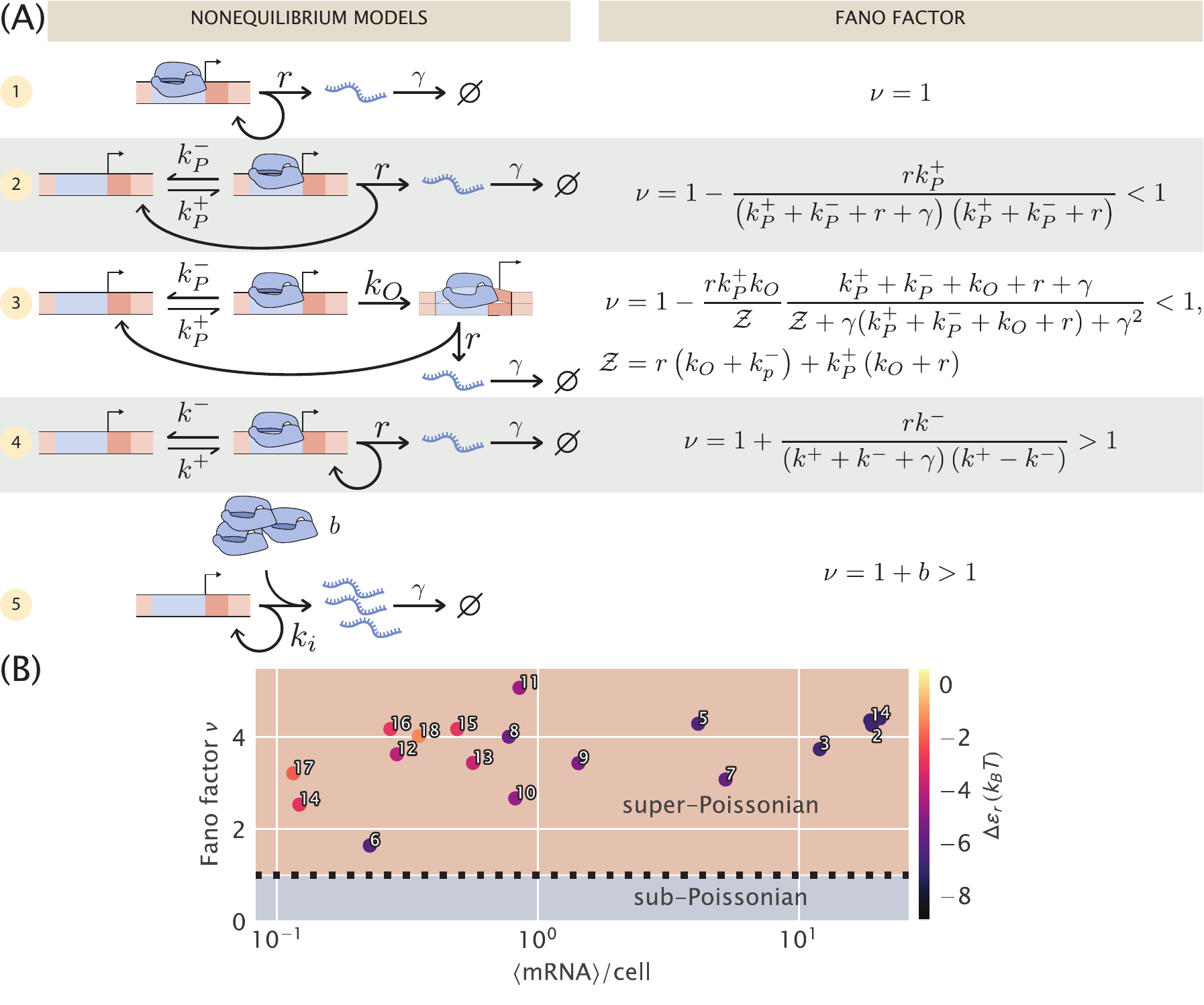}
\caption{\textbf{Comparison of different models for noise in the constitutive promoter.}
(A) The left column depicts various plausible models for the dynamics of
constitutive promoters. In model (1), transcripts are produced in a Poisson
process~\cite{Sanchez2013, Jones2014}. Model (2) features explicit modeling of
RNAP binding/unbinding kinetics~\cite{Phillips2015a}. Model (3) is a more
detailed generalization of model (2), treating transcription initiation as a
multi-step process proceeding through closed and open
complexes~\cite{Mitarai2015}. Model (4) is somewhat analogous to (2) except with
the precise nature of active and inactive states left
ambiguous~\cite{Peccoud1995, Shahrezaei2008, Razo-Mejia2020}. Finally, model (5)
can be viewed as a certain limit of model (4) in which transcripts are produced
in bursts, and initiation of bursts is a Poisson process. 
The right column shows the Fano factor $\nu$ (variance/mean) for each model.
Note especially the crucial diagnostic: (2) and (3) have $\nu$ strictly below 1,
while only for (4) and (5) can $\nu$ exceed 1. Models with Fano factors $\le 1$
cannot produce the single-cell data observed in part (B) without introducing
additional assumptions and model complexity. (B) Data from \cite{Jones2014}.
Mean mRNA count vs. Fano factor (variance/mean) for different promoters as
determined with single-molecule mRNA Fluorescence \textit{in situ}
Hybridization. The colorbar indicates the predicted binding affinity of RNAP to
the promoter sequence as determined in \cite{Brewster2012}. Numbers serve for
cross comparison with data presented in Figure 3.}
\label{fig2:constit_cartoons}
\end{figure}

Before we can tackle simple repression, we need an adequate phenomenological
model of constitutive expression. The literature abounds with options from which
we can choose, and we show several potential kinetic models for constitutive
promoters in Figure~\ref{fig2:constit_cartoons}(A). Let us consider
the suitability of each model for our purposes in turn.

\subsubsection{Noise in the Poisson Promoter Model}

The simplest model of constitutive expression that we can imagine is shown as
model 1 in Figure~\ref{fig2:constit_cartoons}(A) and assumes that transcripts are
produced as a Poisson process from a single promoter state. This is the picture
from Jones et.\ al.~\cite{Jones2014} that was used to interpret a systematic
study of gene expression noise over a series of promoters designed to have
different strengths. This model insists that the ``true'' steady-state mRNA
distribution is Poisson, implying the Fano factor $\nu$ must be 1.
In~\cite{Jones2014}, the authors carefully attribute measured deviations from
Fano = 1 to intensity variability in fluorescence measurements, gene copy number
variation, and copy number fluctuations of the transcription machinery, e.g.,
RNAP itself. In this picture, the master equation makes no appearance, and all
the corrections to Poisson behavior are derived as additive corrections to the
Fano factor. For disproving the ``universal noise curve'' from So et.\
al.~\cite{So2011}, this picture was excellent. It is appealing in its simplicity,
with only two parameters, the initiation rate $r$ and degradation rate $\gamma$.
Since $\gamma$ is independently known from other experiments, and the mean mRNA
copy number is $r/\gamma$, $r$ is easily inferred from data. In other words, the
model is not excessively complex for the data at hand. But for many interesting
questions, for instance in the recent work~\cite{Razo-Mejia2020}, knowledge of
means and variances alone is insufficient, and a description of the full
distribution of molecular counts is necessary. For this we need a (slightly)
more complex model than model 1 that would allow us to incorporate the
non-Poissonian features of constitutive promoters directly into a master
equation formulation.

\subsubsection{Noise in the Two-State Promoter, RNAP Bound or Unbound.}

Our second model of constitutive transcription posits an architecture in which
the promoter is either empty or bound by RNAP ~\cite{Phillips2015a,
Phillips2019}. Here,  as shown in model 2 of Figure~\ref{fig2:constit_cartoons}(A),
transcription initiation results in a state transition from the bound to the
unbound state, reflecting the microscopic reality that an RNAP that has begun to
elongate a transcript is no longer available at the start site to begin another.
As shown in Appendix~\ref{sec:non_bursty}, the Fano factor in this model is
given by
\begin{align}
    \nu = 1 -
        \frac{r\rate{k}{P}{+}}
            {\left(\rate{k}{P}{+} + \rate{k}{P}{-} + r\right)
             \left(\gamma + \rate{k}{P}{+} + \rate{k}{P}{-} + r\right)}.
\label{eq:model2_fano}
\end{align}
The problem with this picture is that the Fano factor is always $<1$. To make
 contact with the experimental reality of $\nu>1$ as shown in
 Figure~\ref{fig2:constit_cartoons}(B), clearly some corrections will be needed.
 While this model adds an appealing element of microscopic reality, we are
 forced to reject it as the additional complexity is unable to capture the
 phenomenology of interest. Obviously the promoter state does in fact proceed
 through cycles of RNAP binding, initiating, and elongating, but it seems that
 the super-Poissonian noise in mRNA copy number we want to model must be
 governed by other features of the system.

\subsubsection{Noise in the Three-State Promoter, Multistep Transcription
Initiation and Escape.} 

How might we remedy the deficits of model 2? It is known~\cite{DeHaseth1998}
that once RNAP initially binds the promoter region, a multitude of distinct
steps occur sequentially before RNAP finally escapes into the elongation phase.
Perhaps adding some of this mechanistic detail as shown in model 3 of
Figure~\ref{fig2:constit_cartoons}(A) might rescue the previous model. The next
simplest refinement of that model could consider open complex formation and
promoter escape as separate steps rather than as a single effective step. In
other words, we construct model 3 by adding a single extra state to model 2, and
we will label the two RNAP-bound states as the closed and open complexes,
despite the true biochemical details certainly being more complex. For example,
earlier work extended this model by adding an additional repressor bound state
and did not explicitly consider the limit with no repressor that we analyze
here~\cite{Mitarai2015}. Again, our goal here is not a complete accounting of
all the relevant biochemical detail; this is an exploratory search for the
important features that a model needs to include to square with the known
experimental reality of constitutive expression.

Unfortunately, as hinted at in earlier work~\cite{Mitarai2015}, this model too
has Fano factor $\nu<1$. We again leave the algebraic details for
Appendix~\ref{sec:non_bursty} and merely state the result that
\begin{align}
\nu = 1 - \frac{r k_P^+ k_O}{\mathcal{Z}}
\frac{k_P^+ + k_P^- + k_O + r + \gamma}
    {\mathcal{Z} + \gamma(k_P^+ + k_P^- + k_O + r) + \gamma^2},
\label{eq:model3_fano}
\end{align}
where we defined $\mathcal{Z} = r(k_O + k_P^-) + k_P^+(k_O + r)$ for notational
tidiness. This is necessarily less than 1 for arbitrary rate constants.

In fact, we suspect \textit{any} model in which transcription proceeds through a
multistep cycle must necessarily have $\nu<1$. The intuitive argument compares
the waiting time distribution to traverse the cycle with the waiting time for a
Poisson promoter (model 1) with the same mean time. The latter is simply an
exponential distribution. The former is a convolution of multiple exponentials,
and intuitively the waiting time distribution for a multistep process
should be more peaked with a smaller fractional width than a single
exponential with the same mean.
A less disperse waiting time distribution means transcription initations are
more uniformly distributed in time relative to a Poisson process. Hence the
distribution of mRNA over a population of cells should be less variable compared
to Poisson, giving $\nu<1$.
(In Appendix~\ref{sec:non_bursty} we present a more precise version
of the intuitive arguments in this paragraph.)
Regardless of the merits of this model in describing the noise properties of
constitutive transcription initiation, it ultimately fails the simplest
quantitative feature of the data, namely that the Fano factor $> 1$ and hence
we must discard this mechanistic picture and search elsewhere.

\subsubsection{Noise in a Two-State Promoter with ``Active'' and ``Inactive''
States}

Inspired by~\cite{Razo-Mejia2020}, we next revisit an analog of model 2 in
Figure~\ref{fig2:constit_cartoons}(A), but as with
the analogous models considered in Section~\ref{section_02_means},
the interpretation of the two
states is changed. Rather than explicitly viewing them as RNAP bound and
unbound, we view them as ``active'' and ``inactive,'' which are able and unable
to initiate transcripts, respectively. We are noncommittal as to the microscopic
details of these states.

One interpretation~\cite{Chong2014, Sevier2016} for the active and inactive
states is that they represent the promoter's supercoiling state: transitions to
the inactive state are caused by accumulation of positive supercoiling, which
inhibits transcription, and transitions back to ``active'' are caused by gyrase
or other topoisomerases relieving the supercoiling. This is an interesting
possibility because it would mean the timescale for promoter state transitions
is driven by topoisomerase kinetics, not by RNAP kinetics. From in vitro
measurements, the former are suggested to be of order minutes~\cite{Chong2014}.
Contrast this with model 2, where the state transitions are assumed to be
governed by RNAP, which, assuming a copy number per cell of order $10^3$, has a
diffusion-limited association rate $k_{on} \sim 10^2~\text{s}^{-1}$ to a target
promoter. Combined with known $K_d$'s of order $\mu$M, this gives an RNAP
dissociation rate $k_{off}$ of order $10^2$~s$^{-1}$. As we will show below,
however, there are some lingering puzzles with interpreting this supercoiling
hypothesis, so we leave it as a speculation and refrain from assigning definite
physical meaning to the two states in this model.

Intuitively one might expect that, since transcripts are produced as a Poisson
process only when the promoter is in one of the two states in this model,
transcription initiations should now be ``bunched'' in time, in contrast to the
``anti-bunching'' of models 2 and 3 above. One might further guess that this
bunching would lead to super-Poissonian noise in the mRNA distribution over a
population of cells.  Indeed, as shown in Appendix~\ref{sec:non_bursty}, a
calculation of the Fano factor produces
\begin{align}
\nu &= 1 + \frac{r k^-}{(k^+ + k^- + \gamma)(k^+ + k^-)},
\label{eq:model4_fano}
\end{align}
which is strictly greater than 1, verifying the above intuition. Note we have
dropped the $P$ label on the promoter switching rates to emphasize that these
very likely do not represent kinetics of RNAP itself. This calculation can also
be sidestepped by noting that the model is mathematically equivalent to the
simple repression model from~\cite{Jones2014}, with states and rates relabeled
and reinterpreted.

How does this model compare to model 1 above? In model 1, all non-Poisson
features of the mRNA distribution were handled as extrinsic corrections. By
contrast, here the 3 parameter model is used to fit the full mRNA distribution
as measured in mRNA FISH experiments. In essence, all variability in the mRNA
distribution is regarded as ``intrinsic,'' arising either from stochastic
initiation or from switching between the two coarse-grained promoter states. The
advantage of this approach is that it fits neatly into the master equation
picture, and the parameters thus inferred can be used as input for more
complicated models with regulation by transcription factors.

While this seems promising, there is a major drawback for our purposes which was
already uncovered by the authors of~\cite{Razo-Mejia2020}: the statistical
inference problem is nonidentifiable, in the sense described in Section 4.3
of~\cite{Gelman2013}. What this means is that it is impossible to infer the
parameters $r$ and $k^-$ from the single-cell mRNA counts data
of~\cite{Jones2014} (as shown in Fig.~S2 of~\cite{Razo-Mejia2020}). Rather, only
the ratio $r/k^-$ could be inferred. In that work, the problem was worked around
with an informative prior on the ratio $k^-/k^+$. That approach is unlikely to
work here, as, recall, our entire goal in modeling constitutive expression is to
use it as the basis for a yet more complicated model, when we add on repression.
But adding more complexity to a model that is already poorly identified is a
fool's errand, so we will explore one more potential model.

\subsubsection{Noise Model for One-State Promoter with Explicit Bursts}

The final model we consider is 
inspired by the failure mode of model 4. The key observation above
was that, as found in~\cite{Razo-Mejia2020}, only two parameters, $k^+$ and the
ratio $r/k^-$, could be directly inferred from the single-cell mRNA data
from~\cite{Jones2014}. So let us take this seriously and imagine a model where
these are the only two model parameters. What would this model look like?

To develop some intuition, consider model 4 in the limit $k^+ \ll k^- \lesssim
r$, which is roughly satisfied by the parameters inferred
in~\cite{Razo-Mejia2020}. In this limit, the system spends the majority of its
time in the inactive state, occasionally becoming active and making a burst of
transcripts. This should call to mind the well-known phenomenon of
transcriptional bursting, as reported in,
e.g.,~\cite{Golding2005,Chong2014,Sevier2016}.
Let us make this correspondence more precise. The mean dwell
time in the active state is $1/k^-$. While in this state, transcripts are
produced at a rate $r$ per unit time. So on average, $r/k^-$ transcripts are
produced before the system switches to the inactive state. Once in the inactive
state, the system dwells there for an average time $1/k^+$ before returning to
the active state and repeating the process. $r/k^-$ resembles an average burst
size, and $1/k^+$ resembles the time interval between burst events. More
precisely, $1/k^+$ is the mean time between the end of one burst and the start
of the next, whereas $1/k^+ + 1/k^-$ would be the mean interval between the
start of two successive burst events, but in the limit $k^+ \ll k^-$, $1/k^+ +
1/k^- \approx 1/k^+$. Note that this limit ensures that the waiting time between
bursts is approximately exponentially distributed, with mean set by the only
timescale left in the problem, $1/k^+$.

Let us now verify this intuition with a precise derivation to check that $r/k^-$
is in fact the mean burst size and to obtain the full burst size distribution.
Consider first a constant, known dwell time $T$ in the active state. Transcripts
are produced at a rate $r$ per unit time, so the number of transcripts $n$
produced during $T$ is Poisson distributed with mean $rT$, i.e.,
\begin{equation}
    P(n\mid T) = \frac{(rT)^n}{n!} \exp(-rT).
\end{equation}
Since the dwell time $T$ is unobservable, we actually want $P(n)$, the dwell
time distribution with no conditioning on $T$. Basic rules of probability theory
tell us we can write $P(n)$ in terms of $P(n\mid T)$ as
\begin{equation}
    P(n) =\int_0^\infty P(n\mid T) P(T) dT.
\end{equation}
But we know the dwell time distribution $P(T)$, which is exponentially
distributed according to
\begin{equation}
    P(T) = k^- \exp(-T k^-),
\end{equation}
so $P(n)$ can be written as
\begin{equation}
    P(n) = k^- \frac{r^n}{n!}
            \int_0^\infty T^n\exp[-(r + k^-)T]\,dT.
\end{equation}
A standard integral table shows $\int_0^\infty x^n e^{-ax}\,dx = n!/a^{n+1}$, so
\begin{equation}
    P(n) = \frac{k^- r^n}{(k^- + r)^{n+1}}
        = \frac{k^-}{k^- + r}
            \left(\frac{r}{k^- + r}\right)^n
        = \frac{k^-}{k^- + r}
            \left(1 - \frac{k^-}{k^- + r}\right)^n,
\end{equation}
which is exactly the geometric distribution with standard parameter
$\theta\equiv k^-/(k^- + r)$ and domain $n \in \{0, 1, 2, \dots\}$.
The mean of the geometric distribution, with this convention, is
\begin{align}
\langle n\rangle = \frac{1 - \theta}{\theta}
        = \left(1 - \frac{k^-}{k^- + r}\right)
                    \frac{k^- + r}{k^-}
        = \frac{r}{k^-},
\end{align}
exactly as we guessed intuitively above.

So in taking the limit $r,k^-\rightarrow\infty$, $r/k^-\equiv b$, we obtain a
model which effectively has only a single promoter state, which initiates bursts
at rate $k^+$ (transitions to the active state, in the model 4 picture). The
master equation for mRNA copy number $m$ as derived in
Appendix~\ref{sec:non_bursty} takes the form
\begin{align}
\begin{split}
{d \over dt}p(m,t) = & (m+1)\gamma p(m+1,t) - m\gamma p(m,t) \\
        &+ \sum_{m'=0}^{m-1} k_i p(m',t) \text{Geom}(m-m';b)
         - \sum_{m'=m+1}^\infty k_i p(m,t) \text{Geom}(m'-m;b),
\end{split}
\end{align}
where we use $k_i$ to denote the burst initiation rate, $\text{Geom}(n;b)$ is the
geometric distribution with mean~$b$, i.e., $\text{Geom}(n;b) =
\frac{1}{1+b}\left(\frac{b}{1+b}\right)^n$ (with domain over nonnegative
integers as above). The first two terms are the usual mRNA degradation terms.
The third term enumerates all ways the system can produce a burst of transcripts
and arrive at copy number $m$, given that it had copy number $m'$ before the
burst. The fourth term allows the system to start with copy number $m$, then
produce a burst and end with copy number $m'$. In fact this last sum has trivial
$m'$ dependence and simply enforces normalization of the geometric distribution.
Carrying it out we have
\begin{equation}
\begin{split}
{d \over dt} p(m,t) = & (m+1)\gamma p(m+1,t) - m\gamma p(m,t) \\
        &+ \sum_{m'=0}^{m-1} k_i p(m',t) \text{Geom}(m-m';b)
            - k_i p(m,t),
\end{split}
\end{equation}
We direct readers again to Appendix~\ref{sec:gen_fcn_appdx} for further details.
This improves on model 4 in that now the parameters are easily inferred, as we
will see later, and have clean interpretations. The non-Poissonian features are
attributed to the empirically well-established phenomenological picture of
bursty transcription.

The big approximation in going from model 4 to 5 is that a burst is produced
instantaneously rather than over a finite time. If the true burst duration is
not short compared to transcription factor kinetic timescales, this could be a
problem in that mean burst size in the presence and absence of repressors could
change, rendering parameter inferences from the constitutive case inappropriate.
Let us make some simple estimates of this.

Consider the time delay between the first and final RNAPs in a burst initiating
transcription (\textit{not} the time to complete transcripts, which potentially
could be much longer.) If this timescale is short compared to the typical search
timescale of transcription factors, then all is well. The estimates from
deHaseth et.\ al.~\cite{DeHaseth1998} put RNAP's diffusion-limited on rate
around $\sim\text{few}\times10^{-2}~\text{nM}^{-1}~\text{s}^{-1}$ and polymerase
loading as high as $1~\text{s}^{-1}$. Then for reasonable burst sizes of $<10$,
it is reasonable to guess that bursts might finish initiating on a timescale of
tens of seconds or less (with another 30-60 sec to finish elongation, but that
does not matter here). A transcription factor with typical copy number of order
10 (or less) would have a diffusion-limited association rate of order
$(10~\text{sec})^{-1}$ \cite{Hammar2014}. Higher copy number TFs tend to have
many binding sites over the genome, which should serve to pull them out of
circulation and keep their effective association rates from rising too large.
Therefore, there is \textit{perhaps} a timescale separation possible between
transcription factor association rates and burst durations, but this assumption
could very well break down, so we will have to keep it in mind when we infer
repressor rates from the Jones et.\ al.\ single-cell mRNA counts data
later~\cite{Jones2014}.

In reflecting on these 5 models, the reader may feel that exploring a multitude
of potential models just to return to a very minimal phenomenological model of
bursty transcription may seem highly pedantic. But the purpose of the exercise
was to examine a host of models from the literature and understand why they are
insufficient, one way or another, for our purposes. Along the way we have
learned that the detailed kinetics of RNAP binding and initiating transcription
are probably irrelevant for setting the population distribution of mRNA. The
timescales are simply too fast, and as we will see later in
Figures~\ref{fig:constit_post_full} and~\ref{fig4:repressed_post_full},
the noise seems to be governed by slower timescales.
Perhaps in hindsight this is not surprising: intuitively, the degradation rate
$\gamma$ sets the fundamental timescale for mRNA dynamics, and any other
processes that substantially modulate the mRNA distribution should not differ
from $\gamma$ by orders of magnitude.
\section{Finding the ``right'' model: Bayesian parameter inference}
\label{section_04_bayesian_inference}

In this section of the paper, we continue our program of providing one complete
description of the entire broad sweep of studies that have been made in the
context of the repressor-operator model, dating all the way back to the original
work of Jacob and Monod and including the visionary quantitative work of
M\"{u}ller-Hill and collaborators~\cite{Oehler1990} and up to more recent
studies \cite{Garcia2011a}. In addition, the aim is to reconcile the equilibrium
and non-equilibrium perspectives that have been brought to bear on this problem.
From Section~\ref{section_02_means}, this reconciliation depends on a key
quantitative question as codified by Eq.~\ref{eq:deltaFR_eq_noneq_equiv}: does
the free energy of repressor binding, as described in the equilibrium models and
indirectly inferred from gene expression measurements, agree with the
corresponding values of repressor binding and unbinding rates in the
non-equilibrium picture, measured or inferred more directly? In this section we
tackle the statistical inference problem of inferring these repressor rates from
single-cell mRNA counts data. But before we can turn to the full case of simple
repression, we must choose an appropriate model of the constitutive promoter and
infer the parameter values in that model. This is the problem we address first.

\subsection{Parameter inference for constitutive promoters}

From consideration of Fano factors in the previous section, we suspect that
model 5 in Figure~\ref{fig2:constit_cartoons}(A), a one-state bursty model of
constitutive promoters, achieves the right balance of complexity and simplicity,
by allowing both Fano factor $\nu>1$, but also by remedying, by design, the
problems of parameter degeneracy that model 4 in
Figure~\ref{fig2:constit_cartoons} suffered~\cite{Razo-Mejia2020}. Does this
stand up to closer scrutiny, namely, comparison to full mRNA distributions
rather than simply their moments? We will test this thoroughly on single-cell
mRNA counts for different unregulated promoters from Jones et.\
al.~\cite{Jones2014}.

It will be instructive, however, to first consider the Poisson promoter, model 1
in Figure~\ref{fig2:constit_cartoons}. As we alluded to earlier, since the
Poisson distribution has a Fano factor $\nu$ strictly equal to 1, and all of the
observed data in Figure~\ref{fig2:constit_cartoons}(B) has Fano factor $\nu>1$,
we might already suspect that this model is incapable of fitting the data. We
will verify that this is in fact the case. Using the same argument we can
immediately rule out models 2 and 3 from Figure~\ref{fig2:constit_cartoons}(A).
These models have Fano factors $\nu\le 1$ meaning they are underdispersed
relative to the Poisson distribution. We will also not explicitly consider model
4 from~\fig{fig2:constit_cartoons}(A) since it was already thoroughly analyzed
in~\cite{Razo-Mejia2020}, and since model 5 can be viewed as a special case of
it.

Our objective for this section will then be to assess whether or not model 5 is
quantitatively able to reproduce experimental data. In other words, if our claim
is that the level of coarse graining in this model is capable of capturing the
relevant features of the data, then we should be able to find values for the
model parameters that can match theoretical predictions with single-molecule
mRNA count distributions. A natural language for this parameter inference
problem is that of Bayesian probability. We will then build a Bayesian inference
pipeline to fit the model parameters to data. To gain intuition on how this
analysis is done we will begin with the ``wrong'' model 1 in
Figure~\ref{fig2:constit_cartoons}(A). We will use the full dataset of
single-cell mRNA counts from~\cite{Jones2014} used in
Figure~\ref{fig2:constit_cartoons}(B).

\subsubsection{Model 1: Poisson promoter}

For this model the master equation of interest is
Eq.~\ref{eq:poisson_promoter_cme} with repressor set to zero, i.e.,
\begin{equation}
{d\over dt}p_U(m)(t) = 
        rp_U(m-1)(t) 
        - rp_U(m)(t)
        + (m+1)\gamma p_U(m+1)(t) 
        - \gamma p_U(m)(t),
\end{equation}
whose steady-state solution is given by a Poisson distribution with parameter
$\lambda \equiv r / \gamma$~\cite{Sanchez2013}. The goal of our inference 
problem is then to find the probability distribution for the parameter value
$\lambda$ given the experimental data. By Bayes' theorem this can be written as
\begin{equation}
p(\lambda \mid D) = {p(D \mid \lambda) p(\lambda) \over p(D)},
\end{equation}
where $D = \{m_1, m_2, \ldots, m_N \}$ are the single-cell mRNA experimental
counts. As is standard we will neglect the denominator $p(D)$ on the right
hand side since it is independent of $\lambda$ and serves only as a
normalization factor.

The steady-state solution for the master equation defines the likelihood term
for a single cell $p(m \mid \lambda)$. What this means is that for a given
choice of parameter $\lambda$, under model 1 of
Figure~\ref{fig2:constit_cartoons}(A), we expect to observe $m$ mRNAs in a
single cell with probability
\begin{equation}
p(m\mid\lambda) = \frac{\lambda^m e^{-\lambda}}{m!}.
\label{eq:poisson_inference010}
\end{equation}
Assuming each cell's mRNA count in our dataset is independent of others, the
likelihood of the full inference problem $p(D\mid\lambda)$ is simply a product
of the single cell likelihoods given by Eq.~\ref{eq:poisson_inference010} above, so
\begin{equation}
p(D\mid\lambda) = \prod_{k=1}^N \frac{\lambda^{m_k}e^{-\lambda}}{m_k!}.
\end{equation}

To proceed we need to specify a prior distribution $p(\lambda)$. In this case we
are extremely data-rich, as the dataset from Jones et.\ al~\cite{Jones2014} has
of order 1000-3000 single-cell measurements for each promoter, so our choice of
prior matters little here, as long as it is sufficiently broad. For details on
the prior selection we refer the reader to Appendix~\ref{sec:bayesian}. For our
purpose here it suffices to specify that we use as prior a Gamma distribution.
This particular choice of prior introduces two new parameters, $\alpha$ and
$\beta$, which parametrize the gamma distribution itself, which we use to encode
the range of $\lambda$ values we view as reasonable. Recall $\lambda$ is the
mean steady-state mRNA count per cell, which \textit{a priori} could plausibly
be anywhere from 0 to a few hundred. $\alpha=1$ and $\beta=1/50$ achieve this,
since the gamma distribution is strictly positive with mean $\alpha/\beta$ and
standard deviation $\sqrt{\alpha}/\beta$.

As detailed in Appendix~\ref{sec:bayesian} this particular choice of prior is
known as the \textit{conjugate} prior for a Poisson likelihood.
Conjugate priors have the convenient properties that a closed form exists for the posterior distribution $p(\lambda \mid D)$ - unusual in Bayesian inference problems - and the closed form posterior takes the same form as the prior. For
our case of a Poisson distribution likelihood with its
Gamma distribution conjugate prior, the posterior distribution is also a Gamma
distribution~\cite{Gelman2013}. Specifically the two parameters $\alpha'$ and
$\beta'$ for this posterior distribution take the form $\alpha' = \alpha +
\bar{m} N$ and $\beta' = \beta + N$, where we defined the sample mean $\bar{m} =
\frac{1}{N}\sum_{k=1}^N m_k$ for notational convenience, and $N$ is the number
of cells in our dataset. Furthermore, given that $N$ is $\mathcal{O}(10^3)$ and
$\langle m\rangle \gtrsim 0.1$ for all promoters measured in~\cite{Jones2014}
our data easily overwhelms the choice of prior, and allows us to approximate the
Gamma distribution with a Gaussian distribution with mean $\bar{m}$ and variance
$\bar{m} / N$ with marginal errors. As an example with real numbers, for the
\textit{lacUV5} promoter, Jones et.\ al~\cite{Jones2014} measured 2648 cells
with an average mRNA count per cell of $\bar{m} \approx 18.7$. For this case our
posterior distribution $P(\lambda \mid D)$ would be a Gaussian distribution with
mean $\mu = 18.7$, and a standard deviation $\sigma \approx 0.08$. This suggests
we have inferred our model's one parameter to a precision of order 1\%.

We remind the reader that we began this section claiming that the Poisson model
was ``wrong'' since it could not reproduce features of the data such as a Fano
factor $> 1$. The fact that we obtain such a narrow posterior distribution for our
parameter $P(\lambda \mid D)$ does not equate to the model being adequate to
describe the data. What this means is that given the data $D$, only values
in a narrow range are remotely plausible for the parameter $\lambda$,
but a narrow posterior distribution does not necessarily
mean the model accurately depicts reality.
As we will see later in Figure~\ref{fig:constit_post_full} after
exploring the bursty promoter model, indeed the correspondence
when contrasting the Poisson model with the experimental data is quite poor.

\subsubsection{Model 5 - Bursty promoter}

Let us now consider the problem of parameter inference for model five
from~\fig{fig1:means_cartoons}(C). As derived in
Appendix~\ref{sec:gen_fcn_appdx}, the steady-state mRNA distribution in this
model is a negative binomial distribution, given by
\begin{equation}
p(m) = \frac{\Gamma(m+k_i)}{\Gamma(m+1)\Gamma(k_i)}
        \left(\frac{1}{1+b}\right)^{k_i}
        \left(\frac{b}{1+b}\right)^m,
\label{eq:neg_bionom}
\end{equation}
where $b$ is the mean burst size and $k_i$ is the burst rate in units of the
mRNA degradation rate $\gamma$. As sketched earlier, to think of the negative
binomial distribution in terms of an intuitive ``story,'' in the precise
meaning of~\cite{Blitzstein2015}, we imagine the arrival of
bursts as a Poisson process with rate $k_i$, where each burst has a
geometrically-distributed size with mean size $b$.

As for the Poisson promoter model, this expression for the steady-state mRNA
distribution is exactly the likelihood we want to use when stating Bayes
theorem. Again denoting the single-cell mRNA count data as $D=\{m_1, m_2,\dots,
m_N\}$, here Bayes' theorem takes the form
\begin{equation}
p(k_i, b \mid D) \propto p(D\mid k_i,b)p(k_i, b).
\end{equation}
We already have our likelihood -- the product of $N$ negative binomials as
Eq.~\ref{eq:neg_bionom} -- so we only need to choose priors on $k_i$ and $b$.
For the datasets from~\cite{Jones2014} that we are analyzing, as for the Poisson
promoter model above we are still data-rich so the prior's influence remains
weak, but not nearly as weak because the dimensionality of our model has
increased from one parameter to two. Details on the arguments behind our prior
distribution selection are left for Appendix~\ref{sec:bayesian}. We state here
that the natural scale to explore these parameters is logarithmic. This is
commonly the case for parameters for which our previous knowledge based on our
domain expertise spans several orders of magnitude. For this we chose log-normal
distributions for both $k_i$ and $b$. Details on the mean and variance of these
distributions can be found in Appendix~\ref{sec:bayesian}.

\begin{figure}
\centering
\includegraphics[width=\textwidth]{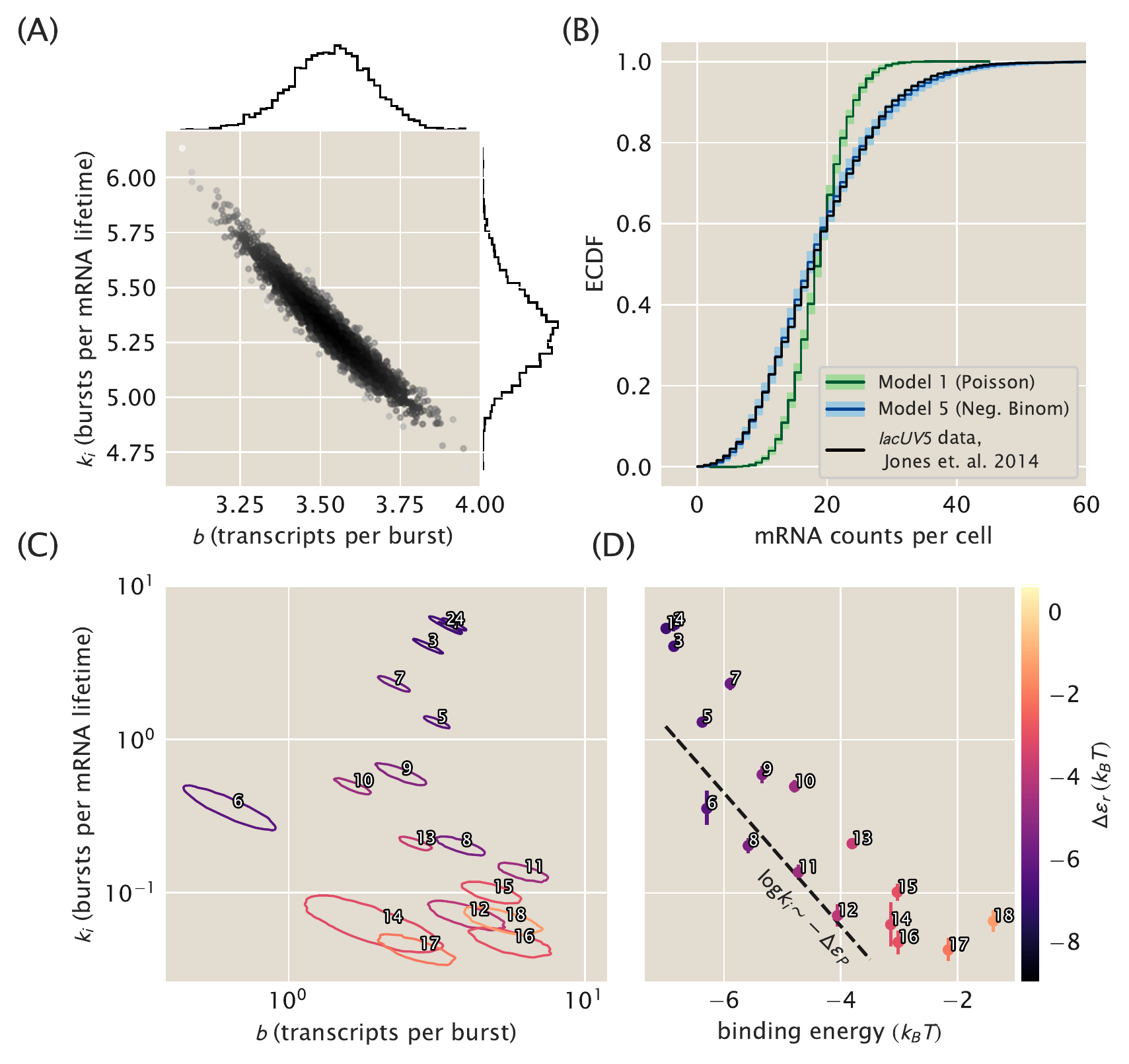}
\caption{\textbf{Constitutive promoter posterior inference and model
comparison.} (A) The joint posterior density of model 5, the bursty promoter
with negative binomially-distributed steady state, is plotted with MCMC samples.
1D marginal probability densities are plotted as flanking histograms. The model
was fit on \textit{lacUV5} data from~\cite{Jones2014}. (B) The empirical
distribution function (ECDF) of the observed population distribution of mRNA
transcripts under the control of a constitutive \textit{lacUV5} promoter is
shown in black. The median posterior predictive ECDFs for models (1), Poisson,
and (5), negative binomial, are plotted in dark green and dark blue,
respectively. Lighter green and blue regions enclose 95\% of all posterior
predictive samples from their respective models. Model (1) is in obvious
contradiction with the data while model (5) is not. Single-cell mRNA count data
is again from~\cite{Jones2014}. (C) Joint posterior distributions for burst rate
$k_i$ and mean burst size $b$ for 18 unregulated promoters
from~\cite{Jones2014}. Each contour indicates the 95\% highest posterior
probability density region for a particular promoter. Note that the vertical
axis is shared with (D). (D) Plots of the burst rate $k_i$ vs.\ the binding
energy for each promoter as predicted in~\cite{Brewster2012}. The dotted line
shows the predicted slope according to Eq.~\ref{eq:bursty_equil_corresp1},
described in text. Each individual promoter is labeled with a unique number in
both (C) and (D) for cross comparison and for comparison with 
Figure~\ref{fig2:constit_cartoons}(B).}
\label{fig:constit_post_full}
\end{figure}

We carried out Markov-Chain Monte Carlo (MCMC) sampling on the posterior of this
model, starting with the constitutive \textit{lacUV5} dataset
from~\cite{Jones2014}. The resulting MCMC samples are shown in
Figure~\ref{fig:constit_post_full}(A). In contrast to the active/inactive
constitutive model considered in~\cite{Razo-Mejia2020} (nonequilibrium model 4
in Figure~\ref{fig2:constit_cartoons}(A)), this model is well-identified with both
parameters determined to a fractional uncertainty of 5-10\%. The strong
correlation reflects the fact that their product sets the mean of the mRNA
distribution, which is tightly constrained by the data, but there is weak
``sloppiness''~\cite{Transtrum2015} along a set of values with a similar
product.

Having found the model's posterior to be well-identified as with the Poisson
promoter, the next step is to compare both models with experimental data. To do
this for the case of the bursty promoter, for each of the parameter samples
shown in Figure~\ref{fig:constit_post_full}(A) we generated negative
bionomial-distributed mRNA counts. As MCMC samples parameter space
proportionally to the posterior distribution, this set of random samples span
the range of possible values that we would expect given the correspondence
between our theoretical model and the experimental data. A similar procedure can
be applied to the Poisson promoter. To compare so many samples with the actual
observed data, we can use empirical cumulative distribution functions (ECDF) of
the distribution quantiles. This representation is shown in
Figure~\ref{fig:constit_post_full}(B). In this example, the median for each
possible mRNA count for the Poisson distribution is shown as a dark green line,
while the lighter green contains 95\% of the randomly generated samples. This
way of representing the fit of the model to the data gives us a sense of the
range of data we might consider plausible, under the assumption that the model
is true. For this case, as we expected given our premise of the Poisson promoter
being wrong, it is quite obvious that the observed data, plotted in black is not
consistent with the Poisson promoter model. An equivalent plot for the bursty
promoter model is shown in blue. Again the darker tone shows the median, while
the lighter color encompasses 95\% of the randomly generated samples. Unlike the
Poisson promoter model, the experimental ECDF closely tracks the posterior
predictive ECDF, indicating this model is actually able to generate the observed
data and increasing our confidence that this model
is sufficient to parametrize the physical reality of the system.

The commonly used promoter sequence \textit{lacUV5} is our primary
target here, since it forms the core of all the
simple repression constructs of~\cite{Jones2014} that we consider in
Section~\ref{sec:rep_kinetics_inference}. Nevertheless, we thought it wise to
apply our bursty promoter model to the other 17 unregulated promoters available
in the single-cell mRNA count dataset from~\cite{Jones2014} as a test that the
model is capturing the essential phenomenology. If the model fit well to all the
different promoters, this would increase our confidence that it would serve well
as a foundation for inferring repressor kinetics later in
Section~\ref{sec:rep_kinetics_inference}. Conversely, were the model to fail on
more than a couple of the other promoters, it would give us pause.

\fig{fig:constit_post_full}(C) shows the results, plotting the posterior
distribution from individually MCMC sampling all 18 constitutive promoter
datasets from~\cite{Jones2014}. To aid visualization, rather than plotting
samples for each promoter's posterior as in \fig{fig:constit_post_full}(A), for
each posterior we find and plot the curve that surrounds the 95\% highest
probability density region. What this means is that each contour 
encloses approximately 95\% of the samples, and thus 95\% of the probability
mass, of its posterior distribution. Theory-experiment comparisons,
shown in Figure~\ref{figS:ppc_unreg} in Appendix~\ref{sec:bayesian},
display a similar level of agreement between data and predictive samples as for
the bursty model with \textit{lacUV5} in \fig{fig:constit_post_full}(B).

One interesting feature from \fig{fig:constit_post_full}(C) is that burst rate
varies far more widely, over a range of $\sim10^2$, than burst size, confined to
a range of $\lesssim10^1$ (and with the exception of promoter 6, just a span of
3 to 5-fold). This suggests that $k_i$, not $b$, is the key dynamic variable that
promoter sequence tunes.

\subsubsection{Connecting inferred parameters to prior work}
It is interesting to connect these inferences on $k_i$ and $b$ to the work
of~\cite{Brewster2012}, where these same 18 promoters were considered through
the lens of the three-state equilibrium model (model 2 in
Figure~\ref{fig1:means_cartoons}(B)) and binding energies $\Delta\varepsilon_P$
were predicted from an energy matrix model derived from~\cite{Kinney2010}. As
previously discussed the thermodynamic models of gene regulation can only make
statements about the mean gene expression. This implies that we can draw the
connection between both frameworks by equating the mean mRNA $\left\langle m
\right\rangle$. This results in
\begin{equation}
\langle m \rangle = \frac{k_i b}{\gamma}
        = \frac{r}{\gamma}
        \frac{\frac{P}{N_{NS}}\exp(-\beta\Delta\varepsilon_P)}
                {1+\frac{P}{N_{NS}}\exp(-\beta\Delta\varepsilon_P)}.
\end{equation}
By taking the weak promoter approximation for the equilibrium model ($P/N_{NS} 
\exp(-\beta\Delta\varepsilon_r) \ll 1$) results in
\begin{equation}
\langle m \rangle = \frac{k_i b}{\gamma}
        = \frac{r}{\gamma} \frac{P}{N_{NS}}\exp(-\beta\Delta\varepsilon_P),
\end{equation}
valid for all the binding energies considered here.

Given this result, how are the two coarse-grainings related? A quick estimate
can shed some light. Consider for instance the \textit{lacUV5} promoter, which
we see from Figure~\ref{fig:constit_post_full}(A) has $k_i/\gamma \sim b \sim
\text{few}$, from Figure~\ref{fig:constit_post_full}(B) has $\langle m \rangle
\sim 20$, and from~\cite{Brewster2012} has $\beta\Delta\varepsilon_P \sim -
6.5$. Further we generally assume $P/N_{NS} \sim 10^{-3}$ since
$N_{NS}\approx4.6\times10^6$ and $P\sim10^3$. After some guess-and-check with
these values, one finds the only association that makes dimensional sense and
produces the correct order-of-magnitude for the known parameters is to take
\begin{equation}
\frac{k_i}{\gamma} = \frac{P}{N_{NS}} \exp(-\beta\Delta\varepsilon_P)
\label{eq:bursty_equil_corresp1}
\end{equation}
and
\begin{equation}
b = \frac{r}{\gamma}.
\label{eq:bursty_equil_corresp2}
\end{equation}
Figure~\ref{fig:constit_post_full}(D) shows that this linear scaling between
$\ln k_i$ and $-\beta\Delta\varepsilon_P$ is approximately true for all 18
constitutive promoters considered. The plotted line is simply
Eq.~\ref{eq:bursty_equil_corresp1} and assumes $P\approx 5000$.

While the associations represented by Eq.~\ref{eq:bursty_equil_corresp1} and
Eq.~\ref{eq:bursty_equil_corresp2} appear to be borne out by the data in
Figure~\ref{fig:constit_post_full}, we do not find the association of parameters
they imply to be intuitive. We are also cautious to ascribe too much physical
reality to the parameters. Indeed, part of our point in comparing the various
constitutive promoter models is to demonstrate that these models each provide an
internally self-consistent framework that adequately describes the data, but
attempting to translate between models reveals the dubious physical
interpretation of their parameters.

We mention one further comparison, between our inferred parameters and the
work of Chong et.\ al.~\cite{Chong2014}, which is interesting and puzzling.
Beautiful experiments in~\cite{Chong2014} convincingly argue that
supercoiling accumulated from the production of mRNA transcripts is key
in setting the burstiness of mRNA production.
In their model, this supercoiling occurs on the scale of $\sim100$~kb
domains of DNA. This suggests that all genes on a given domain should burst
in synchrony, and that the difference between highly and lowly expressed
genes is the \textit{size} of transcriptional bursts, not the
\textit{time between} bursts.
But here, all burst sizes we infer in Figure~\ref{fig:constit_post_full}(C)
are comparable and burst rates vary wildly. It is not immediately clear
how to square this circle. Furthermore, Figure~7E in~\cite{Chong2014}
reports values of the quantity they label $\beta/\alpha$ and we label
$k^+/k^-$ in model 4 from Figure~\ref{fig2:constit_cartoons}. In contrast
to the findings of~\cite{Razo-Mejia2020}, Chong et.\ al.\ do not find
$k^+/k^-\ll1$ for most of the genes they consider. This begs the question:
is the \textit{galK} chromosomal locus used for the reporter constructs
in~\cite{Razo-Mejia2020} and~\cite{Jones2014} merely an outlier,
or is there a deeper puzzle here waiting to be resolved? Without more
apples-to-apples data we can only speculate, and we leave it as an
intriguing open question for the field.

Despite such puzzles, our goal here is not to unravel the mysterious origins
of burstiness in transcription. Our remaining task in this work is a
determination of the physical reality of equilibrium binding energies in
Figure~\ref{fig1:means_cartoons}, as codified by the equilibrium-nonequilibrium
equivalence of Eq.~\ref{eq:deltaFR_eq_noneq_equiv}.
For our phenomenological needs here model 5 in
Figure~\ref{fig2:constit_cartoons} is more than adequate: the posterior
distributions in Figure~\ref{fig:constit_post_full}(C) are cleanly
identifiable and the predictive checks in Figure~\ref{figS:ppc_unreg}
indicate no discrepancies between the model and the mRNA single-moleucle
count data of~\cite{Jones2014}. Of the models we have considered it
is unique in satisfying both these requirements. So we will happily use
it as a foundation to build upon in the next section when we add regulation.

\subsection{Transcription factor kinetics can be inferred from single-cell mRNA
distribution measurements}\label{sec:rep_kinetics_inference}
\subsubsection{Building the model and performing parameter inference}

Now that we have a satisfactory model in hand for constitutive promoters, we
would like to return to the main thread: can we reconcile the equilibrium and
nonequilibrium models by putting to the test
Eq.~\ref{eq:deltaFR_eq_noneq_equiv}, the correspondence between indirectly
inferred equilibrium binding energies and nonequilibrium kinetic rates? To make
this comparison, is it possible to infer repressor binding and unbinding rates
from mRNA distributions over a population of cells as measured by
single-molecule Fluorescence \textit{in situ} Hybridization in~\cite{Jones2014}?
If so, how do these inferred rates compare to direct single-molecule
measurements such as from~\cite{Hammar2014} and to binding energies such as
from~\cite{Garcia2011a} and~\cite{Razo-Mejia2018}, which were inferred under the
assumptions of the equilibrium models in Figure~\ref{fig1:means_cartoons}(B)?
And can this comparison shed light on the unreasonable effectiveness of the
equilibrium models, for instance, in their application in~\cite{Chure2019,
Chure2019a}?

\begin{figure}
\centering
\includegraphics[width=\textwidth]{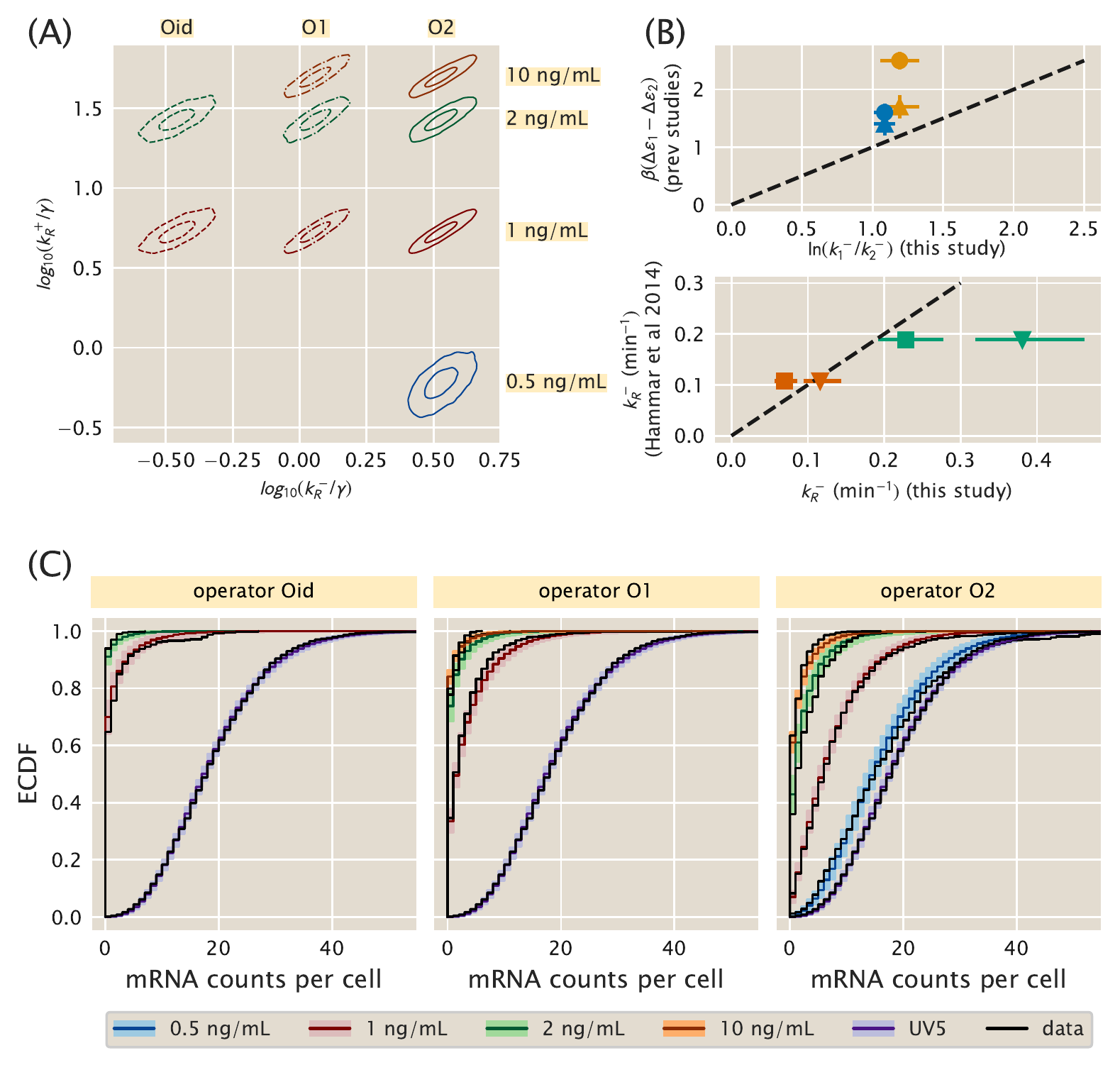}
\caption{\textbf{Simple repression parameter inference and comparison.}
(A) Contours which enclose 50\% and 95\% of the posterior probability mass are
shown for each of several 2D slices of the 9D posterior distribution. The model
assumes one unbinding rate for each operator (Oid, O1, O2) and one binding rate
for each aTc induction concentration (corresponding to an unknown mean repressor
copy number).
(B, upper) Ratios of our inferred unbinding rates are
compared with operator binding energy differences measured by Garcia and
Phillips~\cite{Garcia2011a} (triangles) and Razo-Mejia et.\
al.~\cite{Razo-Mejia2018} (circles). Blue glyphs compare O2-O1, while orange
compare O1-Oid. Points with perfect agreement would lie on the dotted line.
(B, lower) Unbinding rates for O1 (cyan) and Oid (red)
inferred in this work are compared with single-molecule measurements
from Hammar et.\ al.~\cite{Hammar2014}. We plot the comparison assuming illustrative mRNA
lifetimes of $\gamma^{-1}=3$~min (triangles) or $\gamma^{-1}=5$~min
(squares). Dotted line is as in upper panel.
(C) Theory-experiment comparison are shown for each of the datasets
used in the inference of the model in (A). Observed single-molecule mRNA counts
data from~\cite{Jones2014} are plotted as black lines. The median of the randomly
generated samples for each condition is plotted as a dark colored line. Lighter
colored bands enclose 95\% of all samples for a given operator/repressor copy
number pair.
The unregulated promoter, \textit{lacUV5}, is shown with each as a reference.}
\label{fig4:repressed_post_full}
\end{figure}

As we found in Section~\ref{sec:beyond_means}, for our purposes the ``right''
model of a constitutive promoter is the bursty picture, model five in
Figure~\ref{fig2:constit_cartoons}(A). Therefore our starting point here is the
analogous model with repressor added, model 5 in
Figure~\ref{fig1:means_cartoons}(C). For a given repressor binding site and copy
number, this model has four rate parameters to be inferred: the repressor
binding and unbinding rates $k_R^+$, and $k_R^-$, the initiation rate of bursts,
$k_i$, and the mean burst size $b$ (we nondimensionalize all of these by the
mRNA degradation rate $\gamma$).

Before showing the mathematical formulation of our statistical inference model,
we would like to sketch the intuitive structure. The dataset
from~\cite{Jones2014} we consider consists of single-cell mRNA counts data of
nine different conditions, spanning several combinations of three unique
repressor binding sites and four unique repressor copy numbers. We assume that
the values of $k_i$ and $b$ are known, since we have already cleanly inferred
them from constitutive promoter data, and further we assume that these values
are the same across datasets with different repressor binding sites and copy
numbers. In other words, we assume that the regulation of the transcription
factor does not affect the mean burst size nor the burst initiation rate. The
regulation occurs as the promoter is taken away from the transcriptionally
active state when the promoter is bound by repressor. We assume that there is
one unbinding rate parameter for each repressor binding site, and likewise one
binding rate for each unique repressor copy number. This makes our model seven
dimensional, or nine if one counts $k_i$ and $b$ as well. Note that we use only
a subset of the datasets from Jones et.\ al.~\cite{Jones2014}, as discussed more
in Appendix~\ref{sec:bayesian}.

Formally now, denote the set of seven repressor rates to be inferred as
\begin{equation}
\vect{k} =\{k_{Oid}^-, k_{O1}^-, k_{O2}^-,
k_{0.5}^+, k_{1}^+, k_{2}^+, k_{10}^+\},
\end{equation}
where subscripts for dissociation rates $k^-$  indicate the different repressor
binding sites, and subscripts for association rates $k^+$ indicate the
concentration of the small-molecule that controlled the expression of the LacI
repressor (see Appendix~\ref{sec:bayesian}). This is because for this particular
dataset the repressor copy numbers were not measured directly, but it is safe to
assume that a given concentration of the inducer resulted in a specific mean
repressor copy number~\cite{Chure2019a}. Also note that the authors
of~\cite{Jones2014} report estimates of LacI copy number per cell rather than
direct measurements. However, these estimates were made assuming the validity of
the equilibrium models in Figure~\ref{fig1:means_cartoons}, and since testing
these models is our present goal, it would be circular logic if we were to make
the same assumption. Therefore we will make no assumptions about the LacI copy
number for a given inducer concentrations.

Having stated the problem, Bayes' theorem reads
\begin{equation}
p(\vect{k}, k_i, b \mid D)
\propto
p(D \mid\vect{k}, k_i, b) p(\vect{k}, k_i, b),
\end{equation}
where $D$ is again the set of all $N$ observed single-cell mRNA counts
across the various conditions. We assume that individual single-cell
measurements are independent so that the likelihood factorizes as
\begin{equation}
p(D \mid\vect{k}, k_i, b)
= \prod_{j=1}^N p(m\mid \vect{k}, k_i, b)
= \prod_{j=1}^N p(m\mid k_j^+, k_j^-, k_i, b)
\end{equation}
where $k_j^\pm$ represent the appropriate binding and unbinding
rates out of $\vect{k}$ for the $j$-th measured cell. The probability
$p(m\mid k_j^+, k_j^-, k_i, b)$ appearing in the last expression
is exactly Eq.~\ref{eq:p_m_bursty+rep_appdx}, the steady-state
distribution for our bursty model with repression derived in
Section~\ref{sec:gen_fcn_appdx}, which for completeness we reproduce here as
\begin{equation}
\begin{split}
p(m \mid k_R^+, k_R^-, k_i, b) = & ~\frac{
        \Gamma(\alpha + m)\Gamma(\beta + m)\Gamma(k_R^+ + k_R^-)
        }
        {
        \Gamma(\alpha)\Gamma(\beta)\Gamma(k_R^+ + k_R^- + m)
        }
\frac{b^m}{m!}
\\
&\times {_2F_1}(\alpha+m, \beta+m, k_R^++k_R^-+m; -b).
\end{split}
\label{eq:p_m_bursty+rep}
\end{equation}
where $_2F_1$ is the confluent hypergeometric function of the second kind and
$\alpha$ and $\beta$, defined for notational convenience, are
\begin{align}
\begin{split}
\alpha &= \frac{1}{2}
\left(k_i+k_R^-+k_R^+ + \sqrt{(k_i+k_R^-+k_R^+)^2 - 4k_i k_R^-}\right)
\\
\beta &= \frac{1}{2}
\left(k_i+k_R^-+k_R^+ - \sqrt{(k_i+k_R^-+k_R^+)^2 - 4k_i k_R^-}\right).
\end{split}
\end{align}

This likelihood is rather inscrutable. We did not find any of the known
analytical approximations for ${_2F_1}$ terribly useful in gaining intuition, so
we instead resorted to numerics. One insight we found
was that for very strong or very weak repression, the distribution in
Eq.~\ref{eq:p_m_bursty+rep} is well approximated by a negative binomial with
burst size $b$ and burst rate $k_i$ equal to their constitutive \textit{lacUV5}
values, except with $k_i$ multiplied by the fold-change
$\left(1+k_R^+/k_R^-\right)^{-1}$. In other words, once again only the ratio
$k_R^+/k_R^-$ was detectable. But for intermediate repression, the distribution
was visibly broadened with Fano factor greater than $1+b$, the value for the
corresponding constitutive case. This indicates that the repressor rates had
left an imprint on the distribution, and perhaps intuitively, this intermediate
regime occurs for values of $k_R^\pm$ comparable to the burst rate $k_i$. Put
another way, if the repressor rates are much faster or much slower than $k_i$,
then there is a timescale separation and effectively only one timescale remains,
$k_i\left(1+k_R^+/k_R^-\right)^{-1}$. Only when all three rates in the problem
are comparable does the mRNA distribution retain detectable information about
them.

Next we specify priors. As for the constitutive model, weakly informative
log-normal priors are a natural choice for all our rates. We found that if the
priors were too weak, our MCMC sampler would often become stuck in regions of
parameter space with very low probability density, unable to move. We struck a
balance in choosing our prior widths between helping the sampler run while
simultaneously verifying that the marginal posteriors for each parameter were
not artificially constrained or distorted by the presence of the prior.
All details for our prior distributions are listed in 
Appendix~\ref{sec:bayesian}.

We ran MCMC sampling on the full nine dimensional posterior specified by this
model. To attempt to visualize this object, in
Figure~\ref{fig4:repressed_post_full}(A) we plot several two-dimensional slices
as contour plots, analogous to Figure~\ref{fig:constit_post_full}(C). Each of
these nine slices corresponds to the $(k_R^+, k_R^-)$ pair of rates for one of
the conditions from the dataset used to fit the model and gives a
sense of the uncertainty and correlations in the posterior.
We note that the 95\% uncertainties of all the rates span about $\sim0.3$
log units, or about a factor of two, with the exception of $k_{0.5}^+$, the
association rate for the lowest repressor copy number which is somewhat larger.

\subsubsection{Comparison with prior measurements of repressor binding energies}
Our primary goal in this work is to reconcile the kinetic and equilibrium
pictures of simple repression. Towards this end we would like to compare the
repressor kinetic rates we have inferred with the repressor binding energies
inferred through multiple methods in~\cite{Garcia2011a}
and~\cite{Razo-Mejia2018}. If the agreement is close, then it suggests that the
equilibrium models are not wrong and the repressor binding energies they contain
correspond to physically real free energies, not mere fit parameters.

Figure~\ref{fig4:repressed_post_full}(B) shows both comparisons, with the top
panel comparing to equilibrium binding energies and the bottom panel comparing
to single-molecule measurements. First consider the top panel and its comparison
between repressor kinetic rates and binding energies. As described in
section~\ref{section_02_means}, if the equilibrium binding energies
from~\cite{Garcia2011a} and~\cite{Razo-Mejia2018} indeed are the physically real
binding energies we believe them to be, then they should be related to the
repressor kinetic rates via Eq.~\ref{eq:deltaFR_eq_noneq_equiv}, which we
restate here,
\begin{equation}
\Delta F_R = \beta\Delta\varepsilon_R - \log(R/N_{NS})
        = - \log(k_R^+/k_R^-).
\label{eq:deltaFR_eq_noneq_equiv_repeat}
\end{equation}
Assuming mass action kinetics implies that $k_R^+$ is proportional to repressor
copy number $R$, or more precisely, it can be thought of as repressor copy
number times some intrinsic per molecule association rate. But since $R$ is not
directly known for our data from~\cite{Jones2014}, we cannot use this equation
directly. Instead we can consider two different repressor binding sites and
compute the \textit{difference} in binding energy between them, since this
difference depends only on the unbinding rates and not on the binding rates.
This can be seen by evaluating Eq.~\ref{eq:deltaFR_eq_noneq_equiv_repeat} for
two different repressor binding sites, labeled (1) and (2), but with the same
repressor copy number $R$, and taking the difference to find
\begin{equation}
\Delta F_R^{(1)} - \Delta F_R^{(2)}
= \beta\Delta\varepsilon_1 - \beta\Delta\varepsilon_2
= - \log(k_R^+/k_1^-) + \log(k_R^+/k_2^-),
\end{equation}
or simply
\begin{equation}
\beta\Delta\varepsilon_1 - \beta\Delta\varepsilon_2
= \log(k_2^-/k_1^-).
\end{equation}
The left and right hand sides of this equation are exactly the horizontal and
vertical axes of the top panel of Figure~\ref{fig4:repressed_post_full}. Since
we inferred rates for three repressor binding sites (O1, O2, and Oid), there are
only two independent differences that can be constructed, and we arbitrarily
chose to plot O2-O1 and O1-Oid in Figure~\ref{fig4:repressed_post_full}(B).
Numerically, we compute values of $k_{O1}^- / k_{Oid}^-$ and $k_{O2}^- /
k_{O1}^-$ directly from our full posterior samples, which conveniently provides
uncertainties as well, as detailed in Appendix~\ref{sec:bayesian}.
We then compare these log ratios of rates to the
binding energy differences $\Delta\varepsilon_{O1} - \Delta\varepsilon_{Oid}$
and from $\Delta\varepsilon_{O2} - \Delta\varepsilon_{O1}$ as computed from the
values from both~\cite{Garcia2011a} and~\cite{Razo-Mejia2018}. Three of the four
values are within $\sim0.5~k_BT$ of the diagonal representing perfect agreement,
which is comparable to the $\sim\text{few}\times0.1~k_BT$ variability between
the independent determinations of the same quantities between~\cite{Garcia2011a}
and~\cite{Razo-Mejia2018}. The only outlier involves Oid measurements
from~\cite{Razo-Mejia2018}, and as the authors of~\cite{Razo-Mejia2018} note,
this is a difficult measurement of low fluorescence signal against high
background since Oid represses so strongly. We are therefore inclined to regard
the failure of this point to fall near the diagonal as a testament to the
difficulty of the measurement and not as a failure of our theory.

On the whole then, we regard this as striking confirmation of the validity of
the equilibrium models. Their lynchpin parameter is a phenomenological free
energy of repressor binding that has previously only been inferred indirectly.
Our result shows that the microscopic interpretation of this free energy, as the
log of a ratio of transition rates, does indeed hold true to within the inherent
uncertainties that remain in the entire theory-experiment dialogue.

\subsubsection{Comparison with prior measurements of repressor kinetics}
In the previous section we established the equivalence between the equilibrium
models' binding energies and the repressor kinetics we infer from mRNA
population distributions. But one might worry that the repressor rates we infer
from mRNA distributions are \textit{themselves} merely fitting parameters and
that they do not actually correspond to the binding and unbinding rates of the
repressor in vivo. To verify that this is not the case, we next compare our
kinetic rates with a different measurement of the same rates using a radically
different method: single molecule measurements as performed in Hammar et.\
al.~\cite{Hammar2014}. This is plotted in the lower panel of
Figure~\ref{fig4:repressed_post_full}(B).

Since we do not have access to repressor copy number for either the
single-cell mRNA data from~\cite{Jones2014} or the single-molecule data
from~\cite{Hammar2014}, we cannot make an apples-to-apples comparison of
association rates $k_R^+$. Further, while Hammar et.\ al.\ directly measure the
dissociation rates $k_R^-$, our inference procedure returns $k_R^-/\gamma$,
i.e., the repressor dissociation rate nondimensionalized by the mRNA degradation
rate $\gamma$. So to make the comparison, we must make an assumption for the
value of $\gamma$ since it was not directly measured. For most mRNAs in
\textit{E.\ coli}, quoted values for the typical mRNA lifetime $\gamma^{-1}$
range between about 2.5~min~\cite{Chen2015} to 8~min. We chose $\gamma^{-1} =
3$~min and $\gamma^{-1} = 5$~min as representative values and plot a comparison
of $k_{O1}^-$ and $k_{Oid}^-$ from our inference with corresponding values
reported in~\cite{Hammar2014} for both these choices of $\gamma$.

The degree of quantitative agreement in the lower panel of
Figure~\ref{fig4:repressed_post_full}(B) clearly depends on the precise choice
of $\gamma$. Nevertheless we find this comparison very satisfying, when two
wildly different approaches to a measurement of the same quantity yield broadly
compatible results. We emphasize the agreement between our rates and the rates
reported in~\cite{Hammar2014} for any reasonable $\gamma$: values differ by at
most a factor of 2 and possibly agree to within our uncertainties of 10-20\%.
From this we feel confident asserting that the parameters we have inferred from
Jones et.\ al.'s single-cell mRNA counts data do in fact correspond to repressor
binding and unbinding rates, and therefore our conclusions on the agreement of
these rates with binding energies from~\cite{Garcia2011a}
and~\cite{Razo-Mejia2018} are valid.

\subsubsection{Model checking}
In Figure~\ref{fig:constit_post_full}(B) we saw that the simple Poisson model of
a constitutive promoter, despite having a well behaved posterior, was clearly
insufficient to describe the data. It behooves us to carry out a similar check
for our model of simple repression, codified by Eq.~\ref{eq:p_m_bursty+rep} for
the steady-state mRNA copy number distribution. As derived in
Sections~\ref{section_02_means} and~\ref{sec:beyond_means}, we have compelling
theoretical reasons to believe it is a good model, but if it nevertheless turned
out to be badly contradicted by the data we should like to know.

The details are deferred to Appendix~\ref{sec:bayesian}, and here we only
attempt to summarize the intuitive ideas, as detailed at greater length by
Jaynes~\cite{Jaynes2003} as well as Gelman and
coauthors~\cite{Gelman2013,Gelman2013a}. From our samples of the posterior
distribution, plotted in Figure~\ref{fig4:repressed_post_full}(A), we generate
many replicate data using a random number generator. In
Figure~\ref{fig4:repressed_post_full}(C), we plot empirical cumulative
distribution functions of the middle 95\% quantiles of these replicate data with
the actual experimental data from Jones et.\ al.~\cite{Jones2014} overlaid,
covering all ten experimental conditions spanning repressor binding sites and
copy numbers (as well as the constitutive baseline UV5).

The purpose of Figure~\ref{fig4:repressed_post_full}(C) is simply a graphical,
qualitative assessment of the model: do the experimental data systematically
disagree with the simulated data, which would suggest that our model is missing
important features? A further question is not just whether there is a detectable
difference between simulated and experimental data, but whether this difference
is likely to materially affect the conclusions we draw from the posterior in
Figure~\ref{fig4:repressed_post_full}(A). More rigorous and quantitative
statistical tests are possible~\cite{Gelman2013}, but their quantitativeness
does not necessarily make them more useful. As stated in~\cite{Gelman2013a}, we
often find this graphical comparison more enlightening because it better engages
our intuition for the model, not merely telling \textit{if} the model is wrong
but suggesting \textit{how} the model may be incomplete.

Our broad brush takeaway from Figure~\ref{fig4:repressed_post_full}(C) is
overall of good agreement. There some oddities, in particular the long tails in
the data for Oid, 1~ng/mL, and O2, 0.5~ng/mL. The latter is especially odd since
it extends beyond the tail of the unregulated UV5 distribution. This is a
relatively small number of cells, however, so whether this is a peculiarity of
the experimental data, a statistical fluke of small numbers, or a real
biological effect is unclear. It is conceivable that there is some very slow
timescale switching dynamics that could cause this bimodality, although it is
unclear why it would only appear for specific repressor copy numbers. There is
also a small offset between experiment and simulation for O2 at the higher
repressor copy numbers, especially at 2 and 10~ng/mL.
From the estimate of repressor copy numbers from~\cite{Jones2014}, it is
possible that the repressor copy numbers here are becoming large enough to
partially invalidate our assumption of a separation of timescales between
burst duration and repressor association rate. Another possibility is that
the very large number of zero mRNA counts for Oid, 2~ng/mL is skewing its partner
datasets through the shared association rate.
None of these fairly minor quibbles cause us to
seriously doubt the overall correctness of our model, which further validates
its use to compare the equilibrium models' binding energies to the
nonequilibrium models' repressor kinetics, as we originally set out to do.
\section{Discussion and future work}\label{sec:discussion}

The study of gene expression is one of the dominant themes of modern biology,
made all the more urgent by the dizzying pace at which genomes are being
sequenced. But there is a troubling Achilles heel buried in all of that genomic
data, which is our inability to find and interpret regulatory sequence. In
many cases, this is not possible even qualitatively, let alone the possibility
of quantitative dissection of the regulatory parts of genomes in a predictive
fashion. Other recent work has tackled the challenge of finding and annotating
the regulatory part of genomes \cite{Belliveau2018, Ireland2020}. Once we have
determined the architecture of the regulatory part of the genome, we are then
faced with the next class of questions which are sharpened by formulating them
in mathematical terms, namely, what are the input-output properties of these
regulatory circuits and what knobs control them?

The present work has tackled that question in the context of the first
regulatory architecture hypothesized in the molecular biology era, namely, the
repressor-operator model of Jacob and Monod~\cite{Jacob1961}. Regulation in that
architecture is the result of a competition between a repressor which inhibits
transcription and RNAP polymerase which undertakes it. Through the labors of
generations of geneticists, molecular biologists and biochemists, an
overwhelming amount of information and insight has been garnered into this
simple regulatory motif, licensing it as what one might call the ``hydrogen
atom'' of regulatory biology. It is from that perspective that the present paper
explores the extent to which some of the different models that have been
articulated to describe that motif allow us to understand both the average level
of gene expression found in a population of cells, the intrinsic cell-to-cell
variability, and the full gene expression distribution found in such a population
as would be reported in a single molecule mRNA Fluorescence \textit{in situ}
Hybridization experiment, for example.

Our key insights can be summarized as follows. First, as shown in
Figure~\ref{fig1:means_cartoons}, the mean expression in the simple repression
architecture is captured by a master curve in which the action of repressor and
the details of the RNAP interaction with the promoter appear separately and
additively in an effective free energy. Interestingly, as has been shown
elsewhere in the context of the Monod-Wyman-Changeux model, these kinds of
coarse-graining results are an exact mathematical result and do not constitute
hopeful approximations or biological naivete \cite{Razo-Mejia2018, Chure2019}.
To further dissect the relative merits of the different models, we must appeal
to higher moments of the gene expression probability distribution. To that end,
our second set of insights focus on gene expression noise, where it is seen that
a treatment of the constitutive promoter already reveals that some models have
Fano factors (variance/mean) that are less than one, at odds with any and all
experimental data that we are aware of~\cite{So2011, Jones2014}. This theoretical
result allows us to directly discard a subset of the models (models 1-3 in
Figure~\ref{fig2:constit_cartoons}(A)) since they cannot be reconciled with
experimental observations. The two remaining models (models 4 and 5 in
Figure~\ref{fig2:constit_cartoons}) appear to contain enough microscopic realism
to be able to reproduce the data. A previous exploration of model 4 demonstrated
the ``sloppy''~\cite{Transtrum2015} nature of the model
in which data on single-cell mRNA counts alone
cannot constrain the value of all parameters simultaneously
\cite{Razo-Mejia2020}. Here we demonstrate that the proposed one-state bursty
promoter model (model 5 in Figure~\ref{fig2:constit_cartoons}(A)) emerges as a
limit of the commonly used two-state promoter model \cite{Peccoud1995,
Shahrezaei2008, So2011, Sanchez2013, Jones2014}. We put the idea to the test
that this level of coarse-graining is rich enough to reproduce previous
experimental observations. In particular we perform Bayesian inference to
determine the two parameters describing the full steady-state mRNA distribution,
finding that the model is able to provide a quantitative description of a
plethora of promoter sequences with different mean levels of expression and
noise.

With the results of the constitutive promoter in hand, we then fix the
parameters associated with this class of promoters and use them as input for
evaluating the noise in gene expression for the simple repression motif itself.
This allows us to provide a single overarching analysis of both the constitutive
and simple repression architectures using one simple model and corresponding set
of self-consistent parameters, demonstrating not only a predictive framework,
but also reconciling the equilibrium and non-equilibrium views of the same
simple repression constructs. More specifically, we obtained values for the
transcription factor association and dissociation rates by performing Bayesian
inference on the full mRNA distribution for data obtained from simple-repression
promoters with varying number of transcription factors per cell and affinity of
such transcription factors for the binding site. The free energy value obtained
from these kinetic rates -- computed as the log ratio of the rates -- agrees
with previous inferences performed only from mean gene expression measurements,
that assumed an equilibrium rather than a kinetic
framework~\cite{Garcia2011a, Razo-Mejia2018}.

It is interesting to speculate what microscopic details are being coarse-grained
by our burst rate and burst size in Figure~\ref{fig2:constit_cartoons}, model 5.
Chromosomal locus is one possible influence we have not addressed in this work,
as all the single-molecule mRNA data from~\cite{Jones2014} that we considered
was from a construct integrated at the \textit{galK} locus. The results
of~\cite{Chong2014} indicate that transcription-induced supercoiling contributes
substantially in driving transcriptional bursting, and furthermore, their
Figure~7 suggests that the parameters describing the rate, duration, and size of
bursts vary substantially for transcription from different genomic loci.
Although the authors of~\cite{Englaender2017} do not address noise, they note
enormous differences in mean expression levels when an identical construct is
integrated at different genomic loci. The authors of~\cite{Engl2020} attribute
noise and burstiness in their single-molecule mRNA data to the influence of
different sigma factors, which is a reasonable conclusion from their data. Could
the difference also be due to the different chromosomal locations of the two
operons?
What features of different loci are and are not important?
Could our preferred coarse-grained model capture the variability across
different loci? If so, and we were to repeat the parameter inference
as done in this work, is there a simple theoretical model we could build
to understand the resulting parameters?

In summary, this work took up the challenge of exploring the extent to which a
single specific mechanistic model of the simple-repression regulatory
architecture suffices to explain the broad sweep of experimental data for this
system. Pioneering early experimental efforts from the M\"{u}ller-Hill lab
established the simple-repression motif as an arena for the quantitative
dissection of regulatory response in bacteria, with similar beautiful work
emerging in examples such as the \textit{ara} and \textit{gal} operons as
well~\cite{Dunn1984b, Oehler1990, Weickert1993, Oehler1994, Schleif2000,
Semsey2002, SwintKruse2009}. In light of a new generation of precision
measurements on these systems, the definition of what it means to understand
them can now be formulated as a rigorous quantitative question. In particular,
we believe understanding of the simple repression motif has 
advanced sufficiently that the design of new
versions of the architecture is now possible,
based upon predictions about how repressor copy
number and DNA binding site strength control expression. In our view, the next
step in the progression is to first perform similar rigorous analyses of the
fundamental ``basis set'' of regulatory architectures. Natural follow-ups to this
work are explorations of motifs such as simple activation that is
regulated by a single activator binding site, and the repressor-activator
architecture, mediated by the binding of both a single activator and a single
repressor, and beyond. With the individual input-output functions in hand,
similar quantitative dissections including the rigorous analysis of their tuning
parameters can be undertaken for the ``basis set'' of full gene-regulatory
networks such as switches, feed-forward architectures and oscillators for
example, building upon the recent impressive bonanza of efforts from systems
biologists and synthetic biologists~\cite{Milo2002, Alon2007}.

\section{Methods}

All data and custom scripts were collected and stored using Git version control.
Code for Bayesian inference and figure generation is available on the GitHub
repository (\url{https://github.com/RPGroup-PBoC/bursty_transcription}).
\section*{Acknowledgments}

We thank Rob Brewster for providing the raw single-molecule mRNA FISH data. We
thank Justin Bois for his key support with the Bayesian inference section. We
would also like to thank Griffin Chure for invaluable feedback on the
manuscript. This material is based upon work supported by the National Science
Foundation Graduate Research Fellowship under Grant No. DGE‐1745301. This work
was also supported by La Fondation Pierre-Gilles de Gennes, the Rosen Center at
Caltech, and the NIH 1R35 GM118043 (MIRA). M.R.M. was supported by the Caldwell
CEMI fellowship. 
		\printbibliography[segment=\therefsegment]
	\end{refsegment}
\clearpage

\title{\textbf{Supplemental Information for:
Reconciling Kinetic and Equilibrium Models of Bacterial Transcription}}

\maketitle

\addtocontents{toc}{\protect\setcounter{tocdepth}{3}}

    \begin{refsegment}
		\beginsupplement
		\tableofcontents
\section{Derivations for non-bursty promoter models}
\label{sec:non_bursty}

In this section we detail the calculation of mean mRNA levels, fold-changes in
expression, and Fano factors for nonequilibrium promoter models 1 through 4
in Figure~\ref{fig1:means_cartoons}. These are the results that were quoted but
not derived in Sections~\ref{section_02_means} and \ref{sec:beyond_means} of the
main text. In each of these four models, the natural mathematicization of their
cartoons is as a chemical master equation such as Eq.~\ref{eq:2state_rep_cme}
for model 1. Before jumping into the derivations of the general computation of
the mean mRNA level and the Fano factor we will work through the derivation of
an example master equation. In particular we will focus on model 1 from
Figure~\ref{fig1:means_cartoons}(C). The general steps are applicable to all
other chemical master equations in this work.

\subsection{Derivation of chemical master equation}
\label{sec:cme_from_cartoon}

\begin{figure}[h!]
\centering
\includegraphics{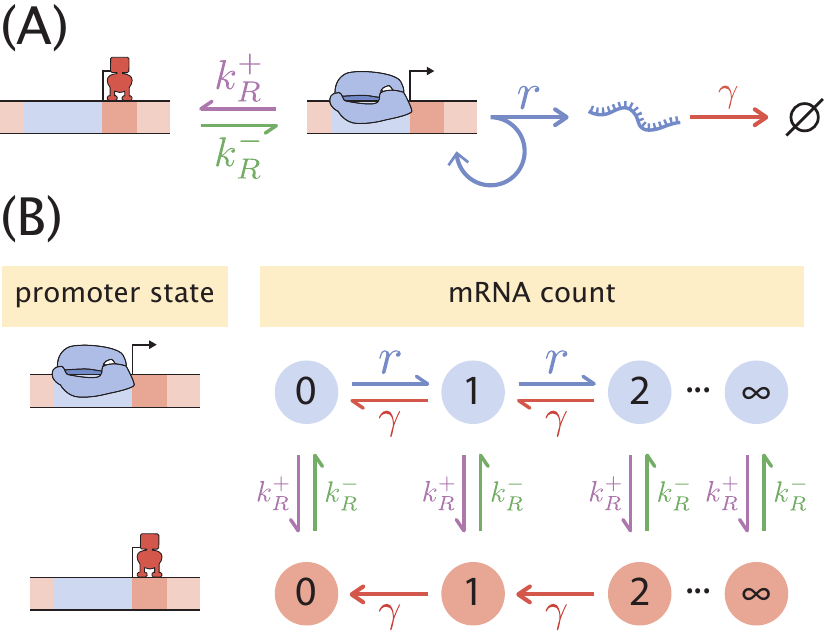}
\caption{
\textbf{Two-state promoter chemical master equation.}
(A) Schematic of the two state promoter simple repression model. Rates $k_R^+$
and $k_R^-$ are the association and dissociation rates of the transcriptional
repressor, respectively, $r$ is the transcription initiation rate, and $\gamma$
is the mRNA degradation rate. (B) Schematic depiction of the mRNA count state
transitions. The model in (A) only allows for jumps in mRNA of size 1. The
production of mRNA can only occur when the promoter is in the transcriptionally
active state.}
\label{fig:two_state}
\end{figure}

The chemical master equation describes the continuous time evolution of a
continuous or discrete probability distribution function. In our specific case
we want to describe the time evolution of the discrete mRNA distribution. What
this means is that we want to compute the probability of having $m$ mRNA
molecules at time $t + \Delta t$, where $\Delta t$ is a sufficiently small time
interval such that only one of the possible reactions take place during that
time interval. For the example that we will work out here in detail we chose the
two-state stochastic simple repression model schematized in
Figure~\ref{fig:two_state}(A). To derive the master equation we will focus more
on the representation shown in Figure~\ref{fig:two_state}(B), where the
transitions between different mRNA counts and promoter states is more explicitly
depicted. Given that the DNA promoter can exist in one of two states --
transcriptionally active state, and with repressor bound -- we will keep track
not only of the mRNA count, but on which state the promoter is. For this we will
keep track of two probability distributions: The probability of having $m$ mRNAs
at time $t$ when the promoter is in the transcriptionally active state $A$,
$p_A(m, t)$, and the equivalent probability but when the promoter is in the
repressor bound state $R$, $p_R(m, t)$.

Since mRNA production can only take place in the transcriptionally active state
we will focus on this function for our derivation. The repressor bound state
will have an equivalent equation without terms involving the production of
mRNAs. We begin by listing the possible state transitions that can occur for a
particular mRNA count with the promoter in the active state. For state changes
in a small time window $\Delta t$ that ``jump into'' state $m$ in the
transcriptionally active state we have
\begin{itemize}
    \item Produce an mRNA, jumping from $m-1$ to $m$.
    \item Degrade an mRNA, jumping from $m+1$ to $m$.
    \item Transition from the repressor bound state $R$ with $m$ mRNAs  to the
    active state $A$ with $m$ mRNAs.
\end{itemize}
Likewise, for state transitions that ``jump out'' of state $m$ in the
transcriptionally inactive state we have
\begin{itemize}
    \item Produce an mRNA, jumping from $m$ to $m+1$.
    \item Degrade an mRNA, jumping from $m$ to $m-1$.
    \item Transition from the active state $A$ with $m$ mRNAs to the repressor
    bound state $R$ with $m$ mRNAs.
\end{itemize}
The mRNA production does not depend on the current number of mRNAs, therefore
these state transitions occur with probability $r\Delta t$. The same is true for
the promoter state transitions; each occurs with probability $k_R^\pm \Delta t$.
As for the mRNA degradation events, these transitions depend on the current
number of mRNA molecules since the more molecules of mRNA there are, the more
will decay during a given time interval. Each molecule has a constant
probability $\gamma \Delta t$ of being degraded, so the total probability for an
mRNA degradation event to occur is computed by multiplying this probability by
the current number of mRNAs.

To see these terms in action let us compute the probability of having $m$ mRNA
at time $t + \Delta t$ in the transcriptionally active state. This takes the 
form
\begin{equation}
\begin{aligned}
p_A(m, t + \Delta t) &= p_A(m, t) \\
&+ \overbrace{(r\Delta t) p_A(m-1, t)}^{m-1 \rightarrow m}
- \overbrace{(r\Delta t) p_A(m, t)}^{m \rightarrow m+1}\\
&+ \overbrace{(m+1)(\gamma \Delta t) p_A(m+1, t)}^{m+1 \rightarrow m}
- \overbrace{m(\gamma \Delta t) p_A(m, t)}^{m \rightarrow m-1}\\
&+\overbrace{(k_R^- \Delta t) p_R(m, t)}^{R \rightarrow A}
-\overbrace{(k_R^+ \Delta t) p_A(m, t)}^{A \rightarrow R},
\end{aligned}
\end{equation}
where the overbrace indicates the corresponding state transitions. Notice that
the second to last term on the right-hand side is multiplied by $p_R(m, t)$
since the transition from state $R$ to state $A$ depends on the probability of
being in state $R$ to begin with. It is through this term that the dynamics of
the two probability distribution functions ($p_R(m,t)$ and $p_A(m,t)$) are
coupled. An equivalent equation can be written for the probability of having $m$
mRNA at time $t + \Delta t$ while in the repressor bound state, the only
difference being that the mRNA production rates are removed, and the sign for
the promoter state transitions are inverted. This is
\begin{equation}
\begin{aligned}
p_R(m, t + \Delta t) &= p_R(m, t) \\
&+ \overbrace{(m+1)(\gamma \Delta t) p_R(m+1, t)}^{m+1 \rightarrow m}
- \overbrace{m(\gamma \Delta t) p_R(m, t)}^{m \rightarrow m-1}\\
&-\overbrace{(k_R^- \Delta t) p_R(m, t)}^{R \rightarrow A}
+\overbrace{(k_R^+ \Delta t) p_A(m, t)}^{A \rightarrow R}.
\end{aligned}
\end{equation}

All we have to do now are simple algebraic steps in order to simplify the 
equations. Let us focus on the transcriptionally active state $A$. First we will
send the term $p_A(m, t)$ to the right-hand side, and then we will divide both
sides of the equation by $\Delta t$. This results in
\begin{equation}
\begin{aligned}
\frac{p_A(m, t + \Delta t) - p_A(m, t)}{\Delta t} &=
r p_A(m-1, t) - r p_A(m, t)\\
&+ (m+1)\gamma p_A(m+1, t)
- m \gamma p_A(m, t)\\
&+k_R^- p_R(m, t)
-k_R^+ p_A(m, t).
\end{aligned}
\end{equation}
Upon taking the limit when $\Delta t \rightarrow 0$ we can transform the 
left-hand side into a derivative, obtaining the chemical master equation
\begin{equation}
\begin{aligned}
\frac{d p_A(m, t)}{dt} &=
r p_A(m-1, t) - r p_A(m, t)\\
&+ (m+1)\gamma p_A(m+1, t)
- m \gamma p_A(m, t)\\
&+k_R^- p_R(m, t)
-k_R^+ p_A(m, t).
\end{aligned}
\end{equation}
Doing equivalent manipulations for the repressor bound state gives an ODE of the
form
\begin{equation}
\begin{aligned}
\frac{d p_R(m, t)}{dt} &=
(m+1)\gamma p_R(m+1, t)
- m \gamma p_R(m, t)\\
&-k_R^- p_R(m, t)
+k_R^+ p_A(m, t).
\end{aligned}
\end{equation}
In the next section we will write these coupled ODEs in a more compact form using
matrix notation.

\subsection{Matrix form of the multi-state chemical master equation}

Having derived an example chemical master equation we now focus on writing a
general matrix form for the kinetic models 1-4 shown in
Figure~\ref{fig1:means_cartoons}(C) in the main text. In each of these four
models, the natural mathematicization of their cartoons is as a chemical master
equation such as Eq.~\ref{eq:2state_rep_cme} for model 1, which for
completeness we reproduce here as
\begin{align}
\begin{split}
\deriv{t}p_R(m,t) =& 
- \overbrace{k_R^- p_R(m,t)}^{R \rightarrow U}
+ \overbrace{k_R^+ p_U(m,t)}^{U \rightarrow R}
+ \overbrace{(m+1)\gamma p_R(m+1,t)}^{m + 1 \rightarrow m}
- \overbrace{\gamma mp_R(m,t)}^{m \rightarrow m - 1}
\\
\deriv{t}p_U(m,t) =&\; 
\overbrace{k_R^- p_R(m,t)}^{R \rightarrow U}
- \overbrace{k_R^+ p_U(m,t)}^{U \rightarrow R}
+ \overbrace{rp_U(m-1,t)}^{m-1 \rightarrow m}
- \overbrace{rp_U(m,t)}^{m \rightarrow m + 1}
\\
&+ \overbrace{(m+1)\gamma p_U(m+1,t)}^{m + 1 \rightarrow m}
- \overbrace{\gamma mp_U(m,t)}^{m \rightarrow m - 1}.
\label{eq:poisson_promoter_cme_appdx}
\end{split}
\end{align}
Here $p_R(m,t)$ and $p_U(m,t)$ are the probabilities of finding the system with
$m$ mRNA molecules at time $t$ either in the repressor bound or unbound states,
respectively. $r$ is the probability per unit time that a transcript will be
initiated when the repressor is unbound, and $\gamma$ is the probability per
unit time for a given mRNA to be degraded. $k_R^-$ is the probability per unit
time that a bound repressor will unbind, while $k_R^+$ is the probability per
unit time that an unbound operator will become bound by a repressor. Assuming
mass action kinetics, $k_R^+$ is proportional to repressor copy number $R$.

Next consider the cartoon for nonequilibrium model 2 in
Figure~\ref{fig1:means_cartoons}(C). Now we must track probabilities $p_R$,
$p_P$, and $p_E$ for the repressor bound, empty, and polymerase bound states,
respectively. By analogy to Eq.~\ref{eq:poisson_promoter_cme_appdx}, the master
equation for model 2 can be written
\begin{align}
\begin{split}
\deriv{t}p_R(m,t) =& 
- \overbrace{k_R^- p_R(m,t)}^{R \rightarrow U}
+ \overbrace{k_R^+ p_E(m,t)}^{U \rightarrow R}
+ \overbrace{(m+1)\gamma p_R(m+1,t)}^{m + 1 \rightarrow m}
- \overbrace{\gamma mp_R(m,t)}^{m \rightarrow m - 1}
\\
\deriv{t}p_E(m,t) =&\; 
  \overbrace{k_R^- p_R(m,t)}^{R \rightarrow U}
- \overbrace{k_R^+ p_E(m,t)}^{U \rightarrow R}
+ \overbrace{(m+1)\gamma p_E(m+1,t)}^{m + 1 \rightarrow m}
- \overbrace{\gamma mp_E(m,t)}^{m \rightarrow m - 1}.
\\
&
+ \overbrace{k_P^- p_P(m,t)}^{A \rightarrow U}
- \overbrace{k_P^+ p_E(m,t)}^{U \rightarrow A}
+ \overbrace{rp_P(m-1,t)}^{m-1 \rightarrow m,\;A \rightarrow U}
\\
\deriv{t}p_P(m,t) =&\; 
- \overbrace{k_P^- p_P(m,t)}^{A \rightarrow U}
+ \overbrace{k_P^+ p_E(m,t)}^{U \rightarrow A}
+ \overbrace{(m+1)\gamma p_P(m+1,t)}^{m + 1 \rightarrow m}
- \overbrace{\gamma mp_P(m,t)}^{m \rightarrow m - 1}.
\\
&- \overbrace{rp_P(m,t)}^{m \rightarrow m + 1,\;A \rightarrow U}.
\label{eq:model2_cme_appdx}
\end{split}
\end{align}
$k_P^+$ and $k_P^-$ are defined in close analogy to $k_R^+$ and $k_R^-$, except
for RNAP binding and unbinding instead of repressor. Similarly $p_P(m,t)$ is
defined for the active (RNAP-bound) state exactly as are $p_R(m,t)$ and
$p_E(m,t)$ for the repressor bound and unbound states, respectively. The new
subtlety Eq.~\ref{eq:model2_cme_appdx} introduces compared to
Eq.~\ref{eq:poisson_promoter_cme_appdx} is that when mRNAs are produced, the
promoter state also changes. This is encoded by the terms involving $r$, the
last term in each of the equations for $p_E$ and $p_P$. The former accounts for
arrivals in the unbound state and the latter accounts for departures from the
RNAP-bound state.

To condense and clarify the unwieldy notation of Eq.~\ref{eq:model2_cme_appdx},
it can be rewritten in matrix form. We define the column vector $\vec{p}(m,t)$
as
\begin{equation}
\vec{p}(m,t)
= \begin{pmatrix} p_R(m,t) \\ p_E(m,t) \\ p_P(m,t) \end{pmatrix}
\end{equation}
to gather, for a given $m$, the probabilities of finding the system in the three
possible promoter states. Then all the transition rates may be condensed into
matrices which multiply this vector. The first matrix is
\begin{equation}
\mathbf{K} = \begin{pmatrix} -k_R^- & k_R^+ & 0 \\
                        k_R^- & -k_R^+ -k_P^+ & k_P^- \\
                        0 & k_P^+ & -k_P^- 
                \end{pmatrix},
\end{equation}
which tracks all promoter state changes in Eq.~\ref{eq:model2_cme_appdx} that
are \textit{not} accompanied by a change in the mRNA copy number. The two terms
accounting for transcription, the only transition that increases mRNA copy
number, must be handled by two separate matrices given by
\begin{equation}
\mathbf{R_A} = \begin{pmatrix}
                0 & 0 & 0 \\ 
                0 & 0 & r \\ 
                0 & 0 & 0
                \end{pmatrix},\
\mathbf{R_D} = \begin{pmatrix}
                0 & 0 & 0 \\ 
                0 & 0 & 0 \\ 
                0 & 0 & r
                \end{pmatrix}.
\end{equation}
$\mathbf{R_A}$ accounts for transitions \textit{arriving} in a given state while
$\mathbf{R_D}$ tracks \textit{departing} transitions. With these definitions, we
can condense Eq.~\ref{eq:model2_cme_appdx} into the single equation
\begin{equation}
\frac{d}{dt} \vec{p}(m,t) =
\left( \mathbf{K} - \mathbf{R_D} - \gamma m \mathbf{I} \right) \vec{p}(m,t)
                + \mathbf{R_A} \vec{p}(m-1,t) +
                \gamma (m+1) \mathbf{I} \vec{p}(m+1,t),
\label{eq:generic_cme_appdx}
\end{equation}
which is just Eq.~\ref{eq:3state_rep_cme} in the main text. Straightforward
albeit tedious algebra verifies that Eqs.~\ref{eq:model2_cme_appdx}
and~\ref{eq:generic_cme_appdx} are in fact equivalent.

Although we derived Eq.~\ref{eq:generic_cme_appdx} for the particular case of
nonequilibrium model 2 in Figure~\ref{fig1:means_cartoons}, in fact the
chemical master equations for all of the nonequilibrium models in
Figure~\ref{fig1:means_cartoons} except for model 5 can be cast in this form.
(We treat model 5 separately in Appendix~\ref{sec:gen_fcn_appdx}.) Model
3 introduces no new subtleties beyond model 2 and
Eq.~\ref{eq:generic_cme_appdx} applies equally well to it, simply with different
matrices of dimension four instead of three. Models 1 and 4 are both
handled by Eq.~\ref{eq:2state_rep_cme} in the main text, which is just
Eq.~\ref{eq:generic_cme_appdx} except in the special case of $\mathbf{R_D} =
\mathbf{R_A} \equiv \mathbf{R}$, since in these two models transcription
initiation events do not change promoter state.

Recalling that our goal in this section is to derive expressions for mean mRNA
and Fano factor for nonequilibrium models 1 through four in
Figure~\ref{fig1:means_cartoons}, and since all four of these models are
described by Eq.~\ref{eq:generic_cme_appdx}, we can save substantial effort by
deriving general formulas for mean mRNA and Fano factor from
Eq.~\ref{eq:generic_cme_appdx} once and for all. Then for each model we can
simply plug in the appropriate matrices for $\mathbf{K}$, $\mathbf{R_D}$, and
$\mathbf{R_A}$ and carry out the remaining algebra.

For our purposes it will suffice to derive the first and second moments of the
mRNA distribution from this master equation, similar to the treatment
in~\cite{Sanchez2011}, but we refer the interested reader
to~\cite{Razo-Mejia2020} for an analogous treatment demonstrating an analytical
solution for arbitrary moments.

\subsection{General forms for mean mRNA and Fano factor}
Our task now is to derive expressions for the first two moments of the
steady-state mRNA distribution from Eq.~\ref{eq:generic_cme_appdx}. Our
treatment of this is analogous to that given in Refs.~\cite{Sanchez2011}
and~\cite{Razo-Mejia2020}. We first obtain the steady-state limit of
Eq.~\ref{eq:generic_cme_appdx} in which the time derivative vanishes, giving
\begin{equation}
0 =
\left( \mathbf{K} - \mathbf{R_D} - \gamma m \mathbf{I} \right) \vec{p}(m)
                + \mathbf{R_A} \vec{p}(m-1) +
                \gamma (m+1) \mathbf{I} \vec{p}(m+1),
\label{eq:generic_cme_ss}
\end{equation}
From this, we want to compute
\begin{equation}
\langle{m}\rangle = \sum_S \sum_{m=0}^\infty m \, p_S(m)
\end{equation}
and
\begin{equation}
\langle{m^2}\rangle = \sum_S \sum_{m=0}^\infty m^2 p_S(m)
\end{equation}
which define the average values of $m$ and $m^2$ at steady state, where the
averaging is over all possible mRNA copy numbers and promoter states $S$. For
example, for model 1 in Figure~\ref{fig1:means_cartoons}(C), the sum on $S$
would cover repressor bound and unbound states ($R$ and $U$ respectively), for
model 2, the sum would cover repressor bound, polymerase bound, and empty
states ($R$, $P$, and $E$), and so on for the other models.

Along the way it will be convenient to define the following
\textit{conditional} moments as
\begin{equation}
\langle\vec{m}\rangle = \sum_{m=0}^\infty m \vec{p}(m),
\end{equation}
and
\begin{equation}
\langle\vec{m}^2\rangle = \sum_{m=0}^\infty m^2 \vec{p}(m).
\end{equation}
These objects are vectors of the same size as $\vect{p}(m)$, and each component
can be thought of as the expected value of the mRNA copy number, or copy number
squared, conditional on the promoter being in a certain state. For example, for
model 1 in Figure~\ref{fig1:means_cartoons} which has repressor bound and
unbound states labeled $R$ and $U$, $\langle\vec{m}^2\rangle$ would be
\begin{equation}
\langle\vec{m}^2\rangle
= \begin{pmatrix} \sum_{m=0}^\infty m^2 p_R(m)
                \\ \sum_{m=0}^\infty m^2 p_U(m) \end{pmatrix}.
\end{equation}
Analogously to $\langle\vec{m}\rangle$ and $\langle\vec{m}^2\rangle$,
it is convenient to define the vector
\begin{equation}
\langle\vec{m}^0\rangle = \sum_{m=0}^\infty \vec{p}(m),
\end{equation}
whose elements are simply the probabilities of finding the system in each of the
possible promoter states. It will be
convenient to denote by $\vec{1}^\dagger$ a row vector of the same length as
$\vec{p}$ whose elements are all 1, such that, for instance, $\vec{1}^\dagger
\cdot \langle\vec{m}^0\rangle = 1$, $\vec{1}^\dagger \cdot \langle\vec{m}\rangle
= \langle{m}\rangle$, etc.

\subsubsection{Promoter state probabilities $\langle\vec{m}^0\rangle$}
\label{sec:m0_appdx}
To begin, we will find the promoter state probabilities
$\langle\vec{m}^0\rangle$ from Eq.~\ref{eq:generic_cme_ss} by summing over all
mRNA copy numbers $m$, producing
\begin{equation}
0 = \sum_{m=0}^\infty \left[
    \left( \mathbf{K} - \mathbf{R_D} - \gamma m \mathbf{I} \right) \vec{p}(m)
                + \mathbf{R_A} \vec{p}(m-1) +
                \gamma (m+1) \mathbf{I} \vec{p}(m+1)
\right]
\end{equation}
Using the definitions of $\langle\vec{m}^0\rangle$ and $\langle\vec{m}\rangle$,
and noting the matrices $\mathbf{K}$, $\mathbf{R_D}$, and $\mathbf{R_A}$
are all independent of $m$ and can be moved outside the sum, this simplifies to
\begin{equation}
0 = (\mathbf{K} - \mathbf{R_D}) \langle\vec{m}^0\rangle
    - \gamma \langle\vec{m}\rangle + \mathbf{R_A} \sum_{m=0}^\infty \vec{p}(m-1)
    + \gamma \sum_{m=0}^\infty (m+1)\vec{p}(m+1).
\label{eq:generic_cme_deriv_020}
\end{equation}
The last two terms can be handled by reindexing the summations, transforming
them to match the definitions of $\langle\vec{m}^0\rangle$ and
$\langle\vec{m}\rangle$. For the first, noting $\vec{p}(-1)=0$ since negative
numbers of mRNA are nonsensical, we have
\begin{equation}
\sum_{m=0}^\infty \vec{p}(m-1)
= \sum_{m=-1}^\infty \vec{p}(m)
= \sum_{m=0}^\infty \vec{p}(m) = \langle\vec{m}^0\rangle.
\end{equation}
Similarly for the second,
\begin{equation}
\sum_{m=0}^\infty (m+1)\vec{p}(m+1)
= \sum_{m=1}^\infty m\vec{p}(m)
= \sum_{m=0}^\infty m\vec{p}(m) = \langle\vec{m}\rangle,
\end{equation}
which holds since in extending the lower limit from $m=1$ to $m=0$, the extra
term we added to the sum is zero. Substituting these back in
Eq.~\ref{eq:generic_cme_deriv_020}, we have
\begin{equation}
0 = (\mathbf{K} - \mathbf{R_D}) \langle\vec{m}^0\rangle
    - \gamma \langle\vec{m}\rangle + \mathbf{R_A} \langle\vec{m}^0\rangle
    + \gamma \langle\vec{m}\rangle,
\end{equation}
or simply
\begin{equation}
0 = (\mathbf{K} - \mathbf{R_D} + \mathbf{R_A}) \langle\vec{m}^0\rangle.
\label{eq:generic_cme_vecm0}
\end{equation}
So given matrices $\mathbf{K}$, $\mathbf{R_D}$, and $\mathbf{R_A}$ describing a
promoter, finding $\langle\vec{m}^0\rangle$ simply amounts to solving this set
of linear equations, subject to the normalization constraint $\vec{1}^\dagger
\cdot \langle\vec{m}^0\rangle = 1$. It will turn out to be the case that, if the
matrix $\mathbf{K} - \mathbf{R_D} + \mathbf{R_A}$ is $n$ dimensional, it will
always have only $n-1$ linearly independent equations. Including the
normalization condition provides the $n$-th linearly independent equation,
ensuring a unique solution. So when using this equation to solve for
$\langle\vec{m}^0\rangle$, we may always drop one row of the matrix equation at
our convenience and supplement the system with the normalization condition
instead.
The reader may find it illuminating to skip ahead and see
Eq.~\ref{eq:generic_cme_vecm0} in use with concrete examples, e.g.,
Eq.~\ref{eq:model1_m0_giver_appdx} and what follows it, before
continuing on through the general formulas for moments.

\subsubsection{First moments $\langle\vec{m}\rangle$ and $\langle{m}\rangle$}
By analogy to the above procedure for finding $\langle\vec{m}^0\rangle$, we may
find $\langle\vec{m}\rangle$ by first multiplying Eq.~\ref{eq:generic_cme_ss} by
$m$ and then summing over $m$. Carrying out this procedure we have
\begin{equation}
0 = \sum_{m=0}^\infty m \left[
\left( \mathbf{K} - \mathbf{R_D} - \gamma m \mathbf{I} \right) \vec{p}(m)
            + \mathbf{R_A} \vec{p}(m-1) +
            \gamma (m+1) \mathbf{I} \vec{p}(m+1)
\right],
\end{equation}
and now identifying $\langle\vec{m}\rangle$ and $\langle\vec{m}^2\rangle$ gives
\begin{equation}
0 = (\mathbf{K} - \mathbf{R_D}) \langle\vec{m}\rangle
    - \gamma \langle\vec{m}^2\rangle + \mathbf{R_A} \sum_{m=0}^\infty m\vec{p}(m-1)
    + \gamma \sum_{m=0}^\infty m(m+1)\vec{p}(m+1).
\label{eq:generic_cme_deriv_030}
\end{equation}
The summations in the last two terms can be reindexed just as we did for
$\langle\vec{m}^0\rangle$, freely adding or removing terms from the sum which
are zero. For the first term we find
\begin{equation}
\sum_{m=0}^\infty m\vec{p}(m-1)
= \sum_{m=-1}^\infty (m+1)\vec{p}(m)
= \sum_{m=0}^\infty (m+1)\vec{p}(m)
= \langle\vec{m}\rangle + \langle\vec{m}^0\rangle,
\end{equation}
and similarly for the second,
\begin{equation}
\sum_{m=0}^\infty m(m+1)\vec{p}(m+1)
= \sum_{m=1}^\infty (m-1)m\vec{p}(m)
= \sum_{m=0}^\infty (m-1) m\vec{p}(m)
= \langle\vec{m}^2\rangle - \langle\vec{m}\rangle.
\end{equation}
Substituting back in Eq.~\ref{eq:generic_cme_deriv_030} then produces
\begin{equation}
0 = (\mathbf{K} - \mathbf{R_D}) \langle\vec{m}\rangle
- \gamma \langle\vec{m}^2\rangle
+ \mathbf{R_A} (\langle\vec{m}\rangle + \langle\vec{m}^0\rangle)
+ \gamma (\langle\vec{m}^2\rangle - \langle\vec{m}\rangle),
\end{equation}
or after simplifying
\begin{equation}
0 = (\mathbf{K} - \mathbf{R_D} + \mathbf{R_A} - \gamma) \langle\vec{m}\rangle
+ \mathbf{R_A} \langle\vec{m}^0\rangle.
\label{eq:generic_cme_deriv_040}
\end{equation}
So like $\langle\vec{m}^0\rangle$, $\langle\vec{m}\rangle$ is also found by
simply solving a set of linear equations after first solving for
$\langle\vec{m}^0\rangle$ from Eq.~\ref{eq:generic_cme_vecm0}.

Next we can find the mean mRNA copy number $\langle{m}\rangle$ from
$\langle\vec{m}\rangle$ according to
\begin{equation}
\langle{m}\rangle = \vec{1}^\dagger\cdot\langle\vec{m}\rangle,
\label{eq:m_from_vecm}
\end{equation}
where $\vec{1}^\dagger$ is a row vector whose elements are all 1.
Eq.~\ref{eq:m_from_vecm} holds since the $i^{th}$ element of the column vector
$\langle\vec{m}\rangle$ is the mean mRNA value conditional on the system
occupying the $i^{th}$ promoter state, so the dot product with $\vec{1}^\dagger$
amounts to simply summing the elements of $\langle\vec{m}\rangle$. Rather than
solving Eq.~\ref{eq:generic_cme_deriv_040} for $\langle\vec{m}\rangle$ and then
computing $\vec{1}^\dagger\cdot\langle\vec{m}\rangle$, we may take a shortcut by
multiplying Eq.~\ref{eq:generic_cme_deriv_040} by $\vec{1}^\dagger$ first. The
key observation that makes this useful is that
\begin{equation}
\vec{1}^\dagger \cdot (\mathbf{K} - \mathbf{R_D} + \mathbf{R_A}) = 0.
\end{equation}
Intuitively, this equality holds because each column of this matrix
represents transitions to and from a given promoter state. In any given column,
the diagonal encodes all departing transitions and off-diagonals encode all
arriving transitions. Conservation of probability means that each column
sums to zero, and summing columns is exactly the operation that multiplying by
$\vec{1}^\dagger$ carries out.

Proceeding then in multiplying Eq.~\ref{eq:generic_cme_deriv_040} by
$\vec{1}^\dagger$ produces
\begin{equation}
0 = -\gamma \vec{1}^\dagger\cdot\langle\vec{m}\rangle
+ \vec{1}^\dagger\cdot\mathbf{R_A}\langle\vec{m}^0\rangle,
\end{equation}
or simply
\begin{equation}
\langle{m}\rangle
= \frac{1}{\gamma} \vec{1}^\dagger\cdot\mathbf{R_A}\langle\vec{m}^0\rangle.
\label{eq:generic_mean_m_appdx}
\end{equation}
We note that the in equilibrium models of transcription such as in
Figure~\ref{fig1:means_cartoons}, it is usually \textit{assumed} that the mean
mRNA level is given by the ratio of initiation rate $r$ to degradation rate
$\gamma$ multiplied by the probability of finding the system in the RNAP-bound
state. Reassuringly, we have recovered exactly this result from the master
equation picture: the product
$\vec{1}^\dagger\cdot\mathbf{R_A}\langle\vec{m}^0\rangle$ picks out the
probability of the active promoter state from $\langle\vec{m}^0\rangle$ and
multiplies it by the initiation rate $r$.

\subsubsection{Second moment $\langle{m}^2\rangle$ and Fano factor $\nu$}
Continuing the pattern of the zeroth and first moments, we now find
$\langle\vec{m}^2\rangle$ by multiplying Eq.~\ref{eq:generic_cme_ss} by $m^2$
and then summing over $m$, which explicitly is
\begin{equation}
0 = \sum_{m=0}^\infty m^2 \left[
\left( \mathbf{K} - \mathbf{R_D} - \gamma m \mathbf{I} \right) \vec{p}(m)
            + \mathbf{R_A} \vec{p}(m-1) +
            \gamma (m+1) \mathbf{I} \vec{p}(m+1)
\right].
\end{equation}
Identifying the moments $\langle\vec{m}^2\rangle$ and $\langle\vec{m}^3\rangle$
in the first term simplifies this to
\begin{equation}
0 = (\mathbf{K} - \mathbf{R_D}) \langle\vec{m}^2\rangle
    - \gamma \langle\vec{m}^3\rangle + \mathbf{R_A} \sum_{m=0}^\infty m^2\vec{p}(m-1)
    + \gamma \sum_{m=0}^\infty m^2(m+1)\vec{p}(m+1).
\label{eq:generic_cme_deriv_050}
\end{equation}
Reindexing the sums of the last two terms proceeds just as it did for the zeroth
and first moments. Explicitly, we have
\begin{equation}
\sum_{m=0}^\infty m^2\vec{p}(m-1)
= \sum_{m=-1}^\infty (m+1)^2\vec{p}(m)
= \sum_{m=0}^\infty (m+1)^2\vec{p}(m)
= \langle\vec{m}^2\rangle + 2\langle\vec{m}\rangle + \langle\vec{m}^0\rangle,
\end{equation}
for the first sum and
\begin{equation}
\sum_{m=0}^\infty m^2(m+1)\vec{p}(m+1)
= \sum_{m=1}^\infty (m-1)^2m\vec{p}(m)
= \sum_{m=0}^\infty (m-1)^2 m\vec{p}(m)
= \langle\vec{m}^3\rangle - 2\langle\vec{m}^2\rangle + \langle\vec{m}\rangle
\end{equation}
for the second. Substituting the results of the sums back in
Eq.~\ref{eq:generic_cme_deriv_050} gives
\begin{equation}
0 = (\mathbf{K} - \mathbf{R_D}) \langle\vec{m}^2\rangle
- \gamma \langle\vec{m}^3\rangle
+ \mathbf{R_A}
    (\langle\vec{m}^2\rangle + 2\langle\vec{m}\rangle + \langle\vec{m}^0\rangle)
+ \gamma
    (\langle\vec{m}^3\rangle - 2\langle\vec{m}^2\rangle + \langle\vec{m}\rangle),
\end{equation}
and after grouping like powers of $m$ we have
\begin{equation}
0 = (\mathbf{K} - \mathbf{R_D} + \mathbf{R_A} - 2\gamma) \langle\vec{m}^2\rangle
+ (2\mathbf{R_A} + \gamma) \langle\vec{m}\rangle
+ \mathbf{R_A} \langle\vec{m}^0\rangle.
\label{eq:generic_cme_deriv_060}
\end{equation}
As we found when computing $\langle{m}\rangle$ from $\langle\vec{m}\rangle$, we
can spare ourselves some algebra by multiplying
Eq.~\ref{eq:generic_cme_deriv_060} by $\vect{1}^\dagger$, which then reduces to
\begin{equation}
0 = - 2\gamma \langle{m}^2\rangle
+ \vec{1}^\dagger\cdot(2\mathbf{R_A} + \gamma) \langle\vec{m}\rangle
+ \vec{1}^\dagger\cdot\mathbf{R_A} \langle\vec{m}^0\rangle,
\end{equation}
and noting from Eq.~\ref{eq:generic_mean_m_appdx} that
$\vec{1}^\dagger\cdot\mathbf{R_A} \langle\vec{m}^0\rangle
= \gamma\langle{m}\rangle$, we have the tidy result
\begin{equation}
\langle{m}^2\rangle
= \langle{m}\rangle + \frac{1}{\gamma}
        \vec{1}^\dagger\cdot\mathbf{R_A} \langle\vec{m}\rangle.
\end{equation}

Finally we have all the preliminary results needed to write a general expression
for the Fano factor $\nu$. The Fano factor is defined as the ratio of variance
to mean, which can be written as
\begin{equation}
\nu = \frac{\langle{m}^2\rangle - \langle{m}\rangle^2}{\langle{m}\rangle}
= \frac{
    \langle{m}\rangle + \frac{1}{\gamma}
        \vec{1}^\dagger\cdot\mathbf{R_A} \langle\vec{m}\rangle
    - \langle{m}\rangle^2
    }{\langle{m}\rangle}
\end{equation}
and simplified to
\begin{equation}
\nu = 1 - \langle{m}\rangle
+ \frac{\vec{1}^\dagger\cdot \mathbf{R_A}\langle\vec{m}\rangle}
        {\gamma \langle{m}\rangle}.
\label{eq:generic_fano_appdx}
\end{equation}
Note a subtle notational trap here: $\langle{m}\rangle = \frac{1}{\gamma}
\vec{1}^\dagger\cdot\mathbf{R_A}\langle\vec{m}^0\rangle$ rather than the by-eye
similar but wrong expression $\langle{m}\rangle \ne \frac{1}{\gamma}
\vec{1}^\dagger\cdot\mathbf{R_A}\langle\vec{m}\rangle$, so the last term in
Eq.~\ref{eq:generic_fano_appdx} is in general quite nontrivial. For a generic
promoter, Eq.~\ref{eq:generic_fano_appdx} may be greater than, less than, or
equal to one, as asserted in Section~\ref{sec:beyond_means}. We have not found
the general form Eq.~\ref{eq:generic_fano_appdx} terribly intuitive and instead
defer discussion to specific examples.

\subsubsection{Summary of general results}
For ease of reference, we collect and reprint here the key results derived in
this section that are used in the main text and subsequent subsections. Mean
mRNA copy number and Fano factor are given by Eqs.~\ref{eq:generic_mean_m_appdx}
and \ref{eq:generic_fano_appdx}, which are
\begin{equation}
\langle{m}\rangle
= \frac{1}{\gamma} \vec{1}^\dagger\cdot\mathbf{R_A}\langle\vec{m}^0\rangle
\label{eq:mean_m_appdx_ref}
\end{equation}
and
\begin{equation}
\nu = 1 - \langle{m}\rangle
+ \frac{\vec{1}^\dagger\cdot \mathbf{R_A}\langle\vec{m}\rangle}
        {\gamma \langle{m}\rangle},
\label{eq:fano_appdx_ref}
\end{equation}
respectively. To compute these two quantities, we need the expressions for
$\langle\vec{m}^0\rangle$ and $\langle\vec{m}\rangle$ given by solving
Eqs.~\ref{eq:generic_cme_vecm0} and \ref{eq:generic_cme_deriv_040},
respectively, which are
\begin{equation}
(\mathbf{K} - \mathbf{R_D} + \mathbf{R_A}) \langle\vec{m}^0\rangle = 0
\label{eq:vecm0_appdx_ref}
\end{equation}
and
\begin{equation}
(\mathbf{K} - \mathbf{R_D}
+ \mathbf{R_A} - \gamma\mathbf{I}) \langle\vec{m}\rangle
= - \mathbf{R_A} \langle\vec{m}^0\rangle.
\label{eq:vecm_appdx_ref}
\end{equation}
Some comments are in order before we consider particular models. First, note
that to obtain $\langle\vec{m}\rangle$ and $\nu$, we need not bother solving for
all components of the vectors $\langle\vec{m}^0\rangle$ and
$\langle\vec{m}\rangle$, but only the components which are multiplied by nonzero
elements of $\mathbf{R_A}$. The only component of $\langle\vec{m}^0\rangle$ that
ever survives is the transciptionally active state, and for the models we
consider here, there is only ever one such state. This will save us some amount
of algebra below.

Also note that we are computing Fano factors to verify the results of
Section~\ref{sec:beyond_means}, concerning the constitutive promoter models in
Figure~\ref{fig2:constit_cartoons} which are analogs of the simple repression
models in Figure~\ref{fig1:means_cartoons}. We can translate the matrices from
the simple repression models to the constitutive case by simply substituting all
occurrences of repressor rates by zero and removing the row and column
corresponding to the repressor bound state. The results for $\langle{m}\rangle$
computed in the repressed case can be easily translated to the constitutive
case, rather than recalculating from scratch, by taking the limit
$k_R^+\rightarrow 0$, since this amounts to sending repressor copy number to
zero.

Finally, we point out that it would be possible to compute
$\langle\vec{m}^0\rangle$ more simply using the diagram methods from King and
Altman~\cite{King1956} (also independently discovered by Hill~\cite{Hill1966}).
But to our knowledge this method cannot be applied to compute
$\langle\vec{m}\rangle$ or $\nu$, so since we would need to resort to solving
the matrix equations anyways for $\langle\vec{m}\rangle$, we choose not to
introduce the extra conceptual burden of the diagram methods simply for
computing $\langle\vec{m}^0\rangle$.

\subsection{Nonequilibrium Model One - Poisson Promoter}
\subsubsection{Mean mRNA}
For nonequilibrium model 1 in Figure~\ref{fig1:means_cartoons}, we have
already shown the full master equation in Eq.~\ref{eq:poisson_promoter_cme} and
Eq.~\ref{eq:poisson_promoter_cme_appdx}, but for completeness we reprint it
again as
\begin{align}
\begin{split}
\deriv{t}p_R(m,t) =& 
- \overbrace{k_R^- p_R(m,t)}^{R \rightarrow U}
+ \overbrace{k_R^+ p_U(m,t)}^{U \rightarrow R}
+ \overbrace{(m+1)\gamma p_R(m+1,t)}^{m + 1 \rightarrow m}
- \overbrace{\gamma mp_R(m,t)}^{m \rightarrow m - 1}
\\
\deriv{t}p_U(m,t) =&\; 
\overbrace{k_R^- p_R(m,t)}^{R \rightarrow U}
- \overbrace{k_R^+ p_U(m,t)}^{U \rightarrow R}
+ \overbrace{rp_U(m-1,t)}^{m-1 \rightarrow m}
- \overbrace{rp_U(m,t)}^{m \rightarrow m + 1}
\\
&+ \overbrace{(m+1)\gamma p_U(m+1,t)}^{m + 1 \rightarrow m}
- \overbrace{\gamma mp_U(m,t)}^{m \rightarrow m - 1}.
\end{split}
\end{align}
This is a direct transcription of the states and rates in
Figure~\ref{fig1:means_cartoons}. This may be converted to the matrix form of
the master equation shown in Eq.~\ref{eq:generic_cme_appdx} with matrices
\begin{equation}
\vec{p}(m) = \begin{pmatrix} p_R(m) \\ p_U(m) \end{pmatrix},\
\mathbf{K} = \begin{pmatrix} -k_R^- & k_R^+ \\ k_R^- & -k_R^+ \end{pmatrix},\
\mathbf{R} = \begin{pmatrix} 0 & 0 \\ 0 & r \end{pmatrix},\
\end{equation}
where $\mathbf{R_A}$ and $\mathbf{R_D}$ are equal, so we drop the subscript and
denote both simply by $\mathbf{R}$. Since our interest is only in steady-state
we dropped the time dependence as well.

First we need $\langle\vec{m}^0\rangle$. Label its components as $p_R$ and
$p_U$, the probabilities of finding the system in either promoter state, and
note that only $p_U$ survives multiplication by $\mathbf{R}$, since
\begin{equation}
\mathbf{R} \langle\vec{m}^0\rangle
= \begin{pmatrix} 0 & 0 \\ 0 & r \end{pmatrix}
    \begin{pmatrix} p_R \\ p_U \end{pmatrix}
= \begin{pmatrix} 0 \\ r p_U \end{pmatrix},
\label{eq:Rm0_model1_appdx}
\end{equation}
so we need not bother finding $p_R$. Then we have
\begin{equation}
(\mathbf{K} - \mathbf{R_D} + \mathbf{R_A}) \langle\vec{m}^0\rangle
= \begin{pmatrix} -k_R^- & k_R^+ \\ k_R^- & -k_R^+ \end{pmatrix}
    \begin{pmatrix} p_R \\ p_U \end{pmatrix} = 0.
\label{eq:model1_m0_giver_appdx}
\end{equation}
As mentioned earlier in Section~\ref{sec:m0_appdx}, the two rows are linearly
dependent, so taking only the first row and using normalization to set $p_R =
1-p_U$ gives
\begin{equation}
-k_R^- (1-p_U) + k_R^+ p_U = 0,
\end{equation}
which is easily solved to find
\begin{equation}
p_U = \frac{k_R^-}{k_R^- + k_R^+}.
\end{equation}
Substituting this into Eq.~\ref{eq:Rm0_model1_appdx}, and the result of that
into Eq.~\ref{eq:mean_m_appdx_ref}, we have
\begin{equation}
\langle{m}\rangle = \frac{r}{\gamma} \frac{k_R^-}{k_R^- + k_R^+}
\end{equation}
as asserted in Eq.~\ref{eq:mean_m_model1} of the main text.

\subsubsection{Fano factor}
To verify that the Fano factor for model 1 in
Figure~\ref{fig2:constit_cartoons}(A) is in fact 1 as claimed in the main text,
note that in this limit $p_U = 1$ and $\langle{m}\rangle = r/\gamma$. All
elements of $\mathbf{K}$ are zero, and $\mathbf{R_A}-\mathbf{R_D} = 0$, so
Eq.~\ref{eq:vecm_appdx_ref} reduces to
\begin{equation}
- \gamma \langle\vec{m}\rangle = - r,
\end{equation}
or, in other words, since there is only one promoter state,
$\langle\vec{m}\rangle = \langle{m}\rangle$. Then it follows that
\begin{equation}
\nu = 1 -\frac{r}{\gamma}
    + \frac{r \langle{m}\rangle}{\gamma \langle{m}\rangle}
= 1
\end{equation}
as claimed.

\subsection{Nonequilibrium Model Two - RNAP Bound and Unbound States}
\subsubsection{Mean mRNA}
As shown earlier, the full master equation for model 2 in
Figure~\ref{fig1:means_cartoons} is
\begin{align}
\begin{split}
\deriv{t}p_R(m,t) =& 
- \overbrace{k_R^- p_R(m,t)}^{R \rightarrow U}
+ \overbrace{k_R^+ p_E(m,t)}^{U \rightarrow R}
+ \overbrace{(m+1)\gamma p_R(m+1,t)}^{m + 1 \rightarrow m}
- \overbrace{\gamma mp_R(m,t)}^{m \rightarrow m - 1}
\\
\deriv{t}p_E(m,t) =&\; 
    \overbrace{k_R^- p_R(m,t)}^{R \rightarrow U}
- \overbrace{k_R^+ p_E(m,t)}^{U \rightarrow R}
+ \overbrace{(m+1)\gamma p_E(m+1,t)}^{m + 1 \rightarrow m}
- \overbrace{\gamma mp_E(m,t)}^{m \rightarrow m - 1}.
\\
&
+ \overbrace{k_P^- p_P(m,t)}^{A \rightarrow U}
- \overbrace{k_P^+ p_E(m,t)}^{U \rightarrow A}
+ \overbrace{rp_P(m-1,t)}^{m-1 \rightarrow m,\;A \rightarrow U}
\\
\deriv{t}p_P(m,t) =&\; 
- \overbrace{k_P^- p_P(m,t)}^{A \rightarrow U}
+ \overbrace{k_P^+ p_E(m,t)}^{U \rightarrow A}
+ \overbrace{(m+1)\gamma p_P(m+1,t)}^{m + 1 \rightarrow m}
- \overbrace{\gamma mp_P(m,t)}^{m \rightarrow m - 1}.
\\
&- \overbrace{rp_P(m,t)}^{m \rightarrow m + 1,\;A \rightarrow U},
\end{split}
\end{align}
which can be condensed to the matrix form of Eq.~\ref{eq:generic_cme_appdx} with
matrices given by
\begin{equation}
\mathbf{K} = \begin{pmatrix} -k_R^- & k_R^+ & 0 \\
                        k_R^- & -k_R^+ -k_P^+ & k_P^- \\
                        0 & k_P^+ & -k_P^- 
                \end{pmatrix},\
\mathbf{R_A} = \begin{pmatrix}
                0 & 0 & 0 \\ 
                0 & 0 & r \\ 
                0 & 0 & 0
                \end{pmatrix},\
\mathbf{R_D} = \begin{pmatrix}
                0 & 0 & 0 \\ 
                0 & 0 & 0 \\ 
                0 & 0 & r
                \end{pmatrix}.
\label{eq:model2_matrices_appdx}
\end{equation}
As for model 1, we must first find $\mathbf{R_A} \langle\vect{m}^0\rangle$.
Denote its components as $p_R$, $p_E$, $p_P$, the probabilities of being found
in repressor bound, empty, or RNAP-bound states, respectively. Only $p_P$ is
necessary to find since
\begin{equation}
\mathbf{R_A} \langle\vec{m}^0\rangle
= \begin{pmatrix} 0 \\ r p_P \\ 0 \end{pmatrix}.
\label{eq:model2_deriv_appdx_020}
\end{equation}
Then Eq.~\ref{eq:vecm0_appdx_ref} for $\langle\vect{m}\rangle$ reads
\begin{equation}
\begin{pmatrix} -k_R^- & k_R^+ & 0 \\
        k_R^- & -k_R^+ -k_P^+ & k_P^- + r\\
        0 & k_P^+ & -k_P^- - r
\end{pmatrix}
\begin{pmatrix}
    p_R \\ p_E \\ p_P
\end{pmatrix}
= 0.
\label{eq:model2_K-R+R_for_m0}
\end{equation}
Discarding the middle row as redundant and incorporating the normalization
condition leads to a set of three linearly independent equations, namely
\begin{align}
-k_R^- p_R + k_R^+ p_E &= 0 \\
k_P^+ p_E + (-k_P^- - r) p_P &= 0 \\
p_R + p_E + p_P &= 1.
\end{align}
Using $p_R = 1 - p_E - p_P$ to eliminate $p_R$ in the first and solving the
resulting equation for $p_E$ gives
$p_E = (1 - p_P){k_R^-}/{(k_R^- + k_R^+)}$.
Substituting this for $p_E$ in the second equation gives an equation in
$p_P$ alone which is
\begin{equation}
k_P^+ k_R^- (1-p_P) - (k_P^- + r)(k_R^+ + k_R^-) p_P = 0
\end{equation}
and solving for $p_P$ gives
\begin{equation}
p_P = \frac{k_P^+ k_R^-}{k_P^+ k_R^- + (k_P^- + r)(k_R^+ + k_R^-)}.
\end{equation}
Substituting this in Eq.~\ref{eq:model2_deriv_appdx_020} and multiplying by
$\mathbf{R_A}$ produces
\begin{equation}
\mathbf{R_A} \langle\vec{m}^0\rangle
= r \frac{k_P^+ k_R^-} {k_P^+ k_R^- + (k_P^- + r)(k_R^+ + k_R^-)}
\begin{pmatrix} 0 \\ 1 \\ 0 \end{pmatrix}
\end{equation}
from which $\langle{m}\rangle$ follows readily,
\begin{equation}
\langle{m}\rangle = \frac{r}{\gamma}
        \frac{k_P^+ k_R^-} {k_P^+ k_R^- + (k_P^- + r)(k_R^+ + k_R^-)},
\label{eq:model2_meanm_appdx}
\end{equation}
as claimed in Eq.~\ref{eq:model2_meanm} in the main text.

\subsubsection{Fano factor}
To compute the Fano factor, we first remove the repressor bound state from the
matrices describing the model, which reduce to
\begin{equation}
\mathbf{K} = \begin{pmatrix}
                -k_P^+ & k_P^- \\
                 k_P^+ &-k_P^- 
                \end{pmatrix},\
\mathbf{R_A} = \begin{pmatrix}
                0 & r \\ 
                0 & 0
                \end{pmatrix},\
\mathbf{R_D} = \begin{pmatrix}
                0 & 0 \\ 
                0 & r
                \end{pmatrix}.    
\end{equation}
Similarly we remove the repressor bound state from $\mathbf{R_A}
\langle\vec{m}^0\rangle$ and take the $k_R^+\rightarrow 0$ limit, which
simplifies to 
\begin{equation}
\mathbf{R_A} \langle\vec{m}^0\rangle
= r \frac{k_P^+ } {k_P^+ + k_P^- + r}
\begin{pmatrix} 1 \\ 0 \end{pmatrix}.
\end{equation}
Then we must compute $\langle\vec{m}\rangle$ from Eq.~\ref{eq:vecm_appdx_ref},
which with these matrices reads
\begin{equation}
(\mathbf{K} - \mathbf{R_D} + \mathbf{R_A} - \gamma\mathbf{I})
    \langle\vec{m}\rangle
= \begin{pmatrix}
    -k_P^+ -\gamma & k_P^- + r\\
    k_P^+          &-k_P^- - r - \gamma
    \end{pmatrix}
    \begin{pmatrix} m_E \\ m_P \end{pmatrix}
= r \frac{k_P^+ } {k_P^+ + k_P^- + r}
    \begin{pmatrix} 1 \\ 0 \end{pmatrix},
\end{equation}
where we labeled the components of $\langle\vec{m}\rangle$ as $m_E$ and $m_P$,
since they are the mean mRNA counts conditional upon the system residing in the
empty or polymerase bound states, respectively. Unlike for
$\langle\vec{m}^0\rangle$, the rows of this matrix are linearly independent so
we simply solve this matrix equation as is. We can immediately eliminate $m_E$
since $m_E = m_P (k_P^- + r + \gamma)/k_P^+$ from the second row, and
substituting into the first row gives an equation for $m_P$ alone, which is
\begin{equation}
\left[-(k_P^+ + \gamma)(k_P^- + r + \gamma) + k_P^+(k_P^- + r)\right] m_P
= - \frac{r (k_P^+)^2}{k_P^+ + k_P^- + r}.
\end{equation}
Expanding the products cancels several terms, and solving for $m_P$ gives
\begin{equation}
m_P = \frac{r (k_P^+)^2}
            {\gamma(k_P^+ + k_P^- + r)(k_P^+ + k_P^- + r + \gamma)}.
\end{equation}
Note then that $\vec{1}^\dagger\cdot\mathbf{R_A}\langle\vec{m}\rangle = rm_P$.
We also need the constitutive limit of $\langle{m}\rangle$ from
Eq.~\ref{eq:model2_meanm_appdx}, again found by taking $k_R^+\rightarrow0$,
which is
\begin{equation}
\langle{m}\rangle = \frac{r}{\gamma} \frac{k_P^+ } {k_P^+ + k_P^- + r}
\end{equation}
and substituting this along with
$\vec{1}^\dagger\cdot\mathbf{R_A}\langle\vec{m}\rangle = rm_P$ into
Eq.~\ref{eq:fano_appdx_ref} for the Fano factor $\nu$, we find
\begin{equation}
\nu = 1 - \frac{r}{\gamma} \frac{k_P^+ } {k_P^+ + k_P^- + r}
    + \frac{r}{\gamma}\frac{r (k_P^+)^2}{\gamma(k_P^+ + k_P^- + r)
                                        (k_P^+ + k_P^- + r + \gamma)}
\left(\frac{r}{\gamma} \frac{k_P^+ } {k_P^+ + k_P^- + r}\right)^{-1}.
\end{equation}
This simplifies to
\begin{equation}
\nu = 1 - \frac{r}{\gamma}
    \left(
        \frac{k_P^+ } {k_P^+ + k_P^- + r}
        - \frac{k_P^+ } {k_P^+ + k_P^- + r + \gamma}
    \right),
\end{equation}
which further simplifies to
\begin{equation}
\nu = 1 - \frac{r k_P^+ } {(k_P^+ + k_P^- + r)(k_P^+ + k_P^- + r + \gamma)},
\end{equation}
exactly Eq.~\ref{eq:model2_fano} in the main text.
    
\subsection{Nonequilibrium Model Three - Multistep Transcription Initiation and
Escape}

\subsubsection{Mean mRNA}
In close analogy to model 2 above, nonequilibrium model 3 from
Figure~\ref{fig1:means_cartoons}(C) can be described by our generic master
equation Eq.~\ref{eq:generic_cme_appdx} with promoter transition matrix given by
\begin{equation}
\mathbf{K} =
\begin{pmatrix} -k_R^- & k_R^+ & 0 & 0\\
        k_R^- & -k_R^+ -k_P^+ & k_P^- & 0 \\
        0 & k_P^+ & -k_P^- - k_O & 0 \\
        0 & 0 & k_O & 0
\end{pmatrix}
\end{equation}
and transcription matrices given by
\begin{equation}
\mathbf{R_A} =
\begin{pmatrix}
        0 & 0 & 0 & 0 \\ 
        0 & 0 & 0 & r \\ 
        0 & 0 & 0 & 0 \\ 
        0 & 0 & 0 & 0
\end{pmatrix},\
\mathbf{R_D} =
\begin{pmatrix}
        0 & 0 & 0 & 0 \\ 
        0 & 0 & 0 & 0 \\ 
        0 & 0 & 0 & 0 \\ 
        0 & 0 & 0 & r
\end{pmatrix}.
\end{equation}
$\langle\vec{m}^0\rangle$ is again given by Eq.~\ref{eq:vecm0_appdx_ref},
which in this case takes the form
\begin{equation}
(\mathbf{K} - \mathbf{R_D} + \mathbf{R_A}) \langle\vec{m}^0\rangle =
\begin{pmatrix} -k_R^- & k_R^+ & 0 & 0\\
    k_R^- & -k_R^+ -k_P^+ & k_P^- & r \\
    0 & k_P^+ & -k_P^- - k_O & 0 \\
    0 & 0 & k_O & - r
\end{pmatrix}
\begin{pmatrix} p_R \\ p_E \\ p_C \\ p_O
\end{pmatrix} = 0,
\end{equation}
where the four components of $\langle\vec{m}^0\rangle$ correspond to the four
promoter states repressor bound, empty, RNAP-bound closed complex, and
RNAP-bound open complex. As explained in Section~\ref{sec:m0_appdx}, we are free
to discard one linearly dependent row from this matrix and replace it with the
normalization condition $p_R + p_E + p_C + p_O = 1$. Using normalization to
eliminate $p_R$ from the first row gives
\begin{equation}
p_E = (1 - p_C - p_O)\frac{k_R^-}{k_R^- + k_R^+}.
\end{equation}
If we substitute this in the third row, then the last two rows constitute two
equations in $p_C$ and $p_O$ given by
\begin{align}
k_P^+k_R^-(1-p_C-p_O) - (k_P^- + k_O)(k_R^+ + k_R^-) p_C &= 0
\\
k_O p_C - r p_O &= 0.
\end{align}
Solving for $p_C = p_O r/k_O$ in the second and substituting into the first
gives us our desired single equation in the single variable $p_O$, which is
\begin{equation}
k_P^+k_R^- - k_P^+k_R^-\left(1 + \frac{r}{k_O}\right)p_O
            - (k_P^- + k_O)(k_R^+ + k_R^-) \frac{r}{k_O}p_O = 0,
\end{equation}
and solving for $p_O$ we find
\begin{equation}
p_O = \frac{k_P^+ k_R^- k_O}{k_P^+ k_R^- k_O + r k_P^+ k_R^- +
                            r (k_P^- + k_O) (k_R^+ + k_R^-)}.
\label{eq:model3_pO}
\end{equation}
Once again $p_O$, the transcriptionally active state, is the only component of
$\langle\vec{m}^0\rangle$ that survives multiplication by $\mathbf{R_A}$, and
$\mathbf{R_A}\langle\vec{m}^0\rangle = r p_O$. So
\begin{equation}
\langle{m}\rangle =
    \frac{1}{\gamma}\vec{1}^\dagger\cdot\mathbf{R_A}\langle\vec{m}^0\rangle
= \frac{r}{\gamma}
    \frac{k_P^+ k_R^- k_O}{k_P^+ k_R^- k_O + r k_P^+ k_R^- +
                            r (k_P^- + k_O) (k_R^+ + k_R^-)},
\end{equation}
which equals Eq.~\ref{eq:model3_mean_m} in the main text.

\subsubsection{Fano factor}
To compute the Fano factor of the analogous constitutive promoter, we first
excise the repressor states and rates from the problem. More precisely, we
construct the matrix $(\mathbf{K} - \mathbf{R_D} + \mathbf{R_A} -
\gamma\mathbf{I})$ and substitute it into Eq.~\ref{eq:vecm_appdx_ref} which is
now
\begin{equation}
(\mathbf{K} - \mathbf{R_D} + \mathbf{R_A} - \gamma\mathbf{I})
    \langle\vec{m}\rangle
= \begin{pmatrix}
    -k_P^+ - \gamma & k_P^- & r \\
     k_P^+ & -k_P^- - k_O - \gamma & 0 \\
     0 & k_O & - r- \gamma
\end{pmatrix}
\begin{pmatrix} m_E \\ m_C \\ m_O \end{pmatrix}
= -r p_O \begin{pmatrix}1 \\ 0 \\ 0 \end{pmatrix}
\end{equation}
where we labeled the unbound, closed complex, and open complex components of
$\langle\vec{m}\rangle$ as $m_E$, $m_C$, and $m_O$, respectively. $p_O$ is given
by the limit of Eq.~\ref{eq:model3_pO} as $k_R^+\rightarrow 0$, which is
\begin{equation}
p_O = \frac{k_P^+  k_O}{k_P^+ (k_O + r) + r (k_P^- + k_O)}
\equiv \frac{k_P^+  k_O}{\mathcal{Z}},
\end{equation}
where we define $\mathcal{Z}$ for upcoming convenience as this sum of terms will
appear repeatedly. We can use the second equation to eliminate $m_E$, finding
$m_E = m_C(k_P^- + k_O + \gamma)/k_P^+$, and the third to eliminate $m_C$, which
is simply $m_C = m_O(r+\gamma)/k_O$. Substituting these both into the first
equation gives a single equation for the variable of interest, $m_O$,
\begin{equation}
-(k_P^+ + \gamma) (k_P^- + k_O + \gamma) (r + \gamma) m_O
    + k_P^- k_P^+ (r + \gamma) m_O + r k_P^+ k_O m_O = - r k_P^+ k_O p_O,
\end{equation}
which is solved for $m_O$ to give
\begin{equation}
m_O = p_O \frac{r k_P^+ k_O}
    {(k_P^+ + \gamma) (k_P^- + k_O + \gamma) (r + \gamma)
        -r k_P^+ k_O - k_P^- k_P^+ (r + \gamma)}.
\end{equation}
Expanding the denominator and canceling terms leads to
\begin{equation}
m_O = p_O \frac{r}{\gamma} \frac{k_P^+ k_O}
    {\mathcal{Z} + \gamma(k_P^+ + k_P^- + k_O + r) + \gamma^2}.
\end{equation}
Now $\vec{1}^\dagger\cdot\mathbf{R_A}\langle\vec{m}\rangle = r m_O$, and
$\langle{m}\rangle = rp_O/\gamma$, so if we substitute these two quantities into
Eq.~\ref{eq:fano_appdx_ref}, we will readily obtain the Fano factor as
\begin{equation}
\nu = 1 - \langle{m}\rangle
    + \frac{\vec{1}^\dagger\cdot\mathbf{R_A}\langle\vec{m}\rangle}
            {\gamma \langle{m}\rangle}
= 1 - \frac{r}{\gamma}p_O + \frac{m_O}{p_O}.
\end{equation}
Substituting, we see that
\begin{equation}
\nu = 1 - \frac{r}{\gamma} \frac{k_P^+ k_O}{\mathcal{Z}}
    + \frac{r}{\gamma}
    \frac{k_P^+ k_O}
            {\mathcal{Z} + \gamma(k_P^+ + k_P^- + k_O + r) + \gamma^2},
\end{equation}
and after simplifying, we obtain
\begin{equation}
\nu = 1 - \frac{r k_P^+ k_O}{\mathcal{Z}}
        \frac{k_P^+ + k_P^- + k_O + r + \gamma}
            {\mathcal{Z} + \gamma(k_P^+ + k_P^- + k_O + r) + \gamma^2},
\end{equation}
as stated in Eq.~\ref{eq:model3_fano} in the main text.

\subsubsection{Generalizing $\nu<1$ to more fine-grained models}
In the main text we argued that the convolution of multiple
exponential distributions should be a distribution with a smaller
fractional width than the corresponding exponential distribution
with the same mean.
This can be made more precise with a result from~\cite{Stewart2007},
who showed that the convolution of multiple gamma distributions (of which the
exponential distribution is a special case) is, to a very good approximation,
also gamma distributed. Using their Eq.~2 for the distribution of the
convolution, with shape parameters set to 1 to give exponential distributions,
the total waiting time distribution has a ratio of variance to squared mean
$\sigma^2/\mu^2 = \sum_i k_i^2/\left(\sum_i k_i\right)^2$, where the $k_i$ are
the rates of the individual steps. Clearly this is less than 1 and therefore the
total waiting time distribution is narrower than the corresponding exponential.

We also claimed in the main text that for a process whose waiting time
distribution is narrower than exponential, i.e., has $\sigma^2/\mu^2<1$,
the distribution of counts should be less variable than a Poisson
distribution, leading to a Fano factor $\nu<1$.
This we argue by analogy to photon statistics where it is known that
``antibunched'' arrivals, in other words more uniformly distributed in
time relative to uncorrelated arrivals, generally gives rise to
sub-Poissonian noise~\cite{Paul1982, Zou1990}. Although loopholes to this
result are known to exist, to our knowledge they appear to arise from
uniquely quantum effects so we do not expect they apply for our problem.
Nevertheless we refrain from elevating this equivalence of
kinetic cycles with sub-Poissonian noise to a ``theorem.''

\subsection{Nonequilibrium Model Four - ``Active'' and ``Inactive'' States}
\subsubsection{Mean mRNA}
The mathematical specification of this model is almost identical to model 2.
The matrix $\mathbf{K}$ is identical, as is $\mathbf{R_D}$. The only difference
is that now $\mathbf{R_A}=\mathbf{R_D}$, i.e., both are diagonal, in contrast to
model 2 where $\mathbf{R_A}$ has an off-diagonal element, as in
Eq.~\ref{eq:model2_matrices_appdx}. Then the analog of
Eq.~\ref{eq:model2_K-R+R_for_m0} for finding $\langle{m}^0\rangle$ is
\begin{equation}
\begin{pmatrix} -k_R^- & k_R^+ & 0 \\
        k_R^- & -k_R^+ -k^+ & k^-\\
        0 & k^+ & -k^-
\end{pmatrix}
\begin{pmatrix}
    p_R \\ p_I \\ p_A
\end{pmatrix}
= 0.
\end{equation}
In fact we need not do this calculation explicitly and can instead recycle the
calculation of mean mRNA $\langle{m}\rangle$ from model 2. The matrices are
identical except for the relabeling $k^- \longleftrightarrow (k_P^- + r)$, and
a careful look through the derivation of $\langle{m}\rangle$ for model 2 shows
that the parameters $k_P^-$ and $r$ only ever appear as the sum $k_P^- + r$. So
taking $\langle{m}\rangle$ from model 2, Eq.~\ref{eq:model2_meanm_appdx}, and
relabeling $(k_P^- + r) \rightarrow k^-$ gives us our answer for model four,
simply
\begin{equation}
\langle{m}\rangle = \frac{r}{\gamma}
        \frac{k^+ k_R^-} {k^+ k_R^- + k^- (k_R^+ + k_R^-)}.
\end{equation}

\subsubsection{Fano factor}
Likewise, for computing the Fano factor of this model we may take a shortcut.
Consider the constitutive model four from Figure~\ref{fig2:constit_cartoons} for
which we want to compute the Fano factor and compare it to nonequilibrium model
one of simple repression in Figure~\ref{fig1:means_cartoons}. Mathematically
these are exactly the same model, just with rates labeled differently and the
meaning of the promoter states interpreted differently. Furthermore,
nonequilibrium model 1 from Figure~\ref{fig1:means_cartoons} was the model
considered by Jones et.\ al.~\cite{Jones2014}, where they derived the Fano
factor for that model to be
\begin{equation}
\nu = 1 + \frac{r k_R^+}{(k_R^+ + k_R^-)(k_R^+ + k_R^- + \gamma)}.
\end{equation}
So recognizing that the relabelings $k_R^+ \rightarrow k^-$ and
$k_R^- \rightarrow k^+$ will translate this result to our model four from
Figure~\ref{fig2:constit_cartoons}, we can immediately write down our Fano
factor as
\begin{equation}
\nu = 1 + \frac{r k^-}{(k^- + k^+)(k^- + k^+ + \gamma)},
\end{equation}
as quoted in Eq.~\ref{eq:model4_fano} and in Figure~\ref{fig2:constit_cartoons}.
\section{Bursty promoter models - generating function solutions and numerics}
\label{sec:gen_fcn_appdx}

\subsection{Constitutive promoter with bursts}

\subsubsection{From master equation to generating function}

\begin{figure}[h!]
\centering
\includegraphics{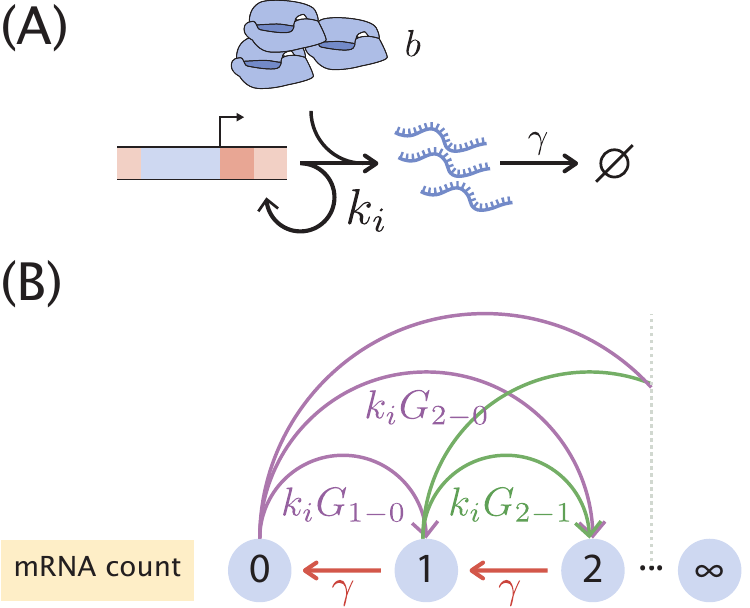}
\caption{
\textbf{Bursty transcription for unregulated promoter.}
(A) Schematic of the one-state bursty transcription model. Rate $k_i$ is the
bursty initiation rate, $\gamma$ is the mRNA degradation rate, and $b$ is the
mean burst size. (B) Schematic depiction of the mRNA count state transitions.
The model in (A) allows for transitions of $> 1$ mRNA counts with probability
$G_{m-m'}$, where the state jumps from having $m'$ mRNA to having $m$ mRNA in a
single burst of gene expression.}
\label{figS1:bursty_one_state}
\end{figure}

The objective of this section is to write down the steady-state mRNA
distribution for model 5 in Figure~\ref{fig2:constit_cartoons}. Our claim is
that this model is rich enough that it can capture the expression pattern of
bacterial constitutive promoters. Figure~\ref{figS1:bursty_one_state} shows two
different schematic representations of the model.
Figure~\ref{figS1:bursty_one_state}(A) shows the promoter cartoon model with
burst initiation rate $k_i$, mRNA degradation rate $\gamma$, and mean burst size
$b$. For our derivation of the chemical master equation we will focus more on
Figure~\ref{figS1:bursty_one_state}(B). This representation is intended to
highlight that bursty gene expression allows transitions between mRNA count $m$
and $m'$ even with $m - m' > 1$.

To derive the master equation we begin by considering the possible state
transitions to ``enter'' state $m$. There are two possible paths to jump from an
mRNA count $m' \neq m$ to a state $m$ in a small time window $\Delta t$:
\begin{enumerate}
        \item By degradation of a single mRNA, jumping from $m+1$ to $m$.
        \item By producing $m-m'$ mRNA for $m' \in \{0, 1, \ldots, m-1\}$.
\end{enumerate}
For the ``exit'' states from $m$ into $m' \neq m$ during a small time window
$\Delta t$ we also have two possibilities:
\begin{enumerate}
        \item By degradation of a single mRNA, jumping from $m$ to $m-1$.
        \item By producing $m'-m$ mRNA for $m'-m \in \{1, 2, \ldots\}$.
\end{enumerate}
This implies that the probability of having $m$ mRNA at time $t + \Delta t$ can
be written as
\begin{equation}
\begin{split}
p(m, t + \Delta t) = &p(m, t)
+ \overbrace{\gamma \Delta t (m + 1) p(m + 1, t)}^{m + 1 \rightarrow m}
- \overbrace{\gamma \Delta t m p(m, t)}^{m \rightarrow m - 1} \\
&+ \overbrace{k_i \Delta t \sum_{m'=0}^{m-1} G_{m-m'} p(m', t)}^
{m'\in \{0, 1, \ldots m-1\} \rightarrow m}
- \overbrace{k_i \Delta t \sum_{m'=m + 1}^{\infty} G_{m'-m} p(m, t)}^
{m \rightarrow m'\in \{m+1, m+2, \ldots\}},
\end{split}
\label{eq:si_master_deltat}
\end{equation}
where we indicate $G_{m'-m}$ as the probability of having a burst of size
$m'-m$, i.e. when the number of mRNAs jump from $m$ to $m' > m$ due to a single
mRNA transcription burst. We suggestively use the letter $G$ as we will assume
that these bursts sizes are geometrically distributed with parameter $\theta$.
This is written as
\begin{equation}
G_{k} = \theta (1 - \theta)^k\; \text{for } k \in \{0, 1, 2, \ldots \}.
\end{equation}
In Section~\ref{sec:beyond_means} of the main text we derive this functional
form for the burst size distribution. An intuitive way to think about it is that
for transcription initiation events that take place instantaneously there are
two competing possibilities: Producing another mRNA with probability $(1 -
\theta)$, or ending the burst with probability $\theta$. What this implies is
that for a geometrically distributed burst size we have a mean burst size $b$ of
the form
\begin{equation}
b \equiv \left\langle m' - m \right\rangle 
= \sum_{k=0}^\infty k \theta (1 - \theta)^k = {1 - \theta \over \theta}.
\end{equation}

To clean up Equation~\ref{eq:si_master_deltat} we can send the first term on the
right hand side to the left, and divide both sides by $\Delta t$. Upon taking
the limit where $\Delta t \rightarrow 0$ we can write
\begin{equation}
{d \over dt}p(m, t) = (m + 1) \gamma p(m + 1, t)
- m \gamma p(m, t)
+ k_i \sum_{m'=0}^{m-1} G_{m-m'} p(m', t) 
- k_i \sum_{m'=m + 1}^{\infty} G_{m'-m} p(m, t).
\end{equation}
Furthermore, given that the timescale for this equation is set by the mRNA
degradation rate $\gamma$ we can divide both sides by this rate, obtaining
\begin{equation}
{d \over d\tau}p(m, \tau) = (m + 1) p(m + 1, \tau)
- m p(m, \tau)
+ \lambda \sum_{m'=0}^{m-1} G_{m-m'} p(m', \tau) 
- \lambda \sum_{m'=m + 1}^{\infty} G_{m'-m} p(m, \tau),
\label{eq:si_master_ode}
\end{equation}
where we defined $\tau \equiv t \times \gamma$, and $\lambda \equiv k_i/\gamma$.
The last term in Eq.~\ref{eq:si_master_ode} sums all burst sizes except for a 
burst of size zero. We can re-index the sum to include this term, obtaining
\begin{equation}
\lambda \sum_{m'=m + 1}^{\infty} G_{m'-m} p(m, \tau) = \lambda p(m, t) \left[
\underbrace{\sum_{m'={m}}^{\infty}G_{m'-m}}
_{\text{re-index sum to include burst size zero}} -
\underbrace{G_0}_{\text{subtract extra added term}}\right].
\end{equation}
Given the normalization constraint of the geometric distribution, adding the
probability of all possible burst sizes -- including size zero since we
re-indexed the sum -- allows us to write
\begin{equation}
\sum_{m'=m}^{\infty}G_{m'-m} - G_0 = 1 - G_0.
\end{equation}
Substituting this into Eq.~\ref{eq:si_master_ode} results in
\begin{equation}
{d \over d\tau}p(m, \tau) = (m + 1) p(m + 1, \tau)
- m p(m, \tau)
+ \lambda \sum_{m'=0}^{m-1} G_{m-m'} p(m', \tau) 
- \lambda p(m, \tau) \left[ 1 - G_0 \right].
\label{eq:si_master_ode_2}
\end{equation}
To finally get at a more compact version of the equation notice that the third
term in Eq.~\ref{eq:si_master_ode_2} includes burst from size $m'-m = 1$ to size
$m' - m = m$. We can include the term $p(m, t) G_0$ in the sum which allows
bursts of size $m' - m = 0$. This results in our final form for the chemical
master equation
\begin{align}
{d \over d\tau}p(m, \tau) = 
(m + 1) p(m+1, \tau)
- m p(m, \tau) - 
\lambda p(m, \tau)
+ \lambda \sum_{m^\prime=0}^m G_{m-m^\prime} p(m^\prime, \tau).
\label{eq:si_master_unreg}
\end{align}

In order to solve Eq.~\ref{eq:si_master_unreg} we will use the generating
function method~\cite{vanKampen2007}. The probability generating function
is defined as
\begin{align}
F(z,t) = \sum_{m=0}^\infty z^m p(m,t),
\end{align}
where $z$ is just a dummy variable that will help us later on to obtain the
moments of the distribution. Let us now multiply both sides of
Eq.~\ref{eq:si_master_unreg} by $z^m$ and sum over all $m$
\begin{equation}
\sum_m z^m {d \over d\tau} p(m, \tau) = 
\sum_m z^m \left[ 
- m p(m, \tau) 
+ (m + 1) p(m + 1, \tau) 
+ \lambda \sum_{m' = 0}^m G_{m-m'} p(m', \tau) - \lambda p(m, \tau)
\right],
\end{equation}
where we use $\sum_m \equiv \sum_{m=0}^\infty$. We can distribute the sum and
use the definition of $F(z, t)$ to obtain
\begin{equation}
{d F(z, \tau) \over d\tau} =
- \sum_m z^m m p(m, \tau)
+ \sum_m z^m (m + 1) p(m + 1, \tau)
+ \lambda \sum_m z^m \sum_{m'=0}^m G_{m-m'} p(m', \tau)
- \lambda F(z, \tau).
\label{eq:si_generating_01}
\end{equation}
We can make use of properties of the generating function to write everything in
terms of $F(z, \tau)$: the first term on the right hand side of
Eq.~\ref{eq:si_generating_01} can be rewritten as
\begin{align}
\sum_{m} z^{m} \cdot m \cdot p(m, \tau) &=
\sum_{m} z \frac{\partial z^{m}}{\partial z} p(m, \tau), \\
&=\sum_{m} z \frac{\partial}{\partial z}\left(z^{m} p(m, \tau)\right), \\
&=z \frac{\partial}{\partial z}\left(\sum_{m} z^{m} p(m, \tau)\right), \\
&= z {\partial F(z, \tau) \over \partial z}.
\end{align}
For the second term on the right hand side of Eq.~\ref{eq:si_generating_01} we
define $k \equiv m + 1$. This allows us to write
\begin{align}
\sum_{m=0}^{\infty} z^{m} \cdot(m+1) \cdot p(m+1, \tau) &=
\sum_{k=1}^{\infty} z^{k-1} \cdot k \cdot p(k, \tau), \\
&=z^{-1} \sum_{k=1}^{\infty} z^{k} \cdot k \cdot p(k, \tau), \\
&=z^{-1} \sum_{k=0}^{\infty} z^{k} \cdot k \cdot p(k, \tau), \\
&=z^{-1} \left(z \frac{\partial F(z)}{\partial z}\right), \\
&=\frac{\partial F(z)}{\partial z}.
\end{align}

\begin{figure}[h!]
\centering
\includegraphics[width=5cm]{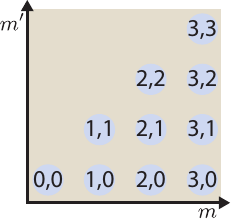}
\caption{\textbf{Reindexing double sum.} Schematic for reindexing the sum
$\sum_{m=0}^\infty \sum_{m'=0}^m$. Blue circles depict the 2D grid of
nonnegative integers restricted to the lower triangular part of the $m, m'$
plane. The trick is that this double sum runs over all $(m, m')$ pairs with
$m'\le m$. Summing $m$ first instead of $m'$ requires determining the
boundary: the upper boundary of the $m'$-first double sum becomes the
lower boundary of the $m$-first double sum.}
\label{figS2:sum_reindex}
\end{figure}

The third term in Eq.~\ref{eq:si_generating_01} is the most trouble. The trick
is to reverse the default order of the sums as
\begin{equation}
\sum_{m=0}^{\infty} \sum_{m'=0}^{m} = \sum_{m'=0}^{\infty} \sum_{m=m'}^{\infty}.
\end{equation}
To see the logic of the sum we point the reader to
Figure~\ref{figS2:sum_reindex}. The key is to notice that the double sum
$\sum_{m=0}^\infty \sum_{m'=0}^m$ is adding all possible pairs $(m, m')$ in the
lower triangle, so we can add the terms vertically as the original sum
indexing suggests, i.e.
\begin{equation}
\sum_{m=0}^{\infty} \sum_{m'=0}^{m} x_{(m, m')}= 
x_{(0, 0)} + x_{(1, 0)} + x_{(1, 1)} + x_{(2, 0)} + x_{(2, 1)} + x_{(2, 2)} + 
\ldots,
\end{equation}
where the variable $x$ is just a placeholder to indicate the order in which the
sum is taking place. But we can also add the terms horizontally as
\begin{equation}
\sum_{m'=0}^{\infty} \sum_{m=m'}^{\infty} x_{(m, m')} =
x_{(0, 0)} + x_{(1, 0)} + x_{(2, 0)} + \ldots + x_{(1,1)} + x_{(2, 1)} + \ldots,
\end{equation}
which still adds all of the lower triangle terms. Applying this reindexing
results in
\begin{align}
\lambda \sum_m z^m \sum_{m'=0}^m G_{m-m'} p(m', \tau) =
\lambda \sum_{m'=0}^{\infty} \sum_{m=m'}^{\infty} z^m 
\theta (1 - \theta)^{m-m'} p(m', \tau),
\end{align}
where we also substituted the definition of the geometric distribution $G_{k} =
\theta (1 - \theta)^k$. Redistributing the sums we can write
\begin{align}
\lambda \sum_{m'=0}^{\infty} \sum_{m=m'}^{\infty} z^m 
\theta (1 - \theta)^{m-m'} p(m', \tau) = 
\lambda \theta \sum_{n=0}^{\infty}(1-\theta)^{m'} P(m', \tau) 
\sum_{m=m'}^{\infty} \left[z (1-\theta)\right]^{m}.
\label{eq:si_generating_02}
\end{align}

The next step requires us to look slightly ahead into what we expect to obtain.
We are working on deriving an equation for the generating function $F(z, \tau)$
that when solved will allow us to compute what we care about, i.e. the
probability function $p(m, \tau)$. Upon finding the function for $F(z, \tau)$,
we will recover this probability distribution by evaluating derivatives
of $F(z, \tau)$ at $z=0$, whereas we can evaluate derivatives of
$F(z, \tau)$ at $z=1$ to instead recover the moments of the
distribution. The point here is that when the dust settles we
will evaluate $z$ to be less than or equal to one.
Furthermore, we know that the parameter of the geometric distribution $\theta$
must be strictly between zero and one. With these two facts we can safely state that
$| z (1 - \theta) | < 1$. Defining $n \equiv m - m'$ we rewrite the last sum in
Eq.~\ref{eq:si_generating_02} as
\begin{align}
\sum_{m=m'}^{\infty} \left[z (1-\theta)\right]^{m} &= 
\sum_{n=0}^{\infty} \left[z (1-\theta)\right]^{n + m'} \\
&= \left[ z (1 - \theta) \right]^{m'} 
\sum_{n=0}^{\infty} \left[ z (1 - \theta) \right]^{n} \\
&= \left[ z (1 - \theta) \right]^{m'} 
\left( {1 \over 1 - z (1 - \theta)} \right),
\end{align}
where we use the geometric series since, as stated before, $| z (1 - \theta) | <
1$. Putting these results together, the PDE for the generating function is
\begin{equation}
{\partial F \over \partial \tau} = 
{\partial F \over\partial z}
- z {\partial F \over \partial z} - \lambda F
+ \frac{\lambda\theta F}{1-z(1-\theta)}.
\end{equation}
Changing variables to $\xi=1-\theta$ and simplifying gives
\begin{align}
{\partial F \over \partial \tau} + (z - 1) {\partial F \over \partial z} = 
\frac{(z-1)\xi}{1-z\xi}\lambda F.
\label{eq:1state_unreg_015}
\end{align}

\subsubsection{Steady-state}

To get at the mRNA distribution at steady state we first must solve
Eq.~\ref{eq:1state_unreg_015} setting the time derivative to zero. At
steady-state, the PDE reduces to the ODE
\begin{align}
\deriv[F]{z} = \frac{\xi}{1-z\xi}\lambda F,
\end{align}
which we can integrate as
\begin{align}
\int \frac{dF}{F} = \int \frac{\lambda\xi dz}{1-\xi z}.
\end{align}
The initial conditions for generating functions can be subtle and confusing. The
key fact follows from the definition
$F(z,t) = \sum_m z^m p(m,t)$.
Clearly normalization of the distribution requires that
$F(z=1, t) = \sum_m p(m,t) = 1$.
A subtlety is that sometimes the generating function may be undefined
\textit{at} $z=1$, in which case the limit as $z$ approaches $1$ from
below suffices to define the normalization condition.
We also warn the reader that, while it is frequently convenient to change
variables from $z$ to a different independent variable, one must
carefully track how the normalization condition transforms.

Continuing on, we evaluate the integrals (producing a constant $c$) which gives
\begin{align}
\ln F &= -\lambda \ln(1-\xi z) + c
\\
F &= \frac{c}{(1-\xi z)^\lambda}.
\end{align}
Only one choice for $c$ can satisfy initial conditions, producing
\begin{align}
F(z) = \left(\frac{1-\xi}{1-\xi z}\right)^\lambda
        = \left(\frac{\theta}{1 - z(1-\theta)}\right)^\lambda,
\label{eq:gen_fn}
\end{align}

\subsubsection{Recovering the steady-state probability distribution}

To obtain the steady state mRNA distribution $p(m)$ we are aiming for we need to
extract it from the generating function
\begin{equation}
F(z) = \sum_m z^m p(m).
\end{equation}
Taking a derivative with respect to $z$ results in
\begin{equation}
{d F(z) \over dz} = \sum_m m z^{m - 1} p(m).
\end{equation}
Setting $z = 0$ leaves one term in the sum when $m = 1$ 
\begin{equation}
\left.\frac{d F(z)}{d z}\right|_{z=0} = 
\left(0 \cdot 0^{-1} \cdot p(0) 
+ 1 \cdot 0^0 \cdot p(1) 
+ 2 \cdot 0^1 \cdot p(2)
+ \cdots \right) = p(1),
\end{equation}
since in the limit $\lim_{x \rightarrow 0^+} x^x = 1$. A second derivative of
the generating function would result in
\begin{equation}
\frac{d^{2} F(z)}{d z^{2}} = \sum_{m=0}^{\infty} m(m-1) z^{m-2} p(m).
\end{equation}
Again evaluating at $z = 0$ gives 
\begin{equation}
\left.\frac{d^{2} F(z)}{d z}\right|_{z=0} = 2 p(z).
\end{equation}
In general any $p(m)$ is obtained from the generating function as
\begin{equation}
p(m) = {1 \over m!} \left. {d^m F(z) \over dz} \right\vert_{z=0}.
\label{eq:prob_from_gen}
\end{equation}

Let's now look at the general form of the derivative for our generating function
in Eq.~\ref{eq:gen_fn}. For $p(0)$ we simply evaluate $F(z=0)$ directly, 
obtaining
\begin{equation}
p(0) = F(z=0) = \theta^{\lambda}.
\end{equation}
The first derivative results in
\begin{equation}
\begin{aligned}
\frac{d F(z)}{d z} &=\theta^{\lambda} \frac{d}{d z}(1-z(1-\theta))^{-\lambda} \\
&=\theta^{\lambda}\left[-\lambda(1-z(1-f))^{-\lambda-1} \cdot(\theta-1)\right]\\
&=\theta^{\lambda}\left[\lambda(1-z(1-\theta))^{-\lambda-1}(1-\theta)\right].
\end{aligned}
\end{equation}
Evaluating this at $z=0$ as required to get $p(1)$ gives
\begin{equation}
\left.\frac{d F(z)}{d z}\right|_{z=0}=\theta^{\lambda} \lambda(1-\theta)
\end{equation}
For the second derivative we find
\begin{equation}
\frac{d^{2} F(z)}{d z^{2}} = \theta^{\lambda}
\left[\lambda(\lambda+1)(1-z(1-\theta))^{-\lambda-2}(1-\theta)^{2}\right].
\end{equation}
Again evaluating $z = 0$ gives
\begin{equation}
\left.\frac{d^{2} F(z)}{d z^{2}}\right|_{z=0} = 
\theta^{\lambda} \lambda(\lambda+1)(1-\theta)^{2}.
\end{equation}
Let's go for one more derivative to see the pattern. The third derivative of the
generating function gives
\begin{equation}
\frac{d^{3} F(z)}{d z^{3}} = 
\theta^{\lambda} 
\left[\lambda(\lambda+1)
(\lambda+2)(1-z(1-\theta))^{-\lambda-3}(1-\theta)^{3}\right],
\end{equation}
which again we evaluate at $z=0$
\begin{equation}
\left.\frac{d^{3} F(z)}{d z^{3}}\right|_{z=1} =
\theta^{\lambda}\left[\lambda(\lambda+1)(\lambda+2)(1-\theta)^{3}\right].
\end{equation}
If $\lambda$ was an integer we could write this as
\begin{equation}
\left.\frac{d^{3} F(z)}{d z^{3}}\right|_{z=0} = 
\frac{(\lambda+2) !}{(\lambda-1) !} \theta^{\lambda}(1-\theta)^{3}.
\end{equation}
Since $\lambda$ might not be an integer we can write this using Gamma functions
as
\begin{equation}
\left.\frac{d^{3} F(z)}{d z^{3}}\right|_{z=0} = 
\frac{\Gamma(\lambda+3)}{\Gamma(\lambda)} \theta^{\lambda}(1-\theta)^{3}.
\end{equation}
Generalizing the pattern we then have that the $m$-th derivative takes the form
\begin{equation}
\left.\frac{d^{m} F(z)}{d z^{m}}\right|_{z=0} =
\frac{\Gamma(\lambda+m)}{\Gamma(\lambda)} \theta^{\lambda}(1-\theta)^{m}.
\end{equation}
With this result we can use Eq.~\ref{eq:prob_from_gen} to obtain the desired
steady-state probability distribution function
\begin{equation}
p(m) = \frac{\Gamma(m+\lambda)}{\Gamma(m+1)\Gamma(\lambda)}
        \theta^\lambda (1-\theta)^m.
\end{equation}
Note that the ratio of gamma functions is often expressed as a binomial
coefficient, but since $\lambda$ may be non-integer, this would be ill-defined.
Re-expressing this exclusively in our variables of interest, burst rate
$\lambda$ and mean burst size $b$, we have
\begin{equation}
p(m) = \frac{\Gamma(m+\lambda)}{\Gamma(m+1)\Gamma(\lambda)}
        \left(\frac{1}{1+b}\right)^\lambda
        \left(\frac{b}{1+b}\right)^m.
\label{eq:nbinom_deriv_final}
\end{equation}

\subsection{Adding repression}
\subsubsection{Deriving the generating function for mRNA distribution}

Let us move from a one-state promoter to a two-state promoter, where one state
has repressor bound and the other produces transcriptional bursts as above.
A schematic of this model is shown as model 5 in
Figure~\ref{fig1:means_cartoons}(C). Although now we have an equation for each
promoter state, otherwise the master equation reads similarly to the one-state
case, except with additional terms corresponding to transitions between promoter
states, namely
\begin{align}
{d\over dt}p_R(m,t) =& 
k_R^+ p_A(m,t) - k_R^- p_R(m,t)
        + (m+1)\gamma p_R(m+1,t) - m\gamma p_R(m,t)
\\
\begin{split}
{d\over dt}p_A(m,t) =& - k_R^+ p_A(m,t) + k_R^- p_R(m,t)
        + (m+1)\gamma p_A(m+1,t) - m\gamma p_A(m,t) 
\\
&- k_i p_A(m,t) + k_i \sum_{m^\prime=0}^m \theta(1-\theta)^{m-m^\prime} p_A(m^\prime,t),
\end{split}
\end{align}
where $p_R(m,t)$ is the probability of the system having $m$ mRNA copies and
having repressor bound to the promoter at time $t$, and $p_A$ is an analogous
probability to find the promoter without repressor bound. $k_R+$ and $k_R^-$
are, respectively, the rates at which repressors bind and unbind to and from the
promoter, and $\gamma$ is the mRNA degradation rate. $k_i$ is the rate at which
bursts initiate, and as before, the geometric distribution of burst sizes has
mean $b=(1-\theta)/\theta$.

\begin{figure}
\centering
\includegraphics{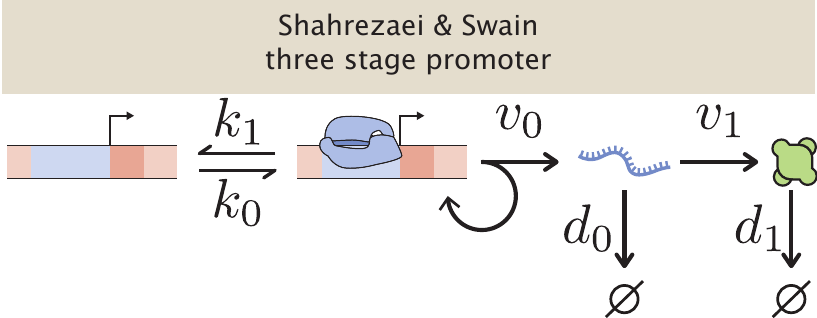}
\caption{\textbf{Schematic of three-stage promoter from~\cite{Shahrezaei2008}.}
Adapted from Shahrezaei \& Swain~\cite{Shahrezaei2008}. In their paper they
derive a closed form solution for the protein distribution. Our two-state bursty
promoter at the mRNA level can be mapped into their solution with some
relabeling.}
\label{fig:shahrezaei}
\end{figure}

Interestingly, it turns out that this problem maps exactly onto the three-stage
promoter model considered by Shahrezaei and Swain in~\cite{Shahrezaei2008}, with
relabelings. Their approximate solution for protein distributions amounts to the
same approximation we make here in regarding the duration of mRNA synthesis
bursts as instantaneous, so their solution for protein distributions also solves
our problem of mRNA distributions. Let us examine the analogy more closely. They
consider a two-state promoter, as we do here, but they model mRNA as being
produced one at a time and degraded, with rates $v_0$ and $d_0$. Then they model
translation as occurring with rate $v_1$, and protein degradation with rate
$d_1$ as shown in Figure~\ref{fig:shahrezaei}. Now consider the limit where
$v_1, d_0\rightarrow\infty$ with their ratio $v_1/d_0$ held constant. $v_1/d_0$
resembles the average burst size of translation from a single mRNA: these are
the rates of two Poisson processes that compete over a transcript, which matches
the story of geometrically distributed burst sizes. In other words, in our 
bursty promoter model we can think of the parameter $\theta$ as determining one
competing process to end the burst and $(1 - \theta)$ as a process wanting to
continue the burst. So after taking this limit, on timescales slow compared to
$v_1$ and $d_0$, it appears that transcription events fire at rate $v_0$ and
produce a geometrically distributed burst of translation of mean size $v_1/d_0$,
which intuitively matches the story we have told above for mRNA with variables
relabeled.

To verify this intuitively conjectured mapping between our problem and the
solution in~\cite{Shahrezaei2008}, we continue with a careful solution for the
mRNA distribution using probability generating functions, following the ideas
sketched in~\cite{Shahrezaei2008}. It is natural to nondimensionalize rates in
the problem by $\gamma$, or equivalently, this amounts to measuring time in
units of $\gamma^{-1}$. We are also only interested in steady state, so we set
the time derivatives to zero, giving
\begin{align}
0 =& k_R^+ p_A(m) - k_R^- p_R(m) + (m+1) p_R(m+1) - m p_R(m)
\\
\begin{split}
0 =& - k_R^+ p_A(m) + k_R^- p_R(m) + (m+1) p_A(m+1) - m p_A(m) 
\\
&- k_i p_A(m) + k_i \sum_{m^\prime=0}^m \theta(1-\theta)^{m-m^\prime} p_A(m^\prime),
\end{split}
\end{align}
where for convenience we kept the same notation for all rates, but these are
now expressed in units of mean mRNA lifetime $\gamma^{-1}$.
        
The probability generating function is defined as before in the constitutive
case, except now we must introduce a generating function for each promoter
state,
\begin{align}
f_A(z) = \sum_{m=0}^\infty z^m p_A(m),
\;
f_R(z) = \sum_{m=0}^\infty z^m p_R(m).
\end{align}
Our real objective is the generating function $f(z)$ that generates the mRNA
distribution $p(m)$, independent of what state the promoter is in. But since
$p(m) = p_A(m) + p_R(m)$, it follows too that $f(z) = f_A(z) + f_R(z)$.

As before we multiply both equations by $z^m$ and sum over all $m$. Each
individual term transforms exactly as did an analogous term in the constitutive
case, so the coupled ODEs for the generating functions read
\begin{align}
0 =& k_R^+ f_A(z) - k_R^- f_R(z) + \pderiv{z} f_R(z) - z \pderiv{z} f_R(z)
\\
\begin{split}
0 =&  - k_R^+ f_A(z) + k_R^- f_R(z) + \pderiv{z} f_A(z) - z \pderiv{z} f_A(z)
\\
&- k_i f_A(z) + k_i \frac{\theta}{1-z(1-\theta)} f_A(z),
\end{split}
\end{align}
and after changing variables $\xi = 1 - \theta$ as before and rearranging, we
have
\begin{align}
0 &= k_R^+ f_A(z) - k_R^- f_R(z) + (1-z) \pderiv{z} f_R(z)
\\
0 &=  - k_R^+ f_A(z) + k_R^- f_R(z) + (1 - z) \pderiv{z} f_A(z)
+ k_i \frac{(z-1)\xi}{1-z\xi} f_A(z),
\end{align}
We can transform this problem from two coupled first-order ODEs to a single
second-order ODE by solving for $f_A$ in the first and plugging into the second,
giving
\begin{align}
\begin{split}
0 = (1&-z) \pderiv[f_R]{z}
+ \frac{1-z}{k_R^+}
        \left(k_R^- \pderiv[f_R]{z} + \pderiv[f_R]{z} +(z-1) \psecderiv[f_R]{z}\right)
\\
&+ \frac{k_i}{k_R^+} \frac{(z-1)\xi}{1-z\xi}
        \left(k_R^- f_R + (z-1) \pderiv[f_R]{z}\right),
\end{split}
\end{align}
where, to reduce notational clutter, we have dropped the explicit $z$ dependence
of $f_A$ and $f_R$. Simplifying we have
\begin{align}
0 = \psecderiv[f_R]{z}
        - \left(\frac{k_i\xi}{1-z\xi}
                + \frac{1 + k_R^- + k_R^+}{1-z}
        \right)\pderiv[f_R]{z}
        + \frac{k_i k_R^- \xi}{(1-z\xi)(1-z)}f_R.
\end{align}
This can be recognized as the hypergeometric differential equation, with
singularities at $z=1$, $z=\xi^{-1}$, and $z=\infty$. The latter can be verified
by a change of variables from $z$ to $x=1/z$, being careful with the chain rule,
and noting that $z=\infty$ is a singular point if and only if $x=1/z=0$ is a
singular point.

The standard form of the hypergeometric differential equation has its
singularities at 0, 1, and $\infty$, so to take advantage of the standard form
solutions to this ODE, we first need to transform variables to put it into a
standard form. However, this is subtle. While any such transformation should
work in principle, the solutions are expressed most simply in the neighborhood
of $z=0$, but the normalization condition that we need to enforce corresponds to
$z=1$. The easiest path, therefore, is to find a change of variables that maps 1
to 0, $\infty$ to $\infty$, and $\xi^{-1}$ to 1. This is most intuitively done
in two steps.

First map the $z=1$ singularity to 0 by the change of variables $v=z-1$, giving
\begin{align}
0 = \psecderiv[f_R]{v}
        + \left(\frac{k_i\xi}{(1+v)\xi - 1}
                + \frac{1 + k_R^- + k_R^+}{v}
        \right)\pderiv[f_R]{v}
        + \frac{k_i k_R^- \xi}{((1+v)\xi - 1)v}f_R.
\end{align}
Now two singularities are at $v=0$ and $v=\infty$. The third is determined by
$(1+v)\xi -1 = 0$, or $v=\xi^{-1} - 1$. We want another variable change that
maps this third singularity to 1 (without moving 0 or infinity). Changing
variables again to $w=\frac{v}{\xi^{-1} - 1} = \frac{\xi}{1-\xi} v$ fits the
bill. In other words, the combined change of variables
\begin{align}
w = \frac{\xi}{1-\xi} (z-1)
\end{align}
maps $z = \{1, \xi^{-1}, \infty\}$ to $w =\{0, 1, \infty\}$ as desired. Plugging
in, being mindful of the chain rule and noting
$(1 + v)\xi - 1 = (1 - \xi)(w - 1)$ gives
\begin{align}
0 = \left(\frac{\xi}{1-\xi}\right)^2 \psecderiv[f_R]{w}
+ \left(
        \frac{\xi k_i}{(1-\xi)(w-1)} + \frac{\xi(1 + k_R^- + k_R^+)}{(1-\xi)w}
\right) \frac{\xi}{1-\xi} \pderiv[f_R]{w}
+ \frac{k_i k_R^- \xi^2}{(1-\xi)^2 w(w-1)}f_R.
\end{align}
This is close to the standard form of the hypergeometric differential equation,
and some cancellation and rearrangement gives
\begin{align}
0 = w(w-1)\psecderiv[f_R]{w}
+ \left(k_i w + (1 + k_R^- + k_R^+)(w-1)\right) \pderiv[f_R]{w}
+ k_i k_R^- f_R.
\end{align}
and a little more algebra produces
\begin{align}
0 = w(1-w)\psecderiv[f_R]{w}
+ \left(1 + k_R^- + k_R^+
        - (1 + k_i + k_R^- + k_R^+)w
\right) \pderiv[f_R]{w}
- k_i k_R^- f_R,
\end{align}
which is the standard form. From this we can read off the solution in terms of
hypergeometric functions ${_2F_1}$ from any standard source,
e.g.~\cite{Abramowitz1964}, and identify the conventional parameters in terms of
our model parameters. We want the general solution in the neighborhood of $w=0$
($z=1$), which for a homogeneous linear second order ODE must be a sum of two
linearly independent solutions. More precisely, 
\begin{align}
f_R(w) = C^{(1)} {_2F_1}(\alpha, \beta, \delta; w)
+ C^{(2)} w^{1-\delta}{_2F_1}(1+\alpha-\delta, 1+\beta-\delta, 2-\delta; w)
\end{align}
with parameters determined by
\begin{align}
\begin{split}
\alpha\beta &= k_i k_R^-
\\
1+\alpha+\beta &= 1+k_i+k_R^-+k_R^+
\\
\delta &= 1 + k_R^- + k_R^+
\end{split}
\end{align}
and constants $C^{(1)}$ and $C^{(2)}$ to be set by boundary conditions. Solving for
$\alpha$ and $\beta$, we find
\begin{align}
\begin{split}
\alpha &= \frac{1}{2}
\left(k_i+k_R^-+k_R^+ + \sqrt{(k_i+k_R^-+k_R^+)^2 - 4k_i k_R^-}\right)
\\
\beta &= \frac{1}{2}
\left(k_i+k_R^-+k_R^+ - \sqrt{(k_i+k_R^-+k_R^+)^2 - 4k_i k_R^-}\right)
\\
\delta &= 1 + k_R^- + k_R^+.
\end{split}
\end{align}
Note that $\alpha$ and $\beta$ are interchangeable in the definition of
${_2F_1}$ and differ only in the sign preceeding the radical.
Since the normalization condition requires that $f_R$ be finite at $w=0$,
we can immediately set $C^{(2)}=0$ to discard the second solution.
This is because all the rate constants are strictly positive,
so $\delta>1$ and therefore $w^{1-\delta}$ blows up as $w\rightarrow0$.
Now that we have $f_R$, we would like to find the generating function
for the mRNA distribution, $f(z) = f_A(z) + f_R(z)$.
We can recover $f_A$ from our solution for $f_R$, namely
\begin{align}
f_A(z) = \frac{1}{k_R^+}\left(k_R^- f_R(z) + (z-1) \pderiv[f_R]{z}\right)
\end{align}
or
\begin{align}
f_A(w) = \frac{1}{k_R^+}\left(k_R^- f_R(w) + w \pderiv[f_R]{w}\right),
\end{align}
where in the second line we transformed our original relation between
$f_R$ and $f_A$ to our new, more convenient, variable $w$.
Plugging our solution for $f_R(w) = C^{(1)}{_2F_1}(\alpha, \beta, \delta; w)$
into $f_A$, we will require the differentiation rule for ${_2F_1}$,
which tells us
\begin{align}
\pderiv[f_R]{w} = C^{(1)}\frac{\alpha\beta}{\delta}
                {_2F_1}(\alpha+1, \beta+1, \delta+1; w),
\end{align}
from which it follows that
\begin{align}
f_A(w) = \frac{C^{(1)}}{k_R^+}
\left(
k_R^- {_2F_1}(\alpha, \beta, \delta; w)
+ w\frac{\alpha\beta}{\delta} {_2F_1}(\alpha+1, \beta+1, \delta+1; w)
\right)
\end{align}
and therefore
\begin{align}
f(w) = C^{(1)}\left(1 + \frac{k_R^-}{k_R^+}\right)
        {_2F_1}(\alpha, \beta, \delta; w)
+ w \frac{C^{(1)}}{k_R^+} \frac{\alpha\beta}{\delta}
        {_2F_1}(\alpha+1, \beta+1, \delta+1; w).
\end{align}
To proceed, we need one of the (many) useful identities known for
hypergeometric functions, in particular
\begin{align}
w\frac{\alpha\beta}{\delta} {_2F_1}(\alpha+1, \beta+1, \delta+1; w)
=
(\delta-1)\left(
{_2F_1}(\alpha, \beta, \delta-1; w) - {_2F_1}(\alpha, \beta, \delta; w)
\right).
\end{align}
Substituting this for the second term in $f(w)$, we find
\begin{align}
f(w) = \frac{C^{(1)}}{k_R^+}
\left[
        \left(k_R^+ + k_R^-\right)
        {_2F_1}(\alpha, \beta, \delta; w)
+ (\delta-1)\left(
        {_2F_1}(\alpha, \beta, \delta-1; w) - {_2F_1}(\alpha, \beta, \delta; w)
        \right)
\right],
\end{align}
and since $\delta-1 = k_R^+ + k_R^-$, the first and third terms cancel,
leaving only
\begin{align}
f(w) = C^{(1)}\frac{k_R^+ + k_R^-}{k_R^+} {_2F_1}(\alpha, \beta, \delta-1; w).
\end{align}
Now we enforce normalization, demanding $f(w=0) = f(z=1) = 1$.
${_2F_1}(\alpha, \beta, \delta-1; 0) = 1$, so we must have
$C^{(1)} = k_R^+ / (k_R^+ + k_R^-)$ and consequently
\begin{align}
f(w) =  {_2F_1}(\alpha, \beta, k_R^+ + k_R^-; w).
\end{align}
Recalling that the mean burst size $b = (1-\theta)/\theta = \xi/(1-\xi)$
and $w = \frac{\xi}{1-\xi} (z-1) = b (z-1)$,
we can transform back to the original variable $z$ to find the tidy result
\begin{align}
f(z) =  {_2F_1}(\alpha, \beta, k_R^+ + k_R^-; b(z-1)),
\end{align}
with $\alpha$ and $\beta$ given above by
\begin{align}
\begin{split}
\alpha &= \frac{1}{2}
\left(k_i+k_R^-+k_R^+ + \sqrt{(k_i+k_R^-+k_R^+)^2 - 4k_i k_R^-}\right)
\\
\beta &= \frac{1}{2}
\left(k_i+k_R^-+k_R^+ - \sqrt{(k_i+k_R^-+k_R^+)^2 - 4k_i k_R^-}\right).
\end{split}
\end{align}
Finally we are in sight of the original goal. We can generate the steady-state
probability distribution of interest by differentiating the generating function,
\begin{align}
p(m) = m! \left.\frac{\partial^m}{\partial z^m} f(z) \right|_{z=0},
\end{align}
which follows easily from its definition. Some contemplation reveals that
repeated application of the derivative rule used above will produce products of
the form $\alpha(\alpha+1)(\alpha+2)\cdots(\alpha+m-1)$ in the expression for
$p(m)$ and similarly for $\beta$ and $\delta$. These resemble ratios of
factorials, but since $\alpha$, $\beta$, and $\delta$ are not necessarily
integer, we should express the ratios using gamma functions instead. More
precisely, one finds
\begin{align}
p(m) = \frac{
        \Gamma(\alpha + m)\Gamma(\beta + m)\Gamma(k_R^+ + k_R^-)
        }
        {
        \Gamma(\alpha)\Gamma(\beta)\Gamma(k_R^+ + k_R^- + m)
        }
\frac{b^m}{m!}{_2F_1}(\alpha+m, \beta+m, k_R^++k_R^-+m; -b)
\label{eq:p_m_bursty+rep_appdx}
\end{align}
which is finally the probability distribution we sought to derive.

\subsection{Numerical considerations and recursion formulas}
\subsubsection{Generalities}
We would like to carry out Bayesian parameter inference on FISH data
from~\cite{Jones2014}, using~\eq{eq:p_m_bursty+rep_appdx} as our
likelihood. This requires accurate (and preferably fast)
numerical evaluation of the hypergeometric function ${_2F_1}$,
which is a notoriously hard problem~\cite{Pearson2017, Gil2007},
and our particular needs here present an especial challenge as we show below.

The hypergeometric function is defined by its Taylor series as
\begin{align}
{_2F_1}(a,b,c;z) 
= \sum_{l=0}^\infty
\frac{\Gamma(a + l)\Gamma(b + l)\Gamma(c)}
        {\Gamma(a)\Gamma(b)\Gamma(c + l)}
\frac{z^l}{l!}
\end{align}
for $|z|<1$, and by analytic continuation elsewhere.
If $z\lesssim1/2$ and $\alpha$ and $\beta$ are not too large
(absolute value below 20 or 30),
then the series converges quickly and an accurate numerical representation is
easily computed by truncating the series after a reasonable number of terms.
Unfortunately, we need to evaluate ${_2F_1}$ over mRNA copy numbers fully out
to the tail of the distribution, which can easily reach 50, possibly 100.
From~\eq{eq:p_m_bursty+rep_appdx}, this means evaluating ${_2F_1}$
repeatedly for values of $a$, $b$, and $c$ spanning the full range
from $\mathcal{O}(1)$ to $\mathcal{O}(10^2)$,
even if $\alpha$, $\beta$, and $\delta$
in~\eq{eq:p_m_bursty+rep_appdx} are small,
with the situation even worse if they are not small.
A naive numerical evaluation of the series definition will be
prone to overflow and, if any of $a,b,c<0$, then some successive terms in the
series have alternating signs which can lead to catastrophic cancellations.

One solution is to evaluate ${_2F_1}$ using arbitrary precision arithmetic
instead of floating point arithmetic,
e.g., using the \texttt{mpmath} library in Python.
This is accurate but incredibly slow computationally.
To quantify how slow, we found that
evaluating the likelihood defined by~\eq{eq:p_m_bursty+rep_appdx} $\sim50$ times
(for a typical dataset of interest from~\cite{Jones2014},
with $m$ values spanning 0 to $\sim50$)
using arbitrary precision arithmetic is 100-1000 fold slower than
evaluating a negative binomial likelihood for the corresponding
constitutive promoter dataset.

To claw back $\gtrsim30$ fold of that slowdown, we can exploit
one of the many catalogued symmetries involving ${_2F_1}$.
The solution involves recursion relations originally explored by Gauss,
and studied extensively in~\cite{Pearson2017, Gil2007}.
They are sometimes known as contiguous relations and relate the values
of any set of 3 hypergeometric functions whose arguments differ by integers.
To rephrase this symbolically, consider a set of hypergeometric functions
indexed by an integer $n$,
\begin{align}
f_n = {_2F_1}(a+\epsilon_i n, b+\epsilon_j n, c+\epsilon_k n; z),
\end{align}
for a fixed choice of $\epsilon_i, \epsilon_j, \epsilon_k \in \{0,\pm 1\}$
(at least one of $\epsilon_i, \epsilon_j, \epsilon_k$ must be nonzero,
else the set of $f_n$ would contain only a single element).
Then there exist known recurrence relations of the form
\begin{align}
A_n f_{n-1} + B_n f_{n} + C_n f_{n+1} = 0,
\end{align}
where $A_n, B_n$, and $C_n$ are some functions of $a,b,c$, and $z$.
In other words, for fixed $\epsilon_i, \epsilon_j, \epsilon_k, a, b,$ and $c$,
if we can merely evaluate ${_2F_1}$ twice, say for $n^\prime$ and $n^\prime-1$,
then we can easily and rapidly generate values for arbitrary $n$.

This provides a convenient solution for our problem: we need repeated
evaluations of ${_2F_1}(a+m, b+m, c+m; z)$
for fixed $a,b$, and $c$ and many integer values of $m$.
They idea is that we can use arbitrary precision arithmetic to evaluate
${_2F_1}$ for just two particular values of $m$ and then generate
${_2F_1}$ for the other 50-100 values of $m$ using the recurrence
relation.
In fact there are even more sophisticated ways of utilizing the recurrence
relations that might have netted another factor of 2 speed-up, and
possibly as much as a factor of 10, but the method described here had
already reduced the computation time to an acceptable
$\mathcal{O}(\text{1 min})$, so these more sophisticated approaches did
not seem worth the time to pursue.

However, there are two further wrinkles.
The first is that
a naive application of the recurrence relation is numerically unstable.
Roughly, this is because the three term recurrence relations,
like second order ODEs, admit two linearly independent solutions.
In a certain eigenbasis, one of these solutions dominates the other
as $n\rightarrow\infty$, and as $n\rightarrow-\infty$,
the dominance is reversed.
If we fail to work in this eigenbasis, our solution of the recurrence relation
will be a mixture of these solutions and rapidly accumulate numerical error.
For our purposes, it suffices to know that the authors of~\cite{Gil2007}
derived the numerically stable solutions (so-called \textit{minimal solutions})
for several possible choices of $\epsilon_i, \epsilon_j, \epsilon_k$.
Running the recurrence in the proper direction using a minimal solution
is numerically robust and can be done entirely in floating point arithmetic, 
so that we only need to evaluate ${_2F_1}$ with arbitrary precision arithmetic
to generate the seed values for the recursion.

The second wrinkle is a corollary to the first.
The minimal solutions are only minimal for certain ranges of the argument $z$,
and not all of the 26 possible recurrence relations
have minimal solutions for all $z$.
This can be solved by using one of the many transformation formulae for
${_2F_1}$ to convert to a different recurrence relation that has
a minimal solution over the required domain of $z$, although
this can require some trial and error to find the right transformation,
the right recurrence relation, and the right minimal solution.

\subsubsection{Particulars}
Let us now demonstrate these generalities for our problem of interest.
In order to evaluate the probability distribution of our
model,~\eq{eq:p_m_bursty+rep_appdx}, we need to evaluate hypergeometric functions
of the form ${_2F_1}(\alpha+m, \beta+m, \delta+m; -b)$
for values of $m$ ranging from $0$ to $\mathcal{O}(100)$.
The authors of~\cite{Gil2007} did not derive a recursion relation
for precisely this case. We could follow their methods and do so ourselves,
but it is much easier to convert to a case that they did consider.
The strategy is to look through the minimal solutions tabulated
in~\cite{Gil2007} and search for a transformation we could apply to
${_2F_1}(\alpha+m, \beta+m, \delta+m; -b)$ that would place the $m$'s
(the variable being incremented by the recursion)
in the same arguments of ${_2F_1}$ as the minimal solution.
After some ``guess and check,'' we found that the transformation
\begin{align}
{_2F_1}(\alpha+m, \beta+m, \delta+m; -b)
=
(1+b)^{-\alpha-m}
        {_2F_1}\left(\alpha+m, \delta-\beta, \delta+m; \frac{b}{1+b}\right),
\label{eq:rec_euler_pretransform}
\end{align}
produces a ${_2F_1}$ on the right hand side that closely resembles
the minimal solutions $y_{3,m}$ and $y_{4,m}$ in Eq.~4.3 in~\cite{Gil2007}.
Explicitly, these solutions are
\begin{align}
y_{3,m}
&\propto
{_2F_1}\left(-\alpha^\prime + \delta^\prime - m,
                -\beta^\prime + \delta^\prime,
                1-\alpha^\prime-\beta^\prime+\delta^\prime-m;
                1-z\right)
\\
y_{4,m}
&\propto
{_2F_1}\left(\alpha^\prime + m,
                \beta^\prime,
                1+\alpha^\prime+\beta^\prime-\delta^\prime+m;
                1-z\right),
\label{eq:minimal_soln_sans_prefac}
\end{align}
where we have omitted prefactors which are unimportant for now.
Which of these two we should use depends on what values $z$ takes on.
Equating $1-z=b/(1+b)$ gives $z=1/(1+b)$, and since $b$ is strictly positive,
$z$ is bounded between 0 and 1.
From Eq.~4.5 in~\cite{Gil2007}, $y_{4,m}$ is the minimal solution
for real $z$ satisfying $0<z<2$, so this is the only minimal solution we need.

Now that we have our minimal solution,
what recurrence relation does it satisfy?
Confusingly, the recurrence relation of which $y_{4,m}$ is a solution
increments different arguments of ${_2F_1}$ that does $y_{4,m}$:
it increments the first only, rather than first and third.
This recurrence relation can be looked up, e.g., Eq.~15.2.10
in~\cite{Abramowitz1964}, which is
\begin{align}
(\delta^\prime - (\alpha^\prime + m)) f_{m-1}
+
(2(\alpha^\prime+m) - \delta^\prime + (\beta^\prime - \alpha^\prime)z)f_m
+ \alpha^\prime(z-1) f_{m+1} = 0.
\label{eq:chosen_rec_rel}
\end{align}
Now we must solve for the parameters appearing in the recurrence relation
in terms of our parameters, namely by setting
\begin{align}
\begin{split}
\alpha^\prime &= \alpha
\\
\beta^\prime &= \delta - \beta
\\
1 + \alpha^\prime + \beta^\prime - \delta^\prime &= \delta
\\
1 - z &= \frac{b}{1+b}
\end{split}
\end{align}
and solving to find
\begin{align}
\begin{split}
\alpha^\prime &= \alpha
\\
\beta^\prime &= \delta - \beta
\\
\delta^\prime &= 1 + \alpha - \beta
\\
z &= \frac{1}{1+b}
.
\end{split}
\end{align}
Finally we have everything we need. The minimal solution
\begin{align}
y_{4,m}
=
\frac{\Gamma(1+\alpha^\prime-\delta^\prime+m)}
        {\Gamma(1+\alpha^\prime+\beta^\prime-\delta^\prime+m)}
\times
{_2F_1}\left(\alpha^\prime + m,
                \beta^\prime,
                1+\alpha^\prime+\beta^\prime-\delta^\prime+m;
                1-z\right),
\end{align}
where we have now included the necessary prefactors,
is a numerically stable solution of the recurrence
relation~\eq{eq:chosen_rec_rel} if the recursion is run
from large $m$ to small $m$.

Let us finally outline the complete procedure as an algorithm to be implemented:
\begin{enumerate}
\item Compute the value of ${_2F_1}$ for the two
largest $m$ values of interest using arbitrary precision arithmetic.
\item Compute the prefactors to construct
$y_{4,\text{max}(m)}$ and $y_{4,\text{max}(m)-1}$.
\item Recursively compute $y_{4,m}$ for all $m$ less than $\text{max}(m)$ down
to $m=0$.
\item Cancel off the prefactors of the resulting values of
$y_{4,m}$ for all $m$ to produce ${_2F_1}$ for all desired $m$ values.
\end{enumerate}

With ${_2F_1}$ computed, the only remaining numerical danger in computing
$p(m)$ in~\eq{eq:p_m_bursty+rep_appdx} is overflow of the gamma functions.
This is easily solved by taking the log of the entire expression
and using standard routines to compute the log of the gamma functions,
then exponentiating the entire expression at the end if $p(m)$
is needed rather than $\log p(m)$.
\section{Bayesian inference}
\label{sec:bayesian}

\subsection{The problem of parameter inference}

One could argue that the whole goal of formulating theoretical models about
nature is to sharpen our understanding from qualitative statements to precise
quantitative assertions about the relevant features of the natural phenomena in
question \cite{Gunawardena2014}. It is in these models that we intend to distill
the essential parts of the object of study. Writing down such models leads to a
propagation of mathematical variables that parametrize our models. By assigning
numerical values to these parameters we can compute concrete predictions that
can be contrasted with experimental data. For these predictions to match the
data the parameter values have to carefully be chosen from the whole parameter
space. But how do we go about assessing the effectiveness of different regions
of parameter space to speak to the ability of our model to reproduce the
experimental observations? The language of probability, and more specifically of
Bayesian statistics is -- we think -- the natural language to tackle this
question.

\subsubsection{Bayes' theorem}

Bayes' theorem is a simple mathematical statement that can apply to \textit{any}
logical conjecture. For two particular events $A$ and $B$ that potentially 
depend on each other Bayes' theorem gives us a recipe for how to update our 
beliefs about one, let us say $B$, given some state of knowledge, or lack thereof, about
$A$. In its most classic form Bayes' theorem is written as
\begin{equation}
P(B \mid A) = {P(A \mid B) P(B) \over P(A)},
\end{equation}
where the vertical line $\mid$ is read as ``given that''. So $P(B \mid A)$ is
read as probability of $B$ given that $A$ took place. $A$ and $B$ can be any
logical assertion. In particular the problem of Bayesian inference focuses on
the question of finding the probability distribution of a particular parameter
value given the data.

For a given model with a set of parameters $\vec{\theta} = (\theta_1, \theta_2,
\ldots, \theta_n)$, the so-called \textit{posterior distribution} 
$P(\vec{\theta} \mid D)$, where $D$ is the experimental data, quantifies the
plausibility of a set of parameter values given our observation of some
particular dataset. In other words, through the application of Bayes' formula we
update our beliefs on the possible values that parameters can take upon learning
the outcome of a particular experiment. We specify the word ``update'' as we
come to every inference problem with prior information about the plausibility of
particular regions of parameter space even before performing any experiment.
Even when we claim as researchers that we are totally ignorant about the values
that the parameters in our models can take, we always come to a problem with
domain expertise that can be exploited. If this was not the case, it is likely
that the formulation of our model is not going to capture the phenomena we claim
to want to understand. This prior information is captured in the \textit{prior
probability} $P(\vec{\theta})$. The relationship between how parameter values
can connect with the data is enconded in the \textit{likelihood function} $P(D
\mid \vec{\theta})$. Our theoretical model, whether deterministic or
probabilistic, is encoded in this term that can be intuitively understood as the
probability of having observed the particular experimental data we have at hand
given that our model is parametrized with the concrete values $\vec{\theta}$. 
Implicitly here we are also conditioning on the fact that our theoretical model
is ``true,'' i.e. the model itself if evaluated or simulated in the computer is
capable of generating equivalent datasets to the one we got to observe in an 
experiment. In this way Bayesian inference consists of applying Bayes' formula 
as 
\begin{equation}
P(\vec{\theta} \mid D) \propto P(D \mid \vec{\theta}) P(\vec{\theta}).
\end{equation}
Notice than rather than writing the full form of Bayes' theorem, we limit 
ourselves to the terms that depend on our quantity of interest -- that is the 
parameter values themselves $\vec{\theta}$ -- as the denominator $P(D)$ only
serves as a normalization constant.

We also emphasize that the dichotomy we have presented between prior and
likelihood is more subtle. Although it is often stated that our prior
knowledge is entirely encapsulated by the obviously named prior
probability $P(\vec{\theta})$, this is usually too simplistic.
The form(s) we choose for our likelihood function
$P(D \mid \vec{\theta})$ also draw heavily on our prior domain expertise
and the assumptions, implicit and explicit, that these choices encode are
at least as important, and often inseparable from,
the prior probability, as persuasively argued in~\cite{Gelman2017}.

\subsubsection{The likelihood function}

As we alluded in the previous section it is through the likelihood function 
$P(D \mid \vec{\theta})$ that we encode the connection between our parameter 
values and the experimental observables. Broadly speaking there are two classes
of models that we might need to encode into our likelihood function:
\begin{itemize}
        \item Deterministic models: Models for which a concrete selection of
        parameter values give a single output. Said differently, models 
        with a one-to-one mapping between inputs and outputs.
        \item Probabilistic models: As the name suggests, models that, rather than
        having a one-to-one input-output mapping, describe the full
        probability distribution of possible outputs.
\end{itemize}
In this paper we focus on inference done with probabilistic models. After all,
the chemical master equations we wrote down describe the time evolutions of the
mRNA probability distribution. So all our terms $P(\vec{\theta} \mid D)$ will be
given by the steady-state solution of the corresponding chemical master equation
in question. This is rather convenient as we do not have to worry about adding a
statistical model on top of our model to describe deviations from the
predictions. Instead our models themselves focus on predicting such variation
in cell count.

\subsubsection{Prior selection}
The different models explored in this work embraced different levels of
coarse-graining that resulted in a diverse number of parameters for different
models. For each of these model configurations Bayes' theorem demands from us to
represent our preconceptions on the possible parameter values in the form of the
prior $P(\vec{\theta})$. Throughout this work for models with $> 1$ parameter we
assign independent priors to each of the parameters; this is
\begin{equation}
P(\vec{\theta}) = \prod_{i=1}^n P(\theta_i).
\end{equation}
Although it is not uncommon practice to use non-informative, or maximally
uninformative priors, we are of the mindset that this is a disservice to the
philosophical and practical implications of Bayes' theorem. It sounds almost
contradictory to claim that can we represent our thinking about a natural
phenomenon in the form of a mathematical model -- in the context of Bayesian
inference this means choosing a form for the likehihoods, and even making
this choice presupposes prior understanding or assumptions as to the
relevant features in the system under study -- but that we have absolutely
no idea what the parameter values could or could not be. We therefore make
use of our own expertise, many times in the form of order-of-magnitude
estimates, to write down weakly-informative prior distributions
for our parameters.

For our particular case all of the datasets from~\cite{Jones2014} used in this
paper have $\mathcal{O}(10^3)$ data points. What this implies is that our
particular choice of priors will not significantly affect our inference as long
as they are broad enough. A way to see why this is the case is to simply look at
Bayes' theorem. For $N ~ 1000-3000$ datum all of the independent of each other
and $n \ll 10^3$ parameters Bayes' theorem reads as
\begin{equation}
P(\vec{\theta} \mid D) \propto \prod_{k=1}^{N} P(d_k \mid \vec{\theta})
\prod_{i=1}^n P(\theta_i),
\end{equation}
where $d_k$ represents the $k$-th datum. That means that if our priors span a
wide range of parameter space, the posterior distribution would be dominated by
the likelihood function.

\subsubsection{Expectations and marginalizations}
For models with more than one or two parameters, it is generally difficult
to visualize or reason about the full joint posterior distribution
$P(\vec{\theta} \mid D)$ directly.
One of the great powers of Bayesian analysis is \textit{marginalization},
allowing us to reduce the dimensionality to only the parameters of
immediate interest by averaging over the other dimensions.
Formally, for a three dimensional model with parameters
$\theta_1$, $\theta_2$, and $\theta_3$, we can for instance
marginalize away $\theta_3$ to produce a 2D posterior as
\begin{equation}
P(\theta_1, \theta_2 \mid D) \propto
        \int_{\theta_3} d\theta_3 \,P(\theta_1, \theta_2, \theta_3 \mid D),
\end{equation}
or we can marginalize away $\theta_1$ and $\theta_3$ to produce the
1D marginal posterior of $\theta_2$ alone, which would be
\begin{equation}
P(\theta_2 \mid D) \propto
        \int_{\theta_1} d\theta_1 \int_{\theta_3} d\theta_3
        \,P(\theta_1, \theta_2, \theta_3 \mid D).
\end{equation}
Conceptually, this is what we did in generating the 2D slices of the
full 9D model in Figure~\ref{fig4:repressed_post_full}(A).
In practice, this marginalization is even easier with Markov Chain Monte Carlo
samples in hand. Since each point is simply a list of parameter values,
we simply ignore the parameters which we want to marginalize
away~\cite{Gelman2013}.
        
\subsubsection{Markov Chain Monte Carlo}
The theory and practice of Bayesian inference with Markov Chain Monte Carlo
(MCMC) is a rich subject with fascinating and deep analogies to statistical
mechanics, even drawing on classical Hamiltonian mechanics and general
relativity in its modern incarnations. We refer the interested reader
to~\cite{Gelman2013} and~\cite{Betancourt2018} for excellent introductions.
Here we merely give a brief summary of
the MCMC computations carried out in this work.

We used the Python package \texttt{emcee} for most of the MCMC sampling
in this work. For the constitutive promoter inference, we also ran
sampling with the excellent Stan modeling language as a check. We did not
use Stan for the inference of the simple repression model because
implementing the gradients of the hypergeometric function ${_2F_1}$
appearing in Eq.~\ref{eq:p_m_bursty+rep_appdx}, the probability
distribution for our bursty model with repression, would have been
an immensely challenging task. \texttt{emcee} was more than adequate for
our purposes, and we were perhaps lucky that the 9-D posterior model
for the model of simple repression with bursty promoter was quite well
behaved and did not require the extra power of the Hamiltonian Monte
Carlo algorithm provided by Stan~\cite{Carpenter2017}.
Source code for all statistical inference will be made available at
\url{https://github.com/RPGroup-PBoC/bursty_transcription}.

\subsection{Bayesian inference on constitutive promoters}
\label{sec:si_bayes_unreg}

Having introduced the ideas behind Bayesian inference we are ready to apply the
theoretical machinery to our non-equilibrium models. In particular in this
section we will focus on model 1 and model 5 in
Figure~\ref{fig2:constit_cartoons}(A). Model 1, the Poisson promoter, will help
us build practical intuition into the implementation of the Bayesian inference
pipeline as we noted in Section~\ref{sec:beyond_means} of the main text that
this model cannot be reconciled with experimental data from observables such as
the Fano factor. In other words, we acknowledge that this model is ``wrong,''
but we still see value in going through the analysis since the simple nature of
the model translates into a neat statistical analysis.

\subsubsection{Model 1 - Poisson promoter}

As specified in the main test, the mRNA steady-state distribution for model 1 in
Figure~\ref{fig2:constit_cartoons}(A) is Poisson with parameter $\lambda$.
Throughout this Appendix we will appeal to the convenient notation for
probability distributions of the form
\begin{equation}
m \sim \text{Poisson}(\lambda),
\end{equation}
where the simbol ``$\sim$'' can be read as \textit{is distributed according to}.
So the previous equation can be read as: the mRNA copy number $m$ is distributed
according to a Poisson distribution with parameter $\lambda$. Our objective then
is to compute the posterior probability distribution $P(\lambda \mid D)$, where,
as in the main text, $D = \{ m_1, m_2, \ldots, m_N \}$ are the data consisting
of single-cell mRNA counts. Since we can assume that each of the cells mRNA
counts are independent of any other cells, our likelihood function $P(D \mid
\lambda)$ consists of the product of $N$ Poisson distributions.

To proceed with the inference problem we need to specify a prior. In this case
we are extremely data-rich, as the dataset from Jones et.\ al~\cite{Jones2014}
has of order 1000-3000 single-cell measurements for each promoter, so our choice
of prior matters little here, as long as it is sufficiently broad. A convenient
choice for our problem is to use a \textit{conjugate} prior. A conjugate prior
is a special prior that causes the posterior to have the same functional form as
the prior, simply with updated model parameters. This makes calculations
analytically tractable and also offers a nice interpretation of the inference
procedure as updating our knowledge about the model parameters. This makes
conjugate priors very useful when they exist. The caveat is that conjugate
priors only exist for a very limited number of likelihoods, mostly with only one
or two model parameters, so in almost all other Bayesian inference problems, we
must tackle the posterior numerically.

But, for the problem at hand, a conjugate prior does in fact exist. For a
Poisson likelihood of identical and identically distributed data, the conjugate
prior is a gamma distribution, as can be looked up in, e.g.,~\cite{Gelman2013},
Section 2.6. Putting a gamma prior on $\lambda$ introduces two new parameters
$\alpha$ and $\beta$ which parametrize the gamma distribution itself, which we
use to encode the range of $\lambda$ values we view as reasonable. Recall
$\lambda$ is the mean steady-state mRNA count per cell, which \textit{a priori}
could plausibly be anywhere from 0 to a few hundred. $\alpha=1$ and $\beta=1/50$
achieve this, since the gamma distribution is strictly positive with mean
$\alpha/\beta$ and standard deviation $\sqrt{\alpha}/\beta$. To be explicit,
then, our prior is
\begin{equation}
\lambda \sim \text{Gamma}(\alpha, \beta)
\end{equation}

As an aside, note that if we did not know that our prior was a conjugate prior,
we could still write down our posterior distribution from its definition as
\begin{equation}
p(\lambda\mid D,\alpha,\beta)
\propto p(D\mid\lambda) p(\lambda \mid\alpha,\beta)
\propto \left(\prod_{k=1}^N \frac{\lambda^{m_k}e^{-\lambda}}{m_k!}\right)
        \frac{\beta}{\Gamma(\alpha)}(\beta\lambda)^{\alpha-1} e^{-\beta\lambda}
.
\end{equation}
Without foreknowledge that this in fact reduces to a gamma distribution, this
expression might appear rather inscrutable. When conjugate priors are
unavailable for the likelihood of interest - which is almost always the case for
models with $>1$ model parameter - this inscrutability is the norm, and making
sense of posteriors analytically is almost always impossible. Fortunately, MCMC
sampling provides us a powerful method of constructing posteriors numerically
which we will make use of extensively.

Since we did use a conjugate prior, we may simply look up our posterior in any
standard reference such as~\cite{Gelman2013}, Section 2.6,
from which we find that
\begin{equation}
\lambda
\sim \text{Gamma}\left(\alpha + \bar{m}N, \beta + N\right),
\end{equation}
where we defined the sample mean $\bar{m} = \frac{1}{N}\sum_k m_k$ for
notational convenience. A glance at the FISH data from~\cite{Jones2014} reveals
that $N$ is $\mathcal{O}(10^3)$ and $\langle m\rangle \gtrsim 0.1$ for all
constitutive strains in~\cite{Jones2014}, so $\bar{m}N \gtrsim 10^2$. Therefore
as we suspected, our prior parameters are completely overwhelmed by the data.
The prior behaves, in a sense, like $\beta$ extra ``data points''
with a mean value of $(\alpha-1)/\beta$~\cite{Gelman2013}, which
gives us some intuition for how much data is needed to overwhelm
the prior in this case: enough data $N$ such that $\beta\ll N$
and $\alpha/\beta \ll \bar{m}$. In
fact, $\bar{m}N$ and $N$ are so large that we can, to an excellent
approximation, ignore the $\alpha$ and $\beta$ dependence and approximate the
gamma distribution as a Gaussian with mean $\bar{m}$ and standard deviation
$\sqrt{\bar{m}/N}$, giving
\begin{equation}
\lambda
\sim \text{Gamma}\left(\alpha + \bar{m}N, \beta + N\right)
\approx \text{Normal}\left(\bar{m}, \sqrt{\frac{\bar{m}}{N}}\right).
\end{equation}
As an example with real numbers, for the \textit{lacUV5} promoter, Jones et.\
al~\cite{Jones2014} measured 2648 cells with an average mRNA count per cell of
$\bar{m} \approx 18.7$. In this case then, our posterior is
\begin{equation}
\lambda
\sim \text{Normal}\left(18.7, 0.08\right),
\label{eq:gauss_posterior}
\end{equation}
which suggests we have inferred our model's one parameter to a precision of
order 1\%.

This is not wrong, but it is not the full story. The model's posterior
distribution is tightly constrained, but is it a good generative model? In other
words, if we use the model to generate synthetic data in the computer does it
generate data that look similar to our actual data, and is it therefore
plausible that the model captures the important features of the data generating
process? This intuitive notion can be codified with \textit{posterior predictive
checks}, or PPCs, and we will see that this simple Poisson model fails badly.

The intuitive idea of posterior predictive checks is simple: 
\begin{enumerate}
\item Make a random draw of the model parameter $\lambda$ from the posterior
distribution.
\item Plug that draw into the likelihood and generate a synthetic dataset
$\{m_k\}$ conditioned on $\lambda$.
\item Repeat many times.
\end{enumerate}
More formally, the posterior predictive distribution can be thought of as the
distribution of future yet-to-be-observed data, conditioned on the data we have
already observed. Clearly if those data appear quite different, the model has a
problem. Put another way, if we suppose the generative model is true, i.e. we
claim that our model explains the process through which our observed
experimental data was generated, then the synthetic datasets we generate should
resemble the actual observed data. If this is not the case, it suggests the
model is missing important features. All the data we consider in this work are
1D (distributions of mRNA counts over a population) so empirical cumulative
distribution functions ECDFs are an excellent visual means of comparing
synthetic and observed datasets. In general for higher dimensional datasets,
much of the challenge is in merely designing good visualizations that can
actually show if synthetic and observed data are similar or not.

For our example Poisson promoter model then, we merely draw many random numbers,
say 1000, from the Gaussian posterior in Eq.~\ref{eq:gauss_posterior}. For each
one of those draws, we generate a dataset from the likelihood, i.e., we draw
2648 (the number of observed cells in the actual dataset) Poisson-distributed
numbers for each of the 1000 posterior draws, for a total of 2648000 samples
from the posterior predictive distribution.

To compare so many samples with the actual observed data, one excellent
visualization for 1D data is ECDFs of the quantiles, as shown for our Poisson
model in~\fig{fig:constit_post_full}(B) in the main text. 

\subsubsection{Model 5 - Bursty promoter}

Let us now consider the problem of parameter inference from FISH data for model
five from~\fig{fig1:means_cartoons}(C). As derived in
Appendix~\ref{sec:gen_fcn_appdx}, the steady-state mRNA distribution in this
model is a negative binomial distribution, given by
\begin{equation}
p(m) = \frac{\Gamma(m+k_i)}{\Gamma(m+1)\Gamma(k_i)}
        \left(\frac{1}{1+b}\right)^{k_i}
        \left(\frac{b}{1+b}\right)^m,
\label{eq:si_neg_bionom}
\end{equation}
where $b$ is the mean burst size and $k_i$ is the burst rate nondimensionalized
by the mRNA degradation rate $\gamma$. As sketched earlier, we can intuitively
think about this distribution through a simple story. The story of this
distribution is that the promoter undergoes geometrically-distributed bursts of
mRNA, where the arrival of bursts is a Poisson process with rate $k_i$ and the
mean size of a burst is $b$.

As for the Poisson promoter model, this expression for the steady-state mRNA
distribution is exactly the likelihood we want to use in Bayes' theorem. Again
denoting the single-cell mRNA count data as $D=\{m_1, m_2,\dots, m_N\}$, here
Bayes' theorem takes the form
\begin{equation}
p(k_i, b \mid D) \propto p(D\mid k_i,b)p(k_i, b),
\end{equation}
where the likelihood $p(D\mid k_i,b)$ is given by the product of $N$ negative
binomials as in Eq.~\ref{eq:si_neg_bionom}. We only need to choose priors on
$k_i$ and $b$. For the datasets from~\cite{Jones2014} that we are analyzing, as
for the Poisson promoter model above we are still data-rich so the prior's
influence remains weak, but not nearly as weak because the dimensionality of our
model has increased from one to two.

We follow the guidance of~\cite{Gelman2013}, Section 2.9 in opting for
weakly-informative priors on $k_i$ and $b$ (conjugate priors do not exist for
this problem), and we find ``street-fighting estimates''~\cite{Mahajan2010} to
be an ideal way of constructing such priors. The idea of weakly informative
priors is to allow all remotely plausible values of model parameters while
excluding the completely absurd or unphysical.

Consider $k_i$. Some of the strongest known bacterial promoters control rRNA
genes and initiate transcripts no faster than $\sim 1/\text{sec}$. It would be
exceedingly strange if any of the constitutive promoters from~\cite{Jones2014}
were stronger than that, so we can take that as an upper bound. For a lower
bound, if transcripts are produced too rarely, there would be nothing to see
with FISH. The datasets for each strain contain of order $10^3$ cells, and if
the $\langle m \rangle = k_i b/\gamma \lesssim 10^{-2}$, then the total number
of expected mRNA detections would be single-digits or less and we would have
essentially no data on which to carry out inference. So assuming $b$ is not too
different from 1, justified next, and an mRNA lifetime of $\gamma^{-1}\sim
3-5~\text{min}$, this gives us soft bounds on $k_i/\gamma$ of perhaps $10^{-2}$
and $3\times 10^1$.

Next consider mean burst size $b$. This parametrization of the geometric
distribution allows bursts of size zero (which could representing aborted
transcripts and initiations), but it would be quite strange for the mean burst
size $b$ to be below $\sim10^{-1}$, for which nearly all bursts would be of size
zero or one. For an upper bound, if transcripts are initiating at a rate
somewhat slower than rRNA promoters, then it would probably take a time
comparable to the lifetime of an mRNA to produce a burst larger than 10-20
transcripts, which would invalidate the approximation of the model that the
duration of bursts are instantaneous compared to other timescales in the
problem. So we will take soft bounds of $10^{-1}$ and $10^1$ for $b$.

Note that the natural scale for these ``street-fighting estimates'' was a log
scale. This is commonly the case that our prior sense of reasonable and
unreasonable parameters is set on a log scale. A natural way to enforce these
soft bounds is therefore to use a lognormal prior distribution, with the soft
bounds set $\pm2$ standard deviations from the mean.

\begin{figure}[p]
\centering
\includegraphics{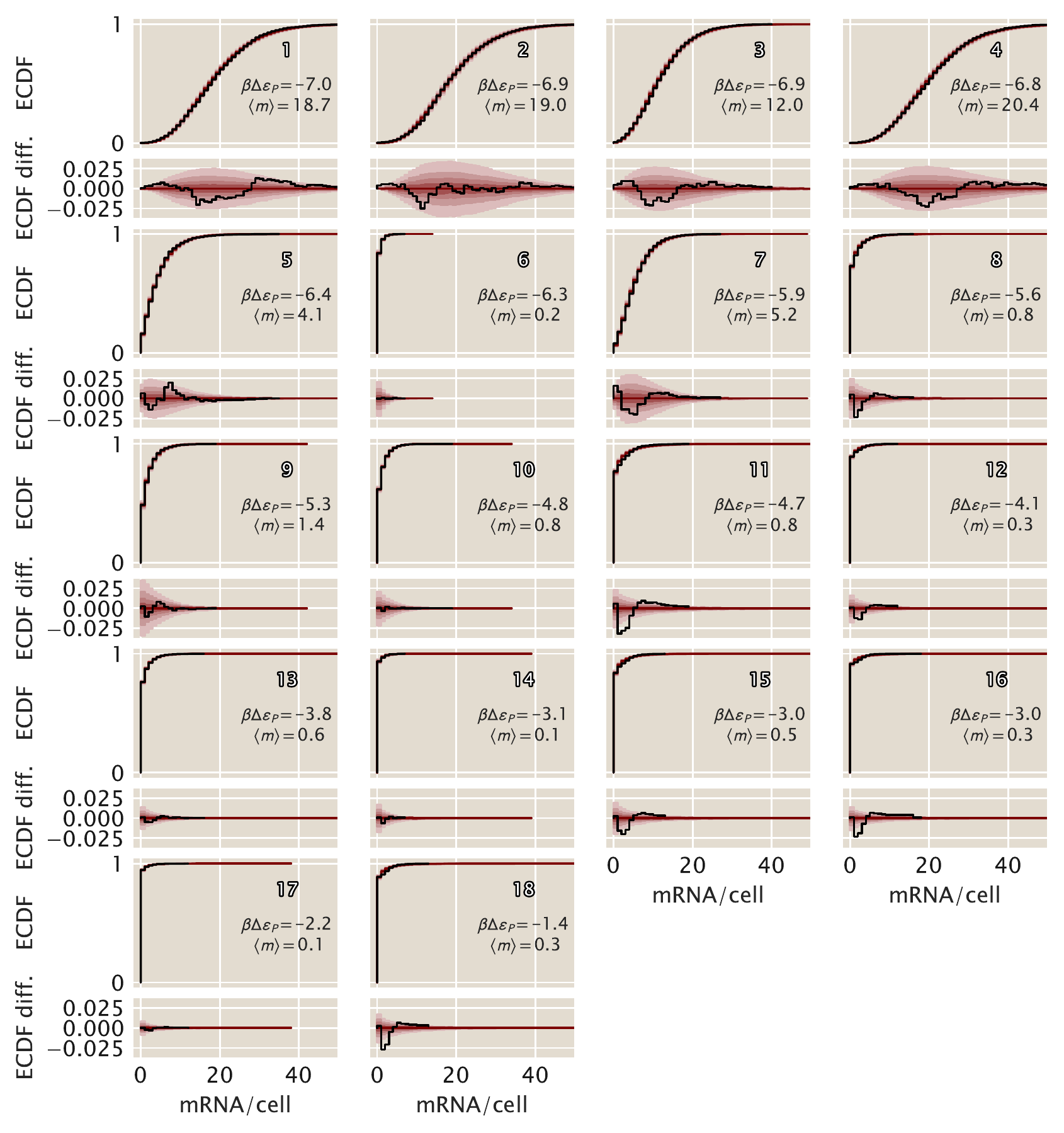}
\caption{\textbf{Theory-data comparison of inference on unregulated promoters.}
Comparison of the inference (red shaded area) vs the experimental measurements
(black lines) for 18 different unregulated promoters with different mean mRNA
expression levels from Ref.~\cite{Jones2014}. Upper panels show the empirical
cumulative distribution function (ECDF), while the lower panels show the
differences with respect to the median of the posterior samples. White numbers
are the same as in Figure~\ref{fig1:means_cartoons} for cross comparison. The
predicted binding energies $\beta\Delta\varepsilon_p$ were obtained from the
energy matrix model in Ref.~\cite{Brewster2012}}
\label{figS:ppc_unreg}
\end{figure}

With this, we are ready to write our full generative model as
\begin{equation}
\begin{split}
\ln k_i \sim \text{Normal}(-0.5, 2),
\\
\ln b \sim \text{Normal}(0.5, 1),
\\
m \sim \text{NBinom}(k_i, b).
\end{split}
\end{equation}
Section~\ref{section_04_bayesian_inference} in the main text details the
results of applying this inference to the single-cell mRNA counts data. There
we show the posterior distribution for the two parameters for different 
promoters. Figure~\ref{figS:ppc_unreg} shows the so-called posterior predictive
checks (see main text for explanation) for all 18 unregulated promoters shown
in the main text.

\subsection{Bayesian inference on the simple-repression architecture}

As detailed in~\ref{section_04_bayesian_inference} in the main text the
inference on the unregulated promoter served as a stepping stone towards our
ultimate goal of inferring repressor rates from the steady-state mRNA
distributions of simple-repression architectures. For this we expand the
one-state bursty promoter model to a two-state promoter as schematized in
Figure~\ref{fig1:means_cartoons}(C) as model 5. This model adds two new
parameters: the repressor binding rate $k^+$, solely function of the repressor
concentration, and the repressor dissociation rate $k^-$, solely a function of
the repressor-DNA binding affinity.

\begin{figure}[h!]
\centering
\includegraphics{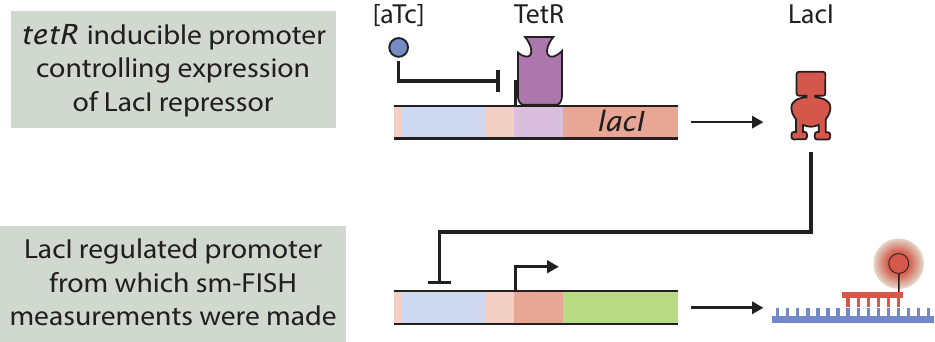}
\caption{\textbf{aTc controlled expression of LacI repressor.} Schematic of the
circuit used in~\cite{Jones2014} to control the expression of the LacI
repressor. The \textit{lacI} gene is under the control of the TetR repressor. As
the TetR repressor is inactivated upon binding of anhydrotetracycline or aTc,
the more aTc added to the media were cells are growing, the less TetR repressors
available to control the expression of the \textit{lacI} gene, resulting in more
LacI repressors per cell. LacI simultaneously controls the expression of the
mRNA on which single-molecule mRNA FISH was performed for gene expression
quantification.}
\label{figS:aTc_circuit}
\end{figure}

The structure of the data in~\cite{Jones2014} for regulated promoters tuned
these two parameters independently. In their work the production of the LacI
repressor was under the control of an inducible promoter regulated by the TetR
repressor as schematized in Figre~\ref{figS:aTc_circuit}. When TetR binds to the
small molecule anhydrotetracycline (aTc), it shifts to an inactive conformation
unable to bind to the DNA. This translates into an increase in gene expression
level. In other words, the higher the concentration of aTc added to the media,
the less TetR repressors that can control the expression of the \textit{lacI}
gene, so the higher the concentration of LacI repressors in the cell. So by 
tuning the amount of aTc in the media where the experimental strains were grown
they effectively tune $k^+$ in our simple theoretical model. On the other hand
to tune $k^-$ the authors swap three different binding sites for the LacI 
repressor, each with different repressor-DNA binding affinities previously 
characterized \cite{Garcia2011a}.

What this means is that we have access to data with different combinations of
$k^-$ and $k^+$. We could naively try to fit the kinetic parameters individually
for each of the datasets, but there is no reason to believe that the binding
site identity for the LacI repressor somehow affects its expression level
controlled from a completely different location in the genome, nor vice versa.
In other words, what makes the most sense it to fit all datasets together to
obtain a single value for each of the association and dissociation rates. What
this means, as described in Section~\ref{section_04_bayesian_inference} of the main text
is that we have a seven dimensional parameter space with four possible
association rates $k^+$ given the four available aTc concentrations, and three
possible dissociation rates $k^-$ given the three different binding sites
available in the dataset.

Formally now, denote the set of seven repressor rates to be inferred as
\begin{equation}
\vect{k} =\{k_{Oid}^-, k_{O1}^-, k_{O2}^-,
k_{0.5}^+, k_{1}^+, k_{2}^+, k_{10}^+\}.
\end{equation}
Note that since the repressor copy numbers are not known directly as explained
before, we label their association rates by the concentration of aTc. Bayes
theorem reads simply
\begin{equation}
p(\vect{k}, k_i, b \mid D)
\propto
p(D \mid\vect{k}, k_i, b) p(\vect{k}, k_i, b),
\end{equation}
where $D$ is the set of all $N$ observed single-cell mRNA counts across the
various conditions. We assume that individual single-cell measurements are
independent so that the likelihood factorizes as
\begin{equation}
p(D \mid\vect{k}, k_i, b)
= \prod_{j=1}^N p(m\mid \vect{k}, k_i, b)
= \prod_{j=1}^N p(m\mid k_j^+, k_j^-, k_i, b)
\end{equation}
where $k_j^\pm$ represent the appropriate binding and unbinding rates for the
$j$-th measured cell. Our likelihood function, previously derived in
Appendix~\ref{sec:gen_fcn_appdx}, is given by the rather complicated result in
Eq.~\ref{eq:p_m_bursty+rep_appdx}, which for completeness we reproduce here as
\begin{equation}
\begin{split}
p(m \mid k_R^+, k_R^-, k_i, b) = & ~\frac{
        \Gamma(\alpha + m)\Gamma(\beta + m)\Gamma(k_R^+ + k_R^-)
        }
        {
        \Gamma(\alpha)\Gamma(\beta)\Gamma(k_R^+ + k_R^- + m)
        }
\frac{b^m}{m!}
\\
&\times {_2F_1}(\alpha+m, \beta+m, k_R^++k_R^-+m; -b).
\end{split}
\label{eq:p_m_bursty+rep_infreprint}
\end{equation}
where $\alpha$ and $\beta$, defined for notational convenience, are
\begin{align}
\begin{split}
\alpha &= \frac{1}{2}
\left(k_i+k_R^-+k_R^+ + \sqrt{(k_i+k_R^-+k_R^+)^2 - 4k_i k_R^-}\right)
\\
\beta &= \frac{1}{2}
\left(k_i+k_R^-+k_R^+ - \sqrt{(k_i+k_R^-+k_R^+)^2 - 4k_i k_R^-}\right).
\end{split}
\end{align}

Next we specify priors. As for the constitutive model, weakly informative
lognormal priors are a natural choice for all our rates. We found that if the
priors were too weak, our MCMC sampler would often become stuck in regions of
parameter space with very low probability density, unable to move. We struck a
balance in choosing our prior widths between helping the sampler run while
simultaneously verifying that the marginal posteriors for each parameter were
not artificially constrained or distorted by the presence of the prior. The only
exception to this is the highly informative priors we placed on $k_i$ and $b$,
since we have strong knowledge of them from our inference of constitutive
promoters above.

With priors and likelihood specified we may write down our complete generative model as
\begin{equation}
\begin{split}
\log_{10}k_i &\sim \text{Normal}(0.725, 0.025)\\
\log_{10}b   &\sim \text{Normal}(0.55, 0.025)\\
\log_{10}k_{0.5}^+ &\sim \text{Normal}(-0.45, 0.3)\\
\log_{10}k_{1}^+   &\sim \text{Normal}(0.6, 0.3)\\
\log_{10}k_{2}^+   &\sim \text{Normal}(1.15, 0.3)\\
\log_{10}k_{10}^+  &\sim \text{Normal}(1.5, 0.3)\\
\log_{10}k_{Oid}^- &\sim \text{Normal}(-0.25, 0.3)\\
\log_{10}k_{O1}^-  &\sim \text{Normal}(0.1, 0.3)\\
\log_{10}k_{O2}^-  &\sim \text{Normal}(0.45, 0.3)\\
m &\sim \text{Likelihood}(k_R^+, k_R^-, k_i, b),
\end{split}
\end{equation}
where the likelihood is specified by Eq.~\ref{eq:p_m_bursty+rep_infreprint}.
We ran MCMC sampling on the full nine dimensional posterior specified
by this generative model.

We found that fitting a single operator/aTc concentration at a time with a
single binding and unbinding rate did not yield a stable inference for most of
the possible operator/aTc combinations. In other words, a single dataset could
not independently resolve the binding and unbinding rates, only their ratio as
set by the mean fold-change in Figure~\ref{fig1:means_cartoons} in the main
text. Only by making the assumption of a single unique binding rate for each
repressor copy number and a single unique unbinding rate for each binding site,
as done in Figure~\ref{fig4:repressed_post_full}(A), was it possible to
independently resolve the rates and not merely their ratios.

We also note that we found it necessary to exclude the very weakly and very
strongly repressed datasets from Jones et.\ al.~\cite{Jones2014}. In both cases
there was, in a sense, not enough information in the distributions for our
inference algorithm to extract, and their inclusion simply caused problems for
the MCMC sampler without yielding any new insight. For the strongly repressed
data (Oid, 10~ng/mL aTc), with $>$ 95\% of cells with zero mRNA, there was quite
literally very little data from which to infer rates. And the weakly repressed
data, all with the repressor binding site O3, had an unbinding rate so fast that
the sampler essentially sampled from the prior; the likelihood had negligible
influence, meaning the data was not informing the sampler in any meaningful way,
so no inference was possible.

		\printbibliography[title={Supplemental References},
		segment=\therefsegment, filter=notother]
	\end{refsegment}
\end{document}